\documentclass[review]{myelsarticle}
\usepackage{amsmath,amsfonts,natbib,hyperref}

\def\p{\partial}
\def\pb{\bar\partial}

\def\half{{1\over 2}}
\newcommand\Id{{\mathbb{I}}}

\def\wh{{\hat w}}
\def\ah{{\hat\alpha}}
\def\bh{{\hat\beta}}
\def\lh{{\hat\lambda}}
\def\bh{{\hat\beta}}
\def\gh{{\hat\gamma}}
\def\dh{{\hat\delta}}
\def\htt{{\hat\theta}}

\def\Str{{{\rm Str}\,}}
\def\tr{{\rm Tr}\,}

\def\a{\alpha}
\def\b{\beta}
\def\g{\gamma}
\def\d{\delta}
\def\e{\epsilon}

\def\l{\lambda}

\def\s{\sigma}
\def\p{\partial}
\def\t{{\theta}}
\def\th{{\hat\theta}}

\def\L{\Lambda}
\def\O{\Omega}

\def\D{\Delta}
\def\ah{{\hat \alpha}}

\def\wb {{\bar{w}}}
\def\zb {{\bar{z}}}
\def\nb {{\bar{\nabla}}}
\def\lnk{{\ln( {|k|^2\over \mu^2} )}}
\def\lfa{{\leftrightarrow}}

\def\ad{{\dot \a}}
\def\bd{{\dot \b}}

\def\aad{{\dot a}}

\def\bbd{{\dot b}}

\def\ZZ{{\mathbb{Z}}}
\def\RR{\mathbb{R}}
\def\cN{{\cal N}}

\def\cH{{\cal H}}
\def\cU{{\cal U}}
\def\cV{{\cal V}}
\def\cO{{\cal O}}
\def\mL{{\lambda_t}}

\def\str{{{\rm Str}\,}}
\def\Tr{{{\rm Tr}}}

\def\bea{\begin{eqnarray}}
\def\eea{\end{eqnarray}}

\newcommand\ba{\begin{array}}
\newcommand\ea{\end{array}}

\def\th{{\hat\theta}}
\def\a{\alpha}
\def\k{\kappa}

\def\ad{{\dot a}}

\def\bd{{\dot b}}
\def\s{\sigma}

\def\e{\epsilon}

\def\lh{{\hat\lambda}}
\def\dh{{\hat\delta}}
\def\gh{{\hat\gamma}}

\def\bh{{\hat\beta}}
\def\gh{{\hat\gamma}}

\def\p{\partial}
\def\pb{\bar\partial}

\def\nb{\bar\nabla}

\def\p{\partial}
\def\pb{\bar\partial}

\def\half{{1\over 2}}
\def\tilde{\widetilde}

\def\Id{{\mathbb{I}}}

\def\ah{{\widehat\alpha}}
\def\lh{{\widehat\lambda}}
\def\bh{{\widehat\beta}}
\def\gh{{\widehat\gamma}}
\def\dh{{\widehat\delta}}

\def\Str{{\textrm{Str}}}
\def\a{\alpha}
\def\b{\beta}
\def\g{\gamma}
\def\d{\delta}
\def\e{\epsilon}

\def\l{\lambda}

\def\s{\sigma}
\def\p{\partial}
\def\t{{\theta}}

\def\L{\Lambda}
\def\O{\Omega}

\def\D{\Delta}
\def\ah{{\hat \alpha}}

\def\ad{{\dot \a}}
\def\bd{{\dot \b}}

\def\ZZ{\ensuremath{\mathbb{Z}}}
\def\RR{\ensuremath{\mathbb{R}}}

\def\M{{\cal M}}

\newcommand{\be}{\begin{equation}}
\newcommand{\ee}{\end{equation}}

\begin{document}


\begin{frontmatter}
\title{Superstrings in AdS}
\author{Luca Mazzucato\fnref{fn1}}
\ead{lmazzucato@scgp.stonybrook.edu}
\address{Simons Center for Geometry and Physics \\
Stony Brook University\\ Stony Brook, NY 11794-3636, USA}
\fntext[fn1]{Now at the Department of Neurobiology and Behavior, Stony Brook University.}
\begin{abstract}
This is a comprehensive review of the worldsheet techniques for the quantization of type IIB superstring theory on the $AdS_5\times S^5$ background, using the pure spinor formalism. Particular emphasis is devoted to AdS/CFT applications, with several examples worked out in details. The review is self-contained and pedagogical. Table of contents: 1) Introduction; 2) Generalities; 3) Supercoset sigma models; 4) Quantum effective action and conformal invariance; 5) String spectrum; 6) Integrability.  \end{abstract}
\end{frontmatter}

\newpage


\tableofcontents


\clearpage

\section{Introduction}

One of the most striking features of gauge theories is the emergence of space-time geometry through the holographic principle. The most celebrated example of such phenomenon is the AdS/CFT correspondence \cite{Maldacena:1997re,Gubser:1998bc,Witten:1998qj}, according to which the maximally supersymmetric four-dimensional gauge theory (CFT) is equivalent to a theory of quantum gravity (type IIB superstring theory) on a space with negative cosmological constant in five dimensions, called Anti-de Sitter space (AdS). In this context, the energy scale in the Conformal Field Theory provides the holographic direction, related to the radial direction of AdS \cite{Douglas:2010rc}. The AdS/CFT is a strong/weak coupling duality and we can use this fact to explore new phenomena on both sides of the correspondence. We can use weakly coupled gravity in AdS to learn about the strongly coupled regime of gauge theories. But one can also use the weakly coupled four-dimensional CFT to study the stringy regime of quantum gravity in AdS. 

A complete solution of the four-dimensional CFT amounts to two steps. Firstly, computing the spectrum of anomalous dimensions of gauge invariant operators, which is equivalent to solving for the gauge theory two point-functions. Anomalous dimensions are pivotal quantities in a quantum field theory. In the theory of strong interactions, they determine the parton distribution functions. The second step in solving the CFT is computing the structure functions, which are encoded in the three-point functions of the gauge invariant field theory operators. We are on the right track to achieve the solution to the first problem: a conjecture has been made for the full spectrum at any value of the coupling, in the planar limit \cite{Gromov:2009tv}. The second problem has received much less attention so far. The aim of this review is to provide the technical tools required to study spectrum and correlation functions of the AdS/CFT system at strong coupling. We hope that future developments, along these lines, allow eventually to solve the problem at any coupling and in the full non-planar theory.
\medskip 

\noindent{\em Spectrum}

\noindent An impressive amount of progress has been achieved on the field theory side of the correspondence at weak gauge coupling, in particular regarding the study of the spectrum of anomalous dimensions of local gauge invariant CFT operators. The gauge theory contains half-BPS operators, whose classical conformal dimension is not renormalized. Most operators however are not protected:   their classical scaling dimension receives quantum corrections. While at weak coupling the anomalous dimensions have been studied since the beginning of quantum field theory, the strong coupling regime is typically very hard to tackle.
\smallskip

\noindent{\em Spinning strings}

\noindent By means of the AdS/CFT duality, we can learn about the strong coupling regime by studying a weakly coupled string theory dual. The first study of non-protected operators at strong coupling was performed for gauge theory operators with very large R-charge, dual to point-like strings rotating fast around the five sphere, the so called BMN sector \cite{Berenstein:2002jq}. The classical string configuration dual to long twist-two operators was studied in \cite{Gubser:2002tv}. Shortly after, one-loop integrability of the dilatation operator on the field theory side was discovered \cite{Minahan:2002ve}. Thanks to the indefatigable work of many research groups, this eventually led to a conjecture for the solution of the anomalous dimensions of all operators, by means of the so-called Y-system \cite{Gromov:2009tv}. Despite an impressive amount of work on the field theory side of the correspondence, it is fair to say that the string theory side still remains obscure.

In order to achieve a proof of the AdS/CFT conjecture, it is necessary to understand superstring theory in AdS. The type IIB supergravity background $AdS_5\times S^5$ is supported by a Ramond-Ramond five-form flux. Due to the presence of such flux, one cannot use the usual Ramond-Neveu-Schwarz (RNS) description of the superstring, but is forced to use a formalism with manifest target space supersymmetry, namely the Green-Schwarz (GS) formalism \cite{Metsaev:2000bj}. This action describes the classical worldsheet theory of superstrings propagating in AdS background and it is an interacting two-dimensional non-linear sigma model, whose target space is a supercoset.

The worldsheet action on such supercoset is classically integrable \cite{Bena:2003wd}. The discovery of integrability led eventually to the complete solution of the classical spectrum of solitonic strings moving in AdS in terms of an algebraic curve that encodes the conserved charges of the soliton \cite{Beisert:2005bm}.

The quantization of the two-dimensional worldsheet theory has been achieved in the sector of semi-classical strings, corresponding to gauge theory operators with very large quantum numbers. This sector can be studied using the GS formalism. Expanding around a solitonic solution of the sigma-model equations of motion, quantization can be achieved in the light-cone gauge. In this way, the spinning strings \cite{Frolov:2002av}, dual to twist-two operators, and the BMN sector \cite{Berenstein:2002jq} have been quantized. The algebraic curve, that encodes the classical spectrum of strings in AdS, can be quantized semi-classically around such solitonic sectors, providing a handle on the quantum properties of strings with very large quantum numbers \cite{Gromov:2011de}. The AdS/CFT correspondence has been tested in this sector to great extent, thanks to the Asymptotic Bethe Ansatz techniques \cite{Beisert:2006ez}.
\smallskip

\noindent{\em Perturbative strings}

\noindent The light-cone gauge for the GS formalism is hard to fix for short strings. This is the sector of perturbative strings with small quantum numbers, corresponding to short operators on the CFT side. It is the curved space analogue of the discrete tower of massive string states in flat space. The prototype example of such operators in ${\cal N}=4$ super Yang-Mills is the Konishi operator, that belongs to a long super-multiplet. If we want to study the quantization of the worldsheet sigma model in this sector, it is convenient to use the pure spinor formalism for the superstring instead \cite{Berkovits:2000fe}. The purpose of this review is to introduce in a pedagogical way the basic tools to achieve such a quantization.\footnote{The authors of \cite{Roiban:2011fe} and \cite{Gromov:2011de} recently extrapolated the spectrum of semi-classical strings all the way to the limit of strings with small quantum numbers, obtaining the same results as using the pure spinor formalism \cite{Vallilo:2011fj} and integrability \cite{Gromov:2009zb}. It is not clear how to generalize this approach beyond the first quantum correction.}

The pure spinor sigma-model is an interacting two-dimensional  conformal field theory, that can be quantized perturbatively  in a covariant way, by means of the standard background field method. The worldsheet metric is in the conformal gauge, which avoids the issues with the light-cone gauge. The fermionic kappa-symmetry of the Green-Schwarz action is replaced in the pure spinor formalism by a BRST symmetry, whose ghosts are target space pure spinors. Hence, the subtleties related to the gauge fixing of kappa-symmetry are absent in the pure spinor formalism. Hence, one can easily prove conformal invariance, vanishing of the central charge and consistency of the action (i.e. absence of gauge or BRST anomalies) at the loop level \cite{Vallilo:2002mh,Mazzucato:2009fv,Berkovits:2004xu}. 
One can show that the AdS/CFT dictionary between the radius of AdS and the gauge theory 't Hooft coupling is not renormalized at strong coupling \cite{Mazzucato:2009fv}.

The spectrum of short strings propagating in AdS can be studied using the usual technology of worldsheet vertex 
operators. The physical string states are primary vertex operators in the cohomology of the BRST charge. The massless closed string vertex operators describe type IIB supergravity in the $AdS_5\times S^5$ background and are dual to the BPS sector of $\cN=4$ SYM. The vertex operators corresponding to massive string states are dual to long supermultiplets on the gauge theory side. The perturbative string spectrum can be conveniently studied by expanding the sigma model around classical configurations with small quantum numbers.

At the classical level, the pure spinor action is integrable \cite{Vallilo:2003nx,Berkovits:2004jw}, just as the GS action is \cite{Bena:2003wd}.
In the pure spinor formalism, one can use the standard techniques of conformal field theory, combined with the power of BRST symmetry, to prove that integrability of the classical action persists quantum mechanically at all loops in the quantum theory. One can then construct a monodromy matrix and study its quantum properties and the associated Yang-Baxter equation \cite{Mikhailov:2007eg}.
\smallskip

\noindent {\em Correlation functions}

\noindent After computing the string spectrum, corresponding to the two-point functions of ${\cal N}=4$ SYM operators, one can go on and study the $n$-point correlator of local gauge invariant SYM operators. Correlation functions on the CFT side are dual to string scattering amplitudes on the AdS side, where each vertex operator insertion is dual to a specific gauge theory operator insertion. In order to compute the correlation functions at strong coupling, we need to use the string side of the correspondence and evaluate the string amplitude of vertex operators on the worldsheet. In the framework of the semi-classical GS formalism, because of the above mentioned difficulties, only correlation functions of very heavy (semi-classical) strings with light strings can been computed \cite{Janik:2010gc,Zarembo:2010rr,Costa:2010rz,Roiban:2010fe}. In the pure spinor formalism, one can compute correlation functions of any kind of strings, in particular of light strings, corresponding to short operators on the gauge theory side. 

\subsection{Guide to the review}

This review is self-contained and it includes all the background materials needed to be up and running with pure spinors in AdS. At the end of each Section, the reader can find a comprehensive set of references in the {\em Guide to the literature}. These include both the original articles where the topics in each Section have been first spelled out, as well as suggestions for the reader who wants to delve deeper into the open problems. The prerequisites are a basic knowledge of the worldsheet description of strings in flat space, in particular some familiarity with conformal field theory techniques, at the level taught in a first year graduate course on string theory. 

In Section~\ref{section:generalities} we introduce the pure spinor formalism for the superstring in the simplest case of flat ten dimensions. We start recalling the worldsheet variables used to describe a manifestly supersymmetric target 
superspace, then we introduce the pure spinor action and BRST charge. We explain the BRST cohomology equations for the massless vertex operators and mention the tree level scattering amplitude prescription and the subtleties related to the fact that the pure spinor conformal field theory describes a curved beta-gamma system. At the end of the Section we recall the GS and pure spinor non-linear sigma models in a curved type II supergravity background and the relation between superspace constraints and consistency of the worldsheet theory. In~\ref{appendix:notations}
the reader can find the conventions used for the ten-dimensional gamma matrices and the supervielbein. \ref{syminsuperspace} contains a very detailed discussion of the relation between the pure spinor BRST cohomology equations and the on-shell description of ten-dimensional SYM theory in superspace.

In Section~\ref{section:supercoset} we introduce the non-linear sigma models on supercosets. We first derive the curved AdS superspace geometry using the supergravity constraints and then discuss the issues related to the choice of the light-cone gauge in the GS formalism, as a motivation to introduce the pure spinor formalism. After a brief introduction to supergroups and supercosets, we construct the GS action on a supercoset that admits a $\ZZ_4$ grading and prove its kappa-symmetry. We finally introduce the pure spinor action in $AdS_5\times S^5$ and briefly mention the generalization to other interesting supercosets.
In~\ref{appendix:supergroups} the reader can find more details on supercosets and the structure constants of the superconformal algebra.

In Section~\ref{section:effectiveaction} we discuss the worldsheet quantum effective action in AdS. We achieve quantization using the standard techniques of the background field method, that we explain at length in the examples of the bosonic and the RNS superstrings in a curved background. We then compute the one-loop quantum effective action for the pure spinor superstring in AdS and show that its divergent part vanishes, proving one-loop conformal invariance. This computation is spelled out in full details and it is a pedagogical way to learn the methods needed for more advanced goals. We then show that the finite renormalization part of the effective action can be removed with local counterterms, which implies that the AdS radius is not renormalized by $\a'$ effects. We show also that the one-loop Weyl central charge vanishes. Then we compute the algebra of OPE of the worldsheet left-invariant currents and find that it is not a chiral algebra, but left- and right-moving currents mix in the OPE. Finally, we extend the proof of conformal invariance and absence of BRST and gauge anomalies to all loops in the sigma model perturbation theory. In~\ref{appendix:worldsheet} we derive several results pertaining this Section, that we omitted from the main discussion to ease the reading.

In Section~\ref{section:spectrum}, we give some examples of physical string states. We start with a short review of some aspects of the AdS/CFT correspondence, related to the spectrum of anomalous dimensions of local gauge invariant operators in the gauge theory. We first discuss the massless vertex operators, corresponding to type IIB supergravity compactified on the five-sphere and dual to the half BPS sector of $\cN=4$ SYM theory. As an example of a massive string state, we derive the energy of a vertex operator at the first massive string level, that is dual to a member of the Konishi multiplet, the simplest example of a long super-multiplet of $\cN=4$ SYM. Since this operator is not protected, its anomalous dimension is non-vanishing. We compute the one-loop corrections to the Virasoro constraint, whose solution determines the string energy, which in turn gives the anomalous dimension. The result confirms its earlier conjectured value, obtained using integrability. Finally, we briefly discuss the $n$-point correlation functions of gauge invariant local operators in $\cN=4$ SYM, dual to the $n$-point function of vertex operators on the string worldsheet. We construct the zero mode measure for the worldsheet variables and show that there is a well-defined higher genus amplitude prescription, which computes $1/N_c$ corrections to the planar limit of the AdS/CFT system. In the course of the discussion we introduce the $b$ antighost (which couples to the worldsheet Beltrami differential) in  AdS, whose construction is significantly easier than its flat space cousin, due to the Ramond-Ramond flux present in the AdS background.

In Section~\ref{section:integrability}, we discuss the integrability of the pure spinor sigma model in AdS. After deriving the Lax representation of the worldsheet equations of motion, we show the interplay between BRST cohomology and higher conserved charges. Then we pass to the quantum theory and prove that the monodromy matrix is not renormalized at one-loop and that the higher charges are conserved at all-loops in the sigma model perturbation theory.

There are several reviews covering general aspects of the pure spinor formalism: \cite{Berkovits:2002zk,Oz:2008zz,Bedoya:2009np} have some overlap with Section~\ref{section:generalities} and \ref{section:supercoset}; \cite{Guttenberg:2008ic} is review of the pure spinor formalism in a generic type II supergravity background, covering in far greater detail the topics in Section~\ref{section:curved}; \cite{Mafra:2009wq} covers in detail string scattering amplitudes in flat space.

\subsection{Open problems}

The pure spinor formulation of superstrings in AdS provides the tools to address a number of long standing problems in the AdS/CFT correspondence. Some of them are:
\begin{itemize}
\item {\em Half-BPS sector:} The most urgent open problem concerns the characterization of the full string spectrum in AdS. There are a number of steps to follow along this line of research. We already know that the worldsheet massless vertex operators correctly describe type IIB supergravity on the $AdS_5\times S^5$ background, as we explain in Section \ref{section:sugra}. However, we would like to rewrite such vertex operators in a way that allows to identify each single AdS supergravity field, together with their KK towers, e.g. as worked out in the classical paper \cite{Kim:1985ez}. In this way we obtain a one-to-one correspondence between each gauge invariant local operators in $\cN=4$ SYM theory and a single string vertex operator. For instance, we want to have a vertex operator that describes the gauge theory chiral primaries $\tr Z^p$ (where $Z=\phi^5+i\phi^6$ and $\phi^i$ are the six scalars in the $\cN=4$ supermultiplet), in the same spirit as in the canonical $AdS_3$ example \cite{Kutasov:1999xu}.
\item {\em Massive spectrum}: After understanding fully the half-BPS sector, the next step is to solve for the massive spectrum. One can easily generalize the example of the Konishi multiplet we discussed at length in Section \ref{massiveK}, to solve for the full massive spectrum in the sector of ``short'' strings, whose energy scales as $\sqrt[4]{\l_t}$. This algorithm can be easily extended to higher loops in the sigma model perturbation theory, implementing the background field method on a computer. 
\item {\em Integrability and the spectrum:} An alternative path to the solution of the spectral problem would make use of integrability techniques, which we do not apply in Section \ref{massiveK}. As explained in Section \ref{section:integrability}, the pure spinor sigma model in AdS is integrable at all-loops and BRST invariant. Understanding the interplay between these two aspects will probably give a handle to achieve the all-loop solution of the spectral problem in an elegant way and, hopefully, derive the Y-system from first principles.
\item {\em Scattering amplitudes:} Once we have some vertex operators at hand, we can stick them into a correlator and compute $n$-point functions at genus $g$, which will give the $1/N_c$ expansion of the $\cN=4$ SYM correlators at strong coupling. The worldsheet prescription for string scattering amplitudes is explained in Section \ref{section:amplitudes}. This line of research is mostly unexplored and will give us new insight into the non-planar features of AdS/CFT.
\item {\em Worldsheet derivation of AdS/CFT:} The open/closed string duality between three-dimensional Chern-Simons theory and topological string theory on a Calabi-Yau manifold, proposed by Gopakumar and Vafa  \cite{Gopakumar:1998ki}, has been derived by worldsheet techniques using the hybrid formalism for the superstring \cite{Ooguri:2002gx,Berkovits:2003pq}. This relies on a formulation of the worldsheet as a gauged linear sigma model. The pure spinor sigma model in AdS admits such a gauged linear sigma model reformulation, in what is believed to be the zero-radius limit of AdS \cite{Berkovits:2007zk,Berkovits:2007rj,Berkovits:2008qc,Linch:2008nt}. This is a very promising idea that may lead to the worldsheet proof of AdS/CFT correspondence using open/closed duality techniques.
\item {\em Deformations of AdS:} There are many different examples of gauge/gravity duality, realized in different ten-dimensional supergravity backgrounds, with less than maximal supersymmetry. An obvious application of the pure spinor formalism in AdS is to construct a worldsheet action for such backgrounds. This has been already done only in the case of the beta-deformation of $\cN=4$ SYM theory, that preserves eight supercharges \cite{Bedoya:2010qz}. 
\item {\em Fermionic T-duality:} An intriguing new symmetry of $\cN=4$ SYM theory, called dual superconformal symmetry, manifests itself on the string theory side as a new form of T-duality, called fermionic T-duality \cite{Berkovits:2008ic,Beisert:2008iq}, which is a manifestation of integrability in disguise. Using the pure spinor formalism, it is possible to study the $1/\sqrt{\l_t}$ and $1/N_c$ corrections to the strong-coupling, planar limit of fermionic T-duality.
Worldsheet fermionic T-duality is applicable to any background with abelian super-isometries and it would be of great interest to study this symmetry in full generality. It has a potential of connecting different string theory backgrounds that were previously regarded as unrelated. \end{itemize}

\subsection{Omitted topics}

Some of the most recent and speculative developments are not discussed in this review.

I am not discussing the new kind of perturbative worldsheet duality, called fermionic T-duality, that has been found in the GS and pure spinor superstring sigma model in AdS \cite{Berkovits:2008ic,Beisert:2008iq}. This is the string theory manifestation of the dual superconformal symmetry, a intriguing property of $\cN=4$ SYM theory scattering amplitudes, related to integrability \cite{Alday:2008yw}.

A second topic I omitted is the recent insight towards a proof of the AdS/CFT conjecture that has been obtained using the pure spinor formalism. In the AdS background, unlike the one in flat space, one can derive the pure spinor sigma model by gauge fixing a classical $G/G$ 
principal chiral model \cite{Berkovits:2008qc}, where $G$ is the $PSU(2,2|4)$ isometry supergroup. Even if naively such model looks trivial, the gauge symmetry group has non-compact directions, whose proper gauge-fixing leads to a non-trivial topological sigma model. This is related to the existence of a truncation of the full superstring sigma model to a topological subsector, conjectured to describe the zero-radius limit of AdS/CFT, namely free SYM theory \cite{Berkovits:2007zk,Berkovits:2007rj,Berkovits:2008qc,Linch:2008nt,Bonelli:2008rv}. This is a very promising avenue towards the final goal of having a worldsheet derivation of AdS/CFT from first principles, along the lines of the Gopakumar-Vafa construction \cite{Ooguri:2002gx,Berkovits:2003pq}.

\clearpage


\section{Generalities}
\label{section:generalities}

In this Section, we introduce the pure spinor formalism for the superstring in the simplest case of flat ten dimensions. We start recalling the worldsheet variables used to describe a manifestly supersymmetric target 
superspace, then we introduce the pure spinor action and BRST charge. We explain the BRST cohomology equations for the massless vertex operators and mention the tree level scattering amplitude prescription and the subtleties related to the fact that the pure spinor conformal field theory describes a curved beta-gamma system. At the end of the Section we recall the GS and pure spinor non-linear sigma models in a curved type II supergravity background and the relation between superspace constraints and consistency of the worldsheet theory. In~\ref{appendix:notations}
the reader can find the conventions used for the ten-dimensional gamma matrices and the supervielbein. \ref{syminsuperspace} contains a very detailed discussion of the relation between the pure spinor BRST cohomology equations and the on-shell description of ten-dimensional SYM theory in superspace.

\subsection{Superstring in flat space}

In this Section we will briefly review the salient features of the pure spinor formulation of the superstring in a flat ten-dimensional
target space.

\subsubsection{Open superstring}

We will start with the construction of
the open pure spinor superstring, or more precisely the holomorphic part of the closed pure spinor superstring.

The matter sector is described by the supermanifold
$(x^m,\theta^{\alpha})$, where $x^m, m=0,...,9$ are commuting
coordinates with the OPE
\be
x^m(z)x^n(0) \sim -\eta^{mn} \log|z|^2 \ ,
\ee
and $\theta^{\alpha}, \alpha=1,...,16$ are real
worldsheet scalars anti-commuting coordinates.
$x^m$ transform in the vector representation of the target space Lorentz group $SO(1,9)$, while
$\theta^{\alpha}$ transform in its ${\bf 16}$ Majorana-Weyl spinor representation. One introduces the worldsheet one forms  $p_{\alpha}$ as conjugate momenta to $\theta^{\alpha}$, with the OPE 
\be
p_{\alpha}(z) \theta^{\beta}(0) \sim \frac{\delta_{\a}^{\b}}{z} \
. \ee
$(p_{\alpha}, \theta^{\beta})$ is a free fermionic $(b,c)$ system of weight $(1,0)$.
$(x^m, p_{\alpha}, \theta^{\beta})$ are the GS variables.

The ghost sector consists of a bosonic complex Weyl spinor ghost $\lambda^{\alpha},\alpha=1,...,16$, which
satisfies the pure spinor constraint 
\be
\lambda^{\alpha}\gamma_{\alpha\beta}^m\lambda^{\beta} = 0 \;\;\;\;\;\; m=0,...,9 \ .
\label{const} \ee
The $\gamma_{\alpha\beta}^m$ are the symmetric
$16\times 16$ Pauli matrices in ten dimensions. The reader can find details of the notations in \ref{appendix:flat}.
A spinor $\lambda^{\alpha}$ that satisfies the constraints (\ref{const}) is called pure spinor.
This set of ten constraints is reducible as we will soon discuss. It reduces the number of degrees
of freedom of $\lambda^{\alpha}$ from sixteen to eleven.
We denote the conjugate momenta to $\lambda^{\alpha}$ by the worldsheet one-form $w_\a$, which is a  bosonic target space complex Weyl spinor of opposite chirality to $\lambda^\a$. Because of the pure spinor
constraints (\ref{const}), the system $(w_\a,\l^\a)$ is a curved beta-gamma system of weights $(1,0)$. If we want to find their free field OPE we have to first solve the pure spinor contraint to go on a patch of the pure spinor manifold, as in (\ref{opepatch}).

The pure spinor constraints imply that $ w_\a$ are defined up to
the gauge transformation \be \label{gaugewalpha}\delta  w_\a = \Lambda^m (\gamma_m
\lambda)_{\alpha} \ . \ee Therefore, $ w_\a$ appear only in gauge
invariant combinations. These are the Lorentz algebra currents
$N_{mn}$, the ghost number current $J_{\lambda}$ which assigns
ghost number $1$ to $\lambda$ and ghost number $-1$ to $ w $ \be
N_{mn} = \frac{1}{2}w\gamma_{mn}\lambda,~~~~~J_{\lambda} =
 w_\a\lambda^{\alpha} \ ,
\label{lorentz}
 \ee and the pure spinor
stress-energy tensor $T_{\lambda}$. In flat space, these are all of the gauge invariant combinations. In AdS, the story will get more interesting.

Unlike the RNS
superstrings, all the variables that we use in the pure spinor
superstring are of integer worldsheet spin.
This is an important property of the formalism, that translates in the simple perturbative expansion of string amplitudes. In the pure spinor formalism, the genus expansion is performed on bosonic Riemann surfaces. The RNS formalism, on the other hand, contains worldsheet spinors and requires that multiloop scattering amplitudes be evaluated on super-Riemann surfaces, which carry the complication of the sum over spin structures. This in turn enforces the GSO projection, which again is absent from the pure spinor formalism.

In the pure spinor formalism, the worldsheet metric is in conformal gauge from the beginning. The conformal gauge fixed worldsheet action of the pure spinor superstring is
\be S =
\int d^2z \left(\frac{1}{2}\p x^m\bar{\p}x_m +
p_{\a}\bar{\p}\theta^{\alpha} -   w_\a\bar{\p} \lambda^{\alpha}
\right) \ , 
\ee
where the last term represents the curved beta-gamma system which describes the pure spinor manifold.
The total central charge of the pure spinor superstring is
\be
c^{tot} = c_{X} + c_{p,\theta} + c_{w,\lambda}  = 10-32+22=0 \ ,
\ee
as required by the absence of a conformal anomaly.

{\em Physical states}

The physical states are defined as the ghost number one cohomology
of the nilpotent BRST operator 
\be Q = \oint dz\,
\lambda^{\alpha}d_{\a} \ , \label{10ld}
\ee 
where 
\be d_{\a} =
p_{\a} - \frac{1}{2}\gamma^m_{\a\b}\theta^{\b}\p x_m
-\frac{1}{8}\gamma^m_{\a\b}\gamma_{m\g\delta}\theta^{\b}
\theta^{\g}\p\theta^{\delta} \ . \ee 

The $d_{\a}$ are the supersymmetric Green-Schwarz constraints.
They are holomorphic and satisfy the OPE
 \be d_{\a}(z) d_{\b}(0)
\label{ope:d}
\sim - \frac{\gamma^m_{\a\b}\Pi_m(0)}{z} \ , 
\ee 
and 
\be d_{\a}(z)
\Pi^m (0) \sim \frac{\gamma^m_{\a\b}\p \theta^{\b}(0)}{z} \ , \ee
where 
\be 
\Pi_m = \p x_m + \frac{1}{2} \theta\gamma_m \p \theta \
, 
\ee 
is the supersymmetric momentum.\footnote{In the GS formalism, $d_\alpha$'s are fermionic constraints on the worldsheet. Half of these constraints is first class and it generates Siegel's kappa-symmetry. The other half of these constraints is second class. Because it is not possible to disentangle in a covariant way the first and second class constraints inside $d_\a$, covariant quantization of the GS string has been problematic to achieve. This is one of the motivations to use the covariant pure spinor formalism.} $d_{\a}$ acts on function on
superspace $F(x^m, \theta^{\a})$ as 
\be d_{\a}(z)
F(x^m(0),\theta^{\b}(0)) \sim \frac{D_{\a}
F(x^m(0),\theta^{\b}(0))}{z} \ , \ee where 
\be\label{susyderi} D_{\a} =
\frac{\p}{\p \theta^{\a}} +
\frac{1}{2}\gamma^m_{\a\b}\theta^{\b}\p_m \ , 
\ee 
is the
supersymmetric derivative in ten dimensions. It is immediate to check the nilpotency of the BRST charge (\ref{10ld}), due to the OPE's (\ref{ope:d}) and the pure spinor constraint (\ref{const}).

Massless states are described by the ghost number one and weight zero
vertex operators 
\be {\cal U}^{(1)} =
\lambda^{\alpha}A_{\alpha}(x,\theta) \ , 
\ee
where $A_{\alpha}(x,\theta)$ is an unconstrained spinor superfield.

The BRST cohomology conditions are
\be
Q{\cal U}^{(1)}=0,\quad \delta{\cal U}^{(1)} =Q \Omega^{(0)} \ ,
\label{coho}
\ee
where $\Omega^{(0)}$ is a real scalar superfield of ghost number zero.
These imply the ten-dimensional field equations and gauge invariance for $A_{\alpha}(x,\theta)$
\be\label{dgammafive}
\gamma^{\a\b}_{mnpqr}D_{\a}A_{\beta}(x,\theta) = 0 \ ,\qquad 
\delta A_{\a} = D_{\a}\Omega^{(0)} \ ,\ee
where we used the fact that 
\be
\l^\a\l^\b = {1\over{1920}}(\l\g^{mnpqr}\l) \g_{mnpqr}^{\a\b}
\ ,
\ee
due to the pure spinor constraint. In \ref{syminsuperspace} we show that these equations are equivalent to the usual linearized equations of motion for the gluon and the gluino. 
These equations imply that $A_{\alpha}$ is
an on-shell super Maxwell spinor superfield in ten dimensions. For instance, in Wess-Zumino gauge $\theta^\a A_\a=0$ we have
\be\label{onshellA}
A_\alpha (x, \theta) = \frac{1}{2} (\gamma^m \theta)_\alpha a_m (x) +
\frac{i}{12} (\theta \gamma^{mnp} \theta) (\gamma_{mnp})_{\alpha
\beta} \psi^\beta (x) + O(\theta^3)
\ee
where $a_m(x)$ is the gauge field and $\psi^{\b}(x)$ is the
gaugino. They satisfy the super Maxwell equations
\be
\partial^m(\p_m a_n -\p_n a_m) = 0,\;\;\;\;\;\;  \gamma_{\a\b}^m \p_m \psi^{\beta} = 0 \ .
\ee

To make contact with the usual string quantization, let us recall how vertex operators look like in the RNS superstring. There, the unintegrated ghost number one weight zero vertex operator for the massless photon in the minus one picture is
\be\label{vectorrns}
\cU^{(1)}=a_m(x)\psi^m ce^{-\phi} \ ,
\ee 
where $\phi$ comes from the bosonization of the superconformal ghosts $\beta\gamma$ and $\psi^m$ are the RNS fermions. By expanding the wavefunction $a_m(x)$ in Fourier modes and taking a constant polarization vector $\epsilon_m$, we arrive at the more familiar form $\epsilon_m\psi^m ce^{-\phi}e^{ikx}$. The vertex operator (\ref{vectorrns}) can be mapped to the pure spinor expression for the massless photon vertex operator $\cU^{(1)}=a_m(x)(\l\g^m \t)$, which is in fact the first term in the expansion (\ref{onshellA}). The relation between integrated vertex operators ${\cal V}^{(0)}$ and unintegrated ones $\cU^{(1)}$ is enforced by the $b$ ghost as
$$
[\oint b(z), {\cal U}^{(1)}(0)]= \cV^{(0)} \ ,
$$
which can be easily checked on (\ref{vectorrns}) and its integrated cousin ${\cal V}^{(0)}=a_m(x)\psi^me^{-\phi}$. By applying the BRST charge and the properties of the stress energy tensor, this relation can be immediately recast into the descent equation
\be
Q\cV^{(0)}=\partial\, \cU^{(1)} \ ,
\ee
which yields
\be
\int dz\, \cV_z^{(0)} = \int dz \left(\p\theta^{\a}A_{\a} +
\Pi^m A_{m} + d_{\a}W^{\a} + \frac{1}{2}M_{mn}F^{mn}\right) \ ,
\ee 
where $W^{\a}$ and $F^{mn}$ are the spinorial and bosonic
field strength, respectively
\be
D_{\a}W^{\b} = \frac{1}{4}(\gamma^{mn})^{\b}_{\a} F_{mn}\;\;\;\;\;\; F_{mn} = \p_m A_n -\p_n A_m,\ .
\ee
 We have $W^{\a} = \psi^{\a} + ...$ and $ F_{mn} = f_{mn} +... $ and
$M_{mn}$ are the spin part of the generators of Lorentz transformations in the matter sector.
When expanded $\cV^{(0)}$ reads
\be
\cV^{(0)} = a_m \p x^m + \frac{1}{2}f_{mn}M^{mn} + \psi^{\a}q_{\a} + ...
\ ,
\ee
where $q_{\a}$ is the space-time supersymmetry current
\be
q_{\a} = p_{\a} +  \frac{1}{2}(\p x^m + \frac{1}{12}\theta\gamma^m\p \theta)(\gamma_m\theta)_{\a} \ .
\ee
Note that, unlike the RNS,  we do not need a GSO projection in order to get a supersymmetric spectrum
and that the Ramond and NS sectors appear on equal footing in the pure spinor formalism.

{\em Scattering Amplitudes}

Consider the tree-level open string scattering amplitudes.
The n-point function $A_n$ on the disk reads
\be\label{treeamplit}
A_n = \langle\, \cU^{(1)}_1(z_1)\cU^{(1)}_2(z_2)\cU^{(1)}_3(z_3)\int dz_4 \cV^{(0)}_4(z_4)...\int dz_n \cV^{(0)}_n(z_n)\, \rangle \ .
\ee
$\cU^{(1)}_i$ are dimension zero, ghost number one vertex operators and $\cV^{(0)}_i$ are
dimension one, ghost number zero vertex operators. There are no moduli on the disk, but only three conformal killing vectors,\footnote{In the RNS formalism, moduli and conformal killing vectors are, respectively, the $b$ and $c$ ghost zero modes.} that generate the $PSL(2,R)$ group as usual. We will briefly comment on the ghost number anomaly in Section~\ref{subsec:purespinormanifold}. Hence, we can use the $PSL(2,R)$ symmetry to fix the worldsheet location of three vertices to their unintegrated form, just as in the bosonic string. Using the free field OPE's we get rid of the non-zero modes and obtain
\be
A_n = \int dz_4 ...\int dz_n \langle \l^{\a}\l^{\b}\l^{\gamma} f_{\a\b\gamma}(z_r,k_r,\theta) \rangle \ ,
\ee
where $k_r$ are the scattering momenta.
Thus, $f_{\a\b\gamma}$ depends only on
the zero modes of $\theta$.
There were eleven bosonic zero modes of $\lambda^{\a}$ and sixteen fermionic zero modes
of $\theta^{\a}$.
One expects eleven of the fermionic zero modes integrals  to cancel the eleven 
bosonic zero modes integrals, leaving five fermionic $\theta$ zero modes.
A Lorentz invariant and supersymmetric prescription for integrating over the remaining five fermionic $\theta$ zero modesis given by picking the component  
\be\label{flatzero}
\langle (\l\g^m\theta)(\l\g^n\theta)(\l\g^p\theta)(\theta\g_{mnp}\theta)\rangle=1 \ .
\ee
One can show that this is the unique ghost number three and weight zero element in the BRST cohomology. For more details about the amplitude prescription, we defer the reader to the references listed at the end of this Section.

\subsubsection{Pure spinor manifold}
\label{subsec:purespinormanifold}

As we anticipated above, the pure spinor variables and their conjugate momenta $(w,\l)$ represent a curved beta-gamma system. The pure spinor
set of constraints (\ref{const}) defines a curved space, which can be covered by sixteen patches
$U_{\alpha}$ on which the $\a$-th component of $\l^{\a}$ is nonvanishing. The set of constraints   (\ref{const}) is reducible. In order to solve it we rotate to Euclidean signature. The pure spinor variables $\lambda^\a$ transform in the ${\bf 16}$ of $SO(10)$. Under $SO(10) \rightarrow U(5) \simeq SU(5)\times U(1)$ we have that ${\bf 16} \rightarrow {\bf 1}_{\frac{5}{2}} \oplus {\bf \bar{10}}_{\frac{1}{2}} \oplus {\bf 5}_{-\frac{3}{2}}$.
We denote the sixteen components of the pure spinor in the
$U(5)$ variables by $\lambda^{\alpha} = \lambda^{+}\oplus\lambda_{ab}\oplus \lambda^a ; a,b=1..5$ with
 $\lambda_{ab}= -\lambda_{ba}$.
In this variables it is easy to solve the pure spinor
set of constraints (\ref{const}) by
\be
\lambda^{+} = e^s,\;\;\;\; \lambda_{ab}= u_{ab},\;\;\;\; \lambda^a = -\frac{1}{8}e^{-s} \varepsilon^{abcde}u_{bc}u_{de} \ .
\label{U5}
\ee

The
system $( w_\a,\lambda^{\alpha})$ is interacting due to the pure
spinor constraints.
It has the central charge $c_{( w ,\lambda)} =
22$, which is twice the complex dimension of the pure spinor space
\be T_{\lambda}(z)T_{\lambda}(0) \sim
\frac{dim_{\mathbb{C}}({\cal M})}{z^4} + ... \ . \ee
This can be computed, for instance,  by introducing
the conjugate momenta to the $U(5)$ variables (\ref{U5}) with the OPE
\be\label{opepatch}
t(z)s(0) \sim \log z,\;\;\;\;\;\;\;\; v^{ab}u_{cd} \sim -\frac{\delta^{ab}_{cd}}{z} \ ,
\ee
with
$\delta^{ab}_{cd} = \frac{1}{2}(\delta^a_c \delta^b_d - \delta^a_d \delta^b_c)$.
The stress energy tensor and the ghost current in $U(5)$ notation read
\be
\label{pureT}
T_{\lambda} = v^{ab}\p u_{ab} + \p t \p s + \p^2 s \ ,
\ee
$$
J_\l=\half u_{ab}v^{ab}+\partial t+3\partial s \ .
$$
The stress energy tensor can be recast in a covariant form by using the Lorentz generators $N^{mn}$ in (\ref{lorentz}) and the ghost current $J_\l$
$$
T_\l={1\over10}N^{mn}N_{mn}-{1\over8}J_\l^2+\partial J_\l \ .
$$
The central charge of the pure spinor beta-gamma system is $22$.\footnote{For a complex $(1,0)$ beta-gamma system one would naively expect a central charge of $44$, since the $\l$'s are complex Weyl spinors. However, the complex conjugate $\bar \l$'s never appear in the formalism, so that we can define the pure spinor as being a hermitian operator $\l^\a=(\l^\a)^\dagger$. Since both the operator and its complex conjugate carry the same ghost number, one avoids any inconsistencies. This ensures that the central charge is $22$.}
The ghost number anomaly reads \be\label{ghostanomaly} J_{\lambda}(z)T_{\lambda}(0) \sim -\frac{8}{z^3} + ... = \frac{c_1({\cal
Q})}{z^3}+ ...\ , \ee where ${c_1({\cal Q})}$ is the first Chern
class of the pure spinor cone base ${\cal Q}$ and 
$J_{\lambda}=w_\a \l^\a$ 
is the ghost number current. The ghost number anomaly is thus $+8$.  

The pure spinor space  ${\cal M}$
is complex eleven-dimensional, which is a cone over
${\cal Q} = \frac{SO(10)}{U(5)}$.
At the origin $\lambda^\a = 0$, both the  pure spinor
set of constraints (\ref{const}) and their derivatives with respect to $\lambda^\a$ vanish.
Thus, the pure spinor space has a singularity at the origin. This singularity is easily seen in the parameterization (\ref{U5}), which naively blows up and requires dealing with multiple patches on the pure spinor manifold. We will briefly mention how to deal with this issue when discussing scattering amplitudes below.

The pure spinor system $(\lambda^{\alpha},w_{\alpha})$ defines a
non-linear $\sigma$-model (also referred to as a curved beta-gamma system) due to the curved nature of the pure
spinor space (\ref{const}). There are global obstructions to
define  the pure spinor system on the worldsheet and on target
space. They are associated
with the need for holomorphic transition functions relating
$(\lambda^{\alpha},w_{\alpha})$ on different patches of the pure
spinor space, which be compatible with their OPE. The need to use multiple patches to chart the pure spinor manifold is evident from the fact, already mentioned, that the $U(5)$ parameterization in (\ref{U5}) is not global, but rather blows up at $\lambda^+=0$. The obstructions are
reflected by quantum anomalies in the worldsheet and target space
(pure spinor space) diffeomorphisms. For example, when passing from one patch to another, one must preserve the simple OPE's (\ref{opepatch}) between the variables and their conjugate momenta, but there may be a topological obstruction to this. The conditions for the
vanishing of these anomalies are the vanishing of the integral
characteristic classes \label{anomalies} \be
\frac{1}{2}c_1(\Sigma)c_1(\M) = 0,~~~~~~\frac{1}{2}p_1(\M)=0 \ ,
\ee $c_1(\Sigma)$ is the first Chern class of the worldsheet
Riemann surface, $c_1(\M)$ is the first Chern class of the pure
spinor space $\M$, and $p_1$ is the first Pontryagin class of the
pure spinor space. The vanishing of $c_1(\M)$ is needed for the
definition of superstring perturbation theory. It is the usual Calabi-Yau condition on this complex eleven-dimensional manifold and it implies the
existence of the nowhere vanishing holomorphic top form
$\Omega(\l)$ on the pure spinor space $\M$
 \be \Omega = \Omega(\lambda)d \l^1\wedge ...\wedge
d\l^{11} \ ,\label{pureform}
 \ee
where overall factor $\Omega(\l)\sim1/\l^3$ has ghost number minus three.

The pure spinor space (\ref{const}) has a singularity at
$\lambda^{\alpha} = 0$. Blowing up the singularity results in an
anomalous theory, as it generates a non-zero first Chern class. However, simply removing the origin leaves a
non-anomalous theory. This means that one should consider the pure
spinor variables as twistor-like variables.

We can reinterpret the tree-level amplitude prescription in light of this discussion of the pure spinor manifold. The volume form $[{\cal D}\l]=\Omega$ on the pure spinor manifold (\ref{pureform}) contains an integration over eleven pure spinors, and it carries ghost number $+8$, which coincides with the ghost number anomaly (\ref{ghostanomaly}). Since the integration over pure spinor zero modes in the scattering amplitude (\ref{treeamplit}) is the naive one, namely $\int d^{11}\l$, we have the schematic relation
$$ A_n\sim\int d^{11}\l\sim\int [{\cal D}\l] \l^3 \ ,$$
which agrees with the insertion of three unintegrated vertex operators in the tree-level amplitude prescription (\ref{treeamplit}). For a precise realization of this relation, which is outside the scope of this review, we refer the reader to the literature cited at the end of this Section.

If we want to compute multiloop scattering amplitudes we need to introduce the $b_{zz}$ antighost, a composite operator of weight two and ghost number minus one, which couples to the worldsheet Beltrami differentials and counts the moduli of the Riemann surface. It is defined such that its anticommutator with the BRST charge gives the total stress tensor: $\{Q,b\}=T$. In a flat background we need to enlarge the field content by introducing a BRST quartet of worldsheet variables to implement such a construction, which is usually referred to as the non-minimal pure spinor formalism. In the $AdS_5\times S^5$ background, however, there is no need to introduce such additional set of fields in order to construct the $b$ antighost. As we will explicitly see in Section~\ref{section:antighost}, this is due to the presence of an invertible RR flux.

\subsubsection{Closed superstring}
\label{section:closed}

The construction of the closed superstring is straightforward.
One introduces the right moving superspace variables
 $(\hat{p}_{{\ah}}, \hat{\theta}^{{\ah}})$, the pure spinor
 system $(\hat{ w }_{{\ah}},\hat{\l}^{{\ah}})$ and
the nilpotent BRST operator \be \hat{Q} = \oint
d\bar{z}\,\hat{\lambda}^{{\ah}}\hat{d}_{{\ah}} \ . \ee
The analysis of the spectrum proceeds by combining the left and
right sectors.
We can describe respectively type IIB or type IIA superstrings by taking the hatted spinor indices to have the same or opposite chirality of the unhatted ones.

Massless states are described by the ghost number two vertex operator
\be
\cU^{(2)} = \lambda^{\a}\hat{\lambda}^{{\ah}}A_{\a{\ah}}(X,\theta, \hat{\theta}) \ .
\ee
The cohomology condition for a physical vertex operator $U$ is that
it satisfies the equations and gauge invariances\footnote{
For the ordinary closed bosonic string with the standard definitions
of $Q$ and $\hat Q$,
these cohomology conditions reproduce the usual
physical spectrum for states with non-zero momentum. For zero momentum
states, there are additional subtleties associated with the $b_0-\overline
b_0$
condition which will be ignored.}
\bea\label{cohomd}
Q\cU^{(2)}= \hat Q \cU^{(2)}=0,\eea
$$
\d \cU^{(2)} = Q\Lambda + \hat Q\widehat\Lambda {\rm ~~~with~~~ }
\hat Q\Lambda = Q\widehat\Lambda=0.$$
Applying these conditions to $\cU^{(2)}=\l^\b\lh^\gh A_{\b\gh}$
where $\Lambda=\lh^\ah \widehat\Omega_\ah$ and
$\widehat\Lambda=\l^\a \Omega_\a$,
and using the fact that pure spinors satisfy
\bea\label{using}\l^\a\l^\b = {1\over{1920}}(\l\g^{mnpqr}\l) \g_{mnpqr}^{\a\b}
{\rm ~~~ and ~~~}
\lh^\ah\lh^\bh = {1\over{1920}}(\lh\g^{mnpqr}\lh) \g_{mnpqr}^{\ah\bh},\eea
one finds that $A_{\b\gh}$ must satisfy the conditions
\bea\label{efms}\g_{mnpqr}^{\a\b} D_\a A_{\b\gh} =
\g_{mnpqr}^{\ah\gh} D_\ah A_{\b\gh} =
0,\eea
$$
\d A_{\b\gh} = D_\b \widehat\Omega_\gh +
D_\gh \Omega_\b
{\rm ~~~with~~~}
\g_{mnpqr}^{\a\b} D_\a \Omega_\b =
\g_{mnpqr}^{\ah\gh}  D_\ah \widehat\Omega_{\gh} = 0,$$
where $D_\a$ and
$
D_\ah$ are the supersymmetric
derivatives of flat type II $D=10$ superspace (\ref{susyderi}).
Note that $Q \Phi =\l^\a D_\a \Phi$ and
$\hat Q \Phi = \lh^\ah D_\ah \Phi$
for any superfield $\Phi(\l,\lh,x,\t,\th)$.

It will now be shown that the conditions of (\ref{efms})
correctly reproduce the Type IIB
supergravity spectrum. The easiest way to check this is
to use the fact that closed superstring vertex operators
can be understood as the left-right product of open superstring
vertex operators. The massless open superstring vertex operator
is described by a spinor superfield $A_\b(x,\t)$ satisfying the conditions (\ref{dgammafive}), that imply the super-Maxwell
spectrum.
Similarly, the equations of (\ref{efms}) imply that there exists a gauge choice
such
that
\bea\label{sugra}A_{\b\gh}(x,\t,\th) = (\g^m\t)_\b (\g^n\th)_\gh h_{mn}(x)
+ (\g^m\t)_\b (\th\g^{pqr}\th) (\g_{pqr}\widehat\psi_m(x))_\gh \eea
$$
+ (\t\g^{pqr}\t)(\g^m\th)_\gh (\g_{pqr}\psi_m(x))_\b
+ (\t\g_{mnp}\t)\g^{mnp}_{\b\a}(\th\g_{qrs}\th)\g^{qrs}_{\gh\dh}
P^{\a\dh}(x)
+ ... $$
where $...$ contains only auxiliary fields and where
$h_{mn}(x)$, $\psi^\a_m(x)$, $\widehat\psi^\ah_m(x)$ and
$P^{\a\dh}(x)$ satisfy the equations
\bea\label{compeq}\p^m(\p_m h_{pn} -\p_p h_{mn})  =
\p^m(\p_m h_{pn} - \p_n h_{pm}) =0,\eea
$$\p^m (\p_m \psi^\a_n - \p_n\psi^\a_m) =
\p^m (\p_m \widehat\psi^\ah_n - \p_n\widehat\psi^\ah_m) = 0,\quad
\p_n (\g^n \psi_m)_\a =
\p_n (\g^n \widehat\psi_m)_\ah = 0,$$
$$\g^n_{\a\b} \p_n P^{\b\gh} = \g^n_{\gh\dh} \p_n P^{\a\dh}=0.$$
The equations of (\ref{compeq}) are those of linearized Type IIB supergravity
where
$h_{mn}$ describes the dilaton, graviton and
anti-symmetric two-form, $\psi_m^\a$ and $\widehat\psi_m^\ah$ describe
the two gravitini and dilatini,
and $P^{\a\dh}$ describes the Ramond-Ramond field
strengths
\bea\label{rrsuperfield}
{\rm IIB:}\qquad{1\over {g_s}} P=&\gamma^{a_1}F_{a_1}+{1\over3!}\gamma^{a_1a_2a_3}F_{a_1a_2a_3}+{1\over2\cdot 5!}\gamma^{a_1\ldots a_5}F_{a_1\ldots a_5} \ , \cr
{\rm IIA:}\qquad{1\over {g_s}} P=& F_0+{1\over2!}\gamma^{a_1a_2}F_{a_1a_2}+{1\over 4!}\gamma^{a_1\ldots a_4}F_{a_1\ldots a_4} \ ,
\eea
where $F_0$ is usually referred to as the Romans mass. We see that all the different RNS sectors (NS,NS), (R,R), (R,NS) and (NS,R) are on equal footing in this representation.
The integrated ghost number zero
vertex operator for the massless states reads \be \int\,d^2z\, {\cal V}^{(0)} = \int
d^2z \left(\p\theta^{\a}A_{\a{\ah}}
\bar{\p}\hat{\theta}^{\ah} + \p\theta^{\a} A_{\a
m}\bar{\Pi}^m
 + ... \right) \ .
\ee

It is sometimes convenient to choose a gauge for physical vertex operators such that
they are dimension zero worldsheet {\em primary fields}, i.e. they have no double poles with the
stress tensor. For the ordinary closed bosonic string, this gauge is implemented by the zero mode of the antighost: $b_0\cU^{(2)} =\bar b_0\cU^{(2)} =
0$. This condition imposes the usual Lorentz gauge on the gauge fields in the massless sector and is generally called Siegel gauge. Since $\{Q,b_0\}=L_0$, this is equivalent to choosing a dimension zero vertex operator. This gauge choice simplifies the computation of scattering amplitudes and, in string field theory, it reduces the string propagator to $b_0/L_0$. In the pure spinor formalism, the $b$ antighost is a complicated composite operator, so one would rather analyze the stress-tensor to find this gauge-fixing condition. 
The left and right-moving stress tensors
are
\bea\label{st}T = \half\p x^m \p x_m + p_\a \p\t^\a + T_\l {\rm~~and~~}
\overline T = \half\overline\p x^m \overline\p x_m +
\widehat p_\ah \widehat\p\t^\ah + \overline T_\lh,\eea
where $T_\l$ and $\overline T_\lh$ are the $c=22$ stress-tensors
constructed from the
pure spinor variables $\l^\a$ and $\lh^\ah$, as in (\ref{pureT}). When acting on the
massless
vertex operator $\cU^{(1,1)}=\l^\a\lh^\bh
A_{\a\bh}(x,\t,\th)$,
the condition of no double poles with $T$ or $\overline T$
implies that $\p_m \p^m A_{\a\bh}=0$. Furthermore, the on-shell BRST cohomology conditions (\ref{efms})
imply that
$\p^m(\p_m A_{n \bh} -\p_n A_{m \bh})=
\p^m(\p_m A_{\a n} -\p_n A_{\a m})= 0$
where $A_{m\gh} = {1\over {16}}\g_m^{\a\b} D_\a A_{\b\gh}$ and
$A_{\a m} = {1\over {16}}\g_m^{\bh\gh} D_\bh A_{\a\gh}$.
So the gauge-fixed equations for $A_{\a\bh}$ are
\bea\label{gf}
\p^m\p_m A_{\a \bh} = 0,\quad
\p^m A_{m \bh}=
\p^m A_{\a m}= 0.\eea

The gauge transformations  (\ref{efms}) on the superfield $A_{\a\ah}$, derived from the requirement that the vertex operator is closed but not exact in the BRST cohomology, include general coordinate invariance, which becomes apparent when considering the component of the supergravity superfield corresponding to the graviton in (\ref{sugra}). So the residual gauge transformations
which leave the gauge-fixed equations of motion (\ref{gf}) invariant reduce to
\bea\label{residf}\d A_{\b\gh} = D_\b \widehat\Omega_\gh +
D_\gh \Omega_\b
{\rm ~~~with~~~}
\p_m\p^m \Omega_\b =
\p_m\p^m \widehat\Omega_\bh =
\p^m \Omega_m =
\p^m\widehat\Omega_m = 0. \nonumber\\\eea

\subsection{Curved backgrounds}
\label{section:curved}

Now that we know how to describe strings in flat space, let us generalize to a curved supergravity background. This is going to be the first step towards our goal of formulating the worldsheet theory in an AdS background in Section \ref{section:supercoset}. Details on the notations used in a curved background can be found in \ref{appendix:curved}.

\subsubsection{Green-Schwarz}

We consider a target superspace with thirty-two supersymmetries in ten dimensions. The conventions we use in a generic background are collected in \ref{appendix:curved}. We denote the curved super-coordinates as $Z^M=\{X^m,\theta^\mu,\hat\theta^{\widehat\mu}\}$, where $m=0,\ldots,9$ and the Grassmann-odd coordinates are sixteen dimensional Majorana-Weyl spinors, of the same chirality for type IIB and of opposite chiralities for type IIA. From now on, we will denote the right-moving spinor variables with a hat and not with a bar, since in a curved background with RR flux the holomorphic and anti-holomorphic separation of the worldsheet variables does not hold anymore. The Green-Schwarz sigma model action is
 \bea\label{typeii}
 S_{GS}={1\over 4\pi\alpha'}\int\,d^2\sigma\,
 \partial_i Z^M\partial_j Z^N\left(\sqrt{g}g^{ij}G_{NM}(Z)+\e^{ij}B_{NM}(Z)\right) \ ,
 \eea
where $G_{NM},B_{NM}$ are background superfields whose lowest components along the directions $G_{mn}(X),B_{mn}(X)$ describe the target space metric and NS-NS two-form potential. The meaning of the other components will be clarified shortly. The last term in the action (\ref{typeii}) is the Wess-Zumino term, that can be written in two dimensional form due to the fact that locally $H=DB$. Let us introduce the target space supervielbeins $E^A_M(Z)$ and denote their pullbacks on the worldsheet as the currents $J^A=dZ^M E^A_M$. In terms of these currents, the action (\ref{typeii}) reads
 \bea\label{taction}
 S={1\over4\pi\alpha'}\int\,d^2\sigma\,
 \left(\sqrt{g}g^{ij}J^a_iJ^b_j\eta_{ab}+\e^{ij}J^A_iJ^B_jB_{BA}\right) \ ,
 \eea
where $G_{MN}=E^a_ME^b_N\eta_{ab}$, $B_{MN}=E^A_ME^B_NB_{BA}$ and $\eta_{ab}$ is the flat Minkowski metric on the bosonic tangent space. In the case of flat backgrounds, $J_i^a=\Pi^a_i$ is the usual supersymmetric momentum, $J^\alpha_i=\partial_i \theta^\alpha$ and $J^\ah_i=\partial_i \theta^\ah$, while the super B-field has non-vanishing components $B_{a\alpha}=(\theta\gamma_a)_\a$, $B_{a\ah}=-(\htt\gamma_a)_{\ah}$ and $B_{\a\ah}=(\theta\gamma_a)_\alpha(\htt\gamma^a)_\ah$. By plugging these expressions into (\ref{taction}) we go back to the type II Green-Schwarz action in a flat background. As we will see in a moment, the spinorial components of the super B-field are related not to background fluxes, but to the torsion.

\subsubsection{Pure spinor}
\label{section:constraints}

The sigma model action for the type II pure spinor superstring in a generic supergravity background is
\bea\label{curvedaction}
S=&{1\over 2\pi\alpha'}\int d^2z[\half\Pi^a\bar\Pi^b\eta_{ab}+\half \Pi^A\bar \Pi^B B_{AB}+d_\a\bar\Pi^\a+\hat d_\ah\hat{\overline{\Pi}}{}^\ah)  \\
&+d_\a\hat d_\ah P^{\a\ah}+\l^\a w_\b \hat d_\gh C_\a{}^{\b\gh}+\hat\l^\ah\hat w_\bh d_\g\tilde C_\ah{}^{\bh\g}+
\l^\a w_\b\hat\l^\ah\hat w_\bh S_{\a\ah}^{\b\bh}\nonumber\\&+w_\a\bar\nabla\l^\a+\hat w_\ah\nabla\hat\l^\ah+\alpha'R\Phi(Z)\nonumber ] \ ,
\eea
The independent worldsheet fields in this action are $Z^M=(X^m,\t^\mu,\th^{\hat\mu})$ and $(d_\a,\hat d_\ah)$ in the matter sector, and $(w_\a,\l^\a,\hat w_\ah,\lh^\ah)$ in the ghost sector. By varying the action with respect to these fields, one can derive their equations of motion. Note that $d_\a$ and ${\hat d}_\ah$
can be treated as independent variables in (\ref{curvedaction})
since $p_\a$ and $\hat p_\ah$ do not appear explicitly.
The worldsheet matter fields are the pullback of the target space super-vielbein $\Pi^A=E^A_M dZ^M$, where $A=(a,\a,\ah)$ is a tangent space superspace index and $M=(m,\mu,\hat\mu)$ a curved superspace index. The ghost content is the same as in flat space and the covariant derivative on $\l$ ($\hat\l$) is defined using the pullback of the left-moving (right-moving) spin connection $\Omega_\a{}^\b=d Z^M\Omega_{M\a}{}^\b$ ($\hat\Omega_\ah{}^\bh=d Z^M\hat\Omega_{M\ah}{}^\bh$) as
$$
(\nabla\l)^\a=\partial\l^\a+\Omega_\b{}^\a\l^\b \ ,\quad (\nabla\hat\l)^\ah=\partial\hat\l^\ah+\hat \Omega_\bh{}^\ah\hat\l^\bh \ .
$$
The background superfields $B,P,C,\tilde C, S, \Phi$ are functions of $X,\t,\th$. The background superfield $B_{AB}$ appearing in (\ref{curvedaction}) is the superspace two-form potential; the lowest components of $C_\a{}^{\b\bh}$ and $\tilde C_\ah{}^{\bh\a}$ are related to the gravitini and dilatini; the lowest component of $P^{\a\ah}$ is the Ramond-Ramond bispinor field strength. In the type II superstring, the dependence of the superfield $P$ on the Ramond-Ramond $p$-form field strengths $F_p$ is (\ref{rrsuperfield}).
For example, in the type IIB $AdS_5\times S^5$ background, $P^{\a\ah} = 
\g_{01234}^{\a\ah}$ whose inverse is $(P^{-1})_{\a\ah}=\g^{01234}_{\a\ah}$. $S_{\a\ah}{}^{\b\bh}$ is related to the Riemann curvature. The first line
of (\ref{curvedaction})
is the standard Type II GS action, but the other lines
are needed for BRST invariance. As will now be shown to lowest order
in $\a'$, nilpotence and holomorphicity
of $\l^\a d_\a$ and nilpotence and antiholomorphicity
of $\lh^\ah {\hat d}_\ah$ imply the 
equations of motion for the background superfields
in (\ref{curvedaction}).

If the Fradkin-Tseytlin term, $\int d^2 z \Phi(Z)R$, is omitted,
(\ref{curvedaction}) is the most general action with classical worldsheet
conformal invariance and zero (left,right)-moving ghost number 
which can be constructed
from the Type II worldsheet variables. Note that $d_\a$ carries
conformal weight $(1,0)$, $\hat d_\ah$ carries conformal weight
$(0,1)$, $\l^\a$ carries ghost number $(1,0)$ and
conformal weight $(0,0)$, 
$\lh^\ah$ carries ghost number $(0,1)$ and
conformal weight $(0,0)$, 
$w_\a$ carries ghost number $(-1,0)$ and
conformal weight $(1,0)$, and
$\hat w_\ah$ carries ghost number $(0,-1)$ and
conformal weight $(0,1)$.
Since 
$w_\a$ and $\hat w_\ah$ can only appear in combinations which commute with
the pure spinor constraints, i.e. are invariant under the gauge transformations (\ref{gaugewalpha}), the background superfields
must satisfy 
\bea\label{conbaII}
(\g^{bcde})_\b^\a
\Omega_{M\a}{}^\b = 
(\g^{bcde})_\bh^\ah
\hat\Omega_{M\ah}{}^\bh = 
(\g^{bcde})_\b^\a C_{\a}^{\b\gh}=0\\
(\g^{bcde})_\bh^\ah \hat C_{\ah}^{\bh\g} =
(\g^{bcde})_\b^\a S_{\a\gh}^{\b\dh} =
(\g^{bcde})_\dh^\gh S_{\a\gh}^{\b\dh} = 0,\nonumber
\eea
and the different components of the spin connections will be defined as
\bea
\Omega_{M\a}{}^\b= \Omega_M^{(s)} \d_\a^\b +\half
\Omega_M^{cd} (\g_{cd})_\a{}^\b,
\quad
\hat\Omega_{M\ah}{}^\bh=\hat\Omega_M^{(s)} \d_\ah^\bh +\half
\hat\Omega_M^{cd} (\g_{cd})_\ah{}^\bh. 
\label{OmegaII}
\eea

Since there are now two independent pure spinors, so one has two
independent fermionic structure groups, each consisting of the 
spin group times scale transformations. One therefore has two
independent sets of spin connections and scale connections,
$(\Omega_M^{(s)},\Omega_M^{ab})$
and $(\hat\Omega_M^{(s)},\hat\Omega_M^{ab})$, which appear explicitly
in the Type II sigma model action. 

In addition to being target-space super-reparameterization invariant,
the action of (\ref{curvedaction}) is invariant under the local
gauge transformations
\bea
\d E_M^b =
\eta_{cd}\Lambda^{bc} E_M^d, \quad \d E^\a_M =\Sigma^\a_\b E^\b_M,\quad
\d E^\ah_M =\hat\Sigma^\ah_\bh E^\bh_M,
\label{localtwo}
\eea
$$
\d \Omega_{M\a}{}^\b = \p_M\Sigma_\a^\b +
\Sigma^\g_\a \Omega_{M\g}{}^\b -
\Sigma^\b_\g \Omega_{M\a}{}^\g,
\quad \d \hat\Omega_{M\ah}{}^\bh = \p_M\hat\Sigma_\ah^\bh +
\hat\Sigma^\gh_\ah \hat\Omega_{M\gh}{}^\bh -
\hat\Sigma^\bh_\gh \hat\Omega_{M\ah}{}^\gh,$$
$$\d \l^\a =
\Sigma^\a_\g \l^\g,\quad
\d w_\a =
-\Sigma^\g_\a w_\g, \quad
\d \lh^\ah =
\hat\Sigma^\ah_\gh \lh^\gh,\quad
\d \hat w_\ah =
-\hat\Sigma^\gh_\ah w_\gh,$$
where $\Sigma_\a^\b
= \Sigma^{(s)}\d_\a^\b +\half \Sigma^{bc}(\g_{bc})_\a{}^\b$, 
$\hat\Sigma_\ah^\bh
= \hat\Sigma^{(s)}\d_\ah^\bh +\half \hat\Sigma^{bc}(\g_{bc})_\ah{}^\bh$, 
$[\Lambda^{bc} ,\Sigma^{bc},\hat\Sigma^{bc}]$
parameterize independent local Lorentz transformations
on the [vector, unhatted spinor, hatted spinor] indices, $\Sigma^{(s)}$
and $\hat\Sigma^{(s)}$
parameterize independent local scale transformations on the unhatted and
hatted spinor indices, and the background superfields $[P^{\a\ah},
C_\a^{\b\gh},\hat C_{\ah}^{\bh\g}, S_{\a\gh}^{\b\dh}]$ transform
according to their spinor indices.

Furthermore, the action of (\ref{curvedaction}) and the BRST operators (\ref{brst})
are invariant under a local shift transformations, which we omit. 

The left- and right- moving BRST charges are
\bea\label{brst}
Q=\oint dz \l^\a d_\a \ ,\qquad \hat Q=\oint d\bar z\hat\l^\ah \hat d_\ah \ ,
\eea
where $d$ and $\hat d$ are the pullback of the spacetime supersymmetric derivatives. Conservation of $Q$ and $\hat Q$ and nilpotency of $Q+\hat Q$ imply a set of type IIA/B supergravity constraints, that put the background onshell. One-loop conformal invariance of the worldsheet action is implied by such constraints. In the following we will recall some of those constraints when needed.

The stress tensor for the pure spinor action in a generic type II supergravity background reads 
\bea\label{stressca}
T=-\half\Pi^a\Pi^b\eta_{ab}-d_\a\Pi^\a -w_\a(\nabla\l)^\a \ ,
\eea

\subsubsection{Type II supergravity constraints}

We will derive the contraints on the background superfields, coming from the requirement that the worldsheet BRST operator be nilpotent and conserved.

{\em Nilpotency constraints}

The conditions implied by nilpotency of
$Q=\oint  \l^\a d_\a$ and $\hat Q=\oint
\lh^\ah {\hat d}_\ah$ are obtained imposing that 
$$\{Q,Q\}=\{\hat Q,\hat Q\}=\{Q,\hat Q\}=0 \ ,
$$
which imply that 
\bea
\l^\a \l^\b T_{\a\b}{}^C = \l^\a \l^\b H_{\a\b B}= \l^\a\l^\b
\hat R_{\a\b\gh}{}^\dh = \l^\a\l^\b\l^\g R_{\a\b\g}{}^\d =0,
\label{nilII}
\eea
$$\lh^\ah \lh^\bh T_{\ah\bh}{}^C = \lh^\ah \lh^\bh H_{\ah\bh B}= \lh^\ah\lh^\bh
\hat R_{\ah\bh\g}{}^\d = \lh^\a\lh^\b\lh^\gh R_{\ah\bh\gh}{}^\dh =0,$$
$$
T_{\a\bh}{}^C = H_{\a\bh B}= \l^\a \l^\b R_{\a\gh\b}{}^\d=
 \hat\l^\ah \hat\l^\bh \hat R_{\g\ah\bh}{}^\dh= 0,$$
for any pure spinors $\l^\a$ and $\lh^\ah$, where 
$T_{AB}{}^\a$ and $R_{AB\b}{}^\g$ are defined 
using the $\Omega_{M\b}{}^\g$ spin connection,
and $T_{AB}{}^\ah$ and $\hat R_{AB\bh}{}^\gh$ are defined 
using the $\hat\Omega_{M\bh}{}^\gh$ spin connection and $H=dB$.
The nilpotency constraints on $R_{ABC}{}^D$
are implied through Bianchi identities by the nilpotency constraints
on $T_{AB}{}^C$.

{\em Holomorphicity constraints }

The BRST charge must be conserved on the worldsheet, i.e. the right-moving BRST current must be holomorphic: $\pb (\l^\a d_\a)=0$; and the left-moving one anti-holomorphic. We need the equations of motion for $\l$ and $d$ coming from the action (\ref{curvedaction}).
To derive the constraints coming from holomorphicity of
$\l^\a d_\a$ and antiholomorphicity of $\lh^\a {\hat d}_\a$,
we vary $\l^\a$, $w_\a$, $\lh^\ah$ and we compute $E_\a^P (\d S/\d Z^P)$, obtaining an expression for $\bar\p(\l^\a d_\a)$.

Plugging into this expression
the equations of motion coming from
varying $d_\a$ and ${\hat d}_\ah$,
\bea
\bar\Pi^\a =  -P^{\a\bh}
\hat d_\bh- \hat C_\bh^{\gh\a}\lh^\bh \hat w_\gh, \quad
\Pi^\ah =  P^{\b\ah}
d_\b-  C_\b^{\g \ah} \l^\b w_\g,
\label{dbarZII}
\eea
one finds that holomorphicity of
$\l^\a d_\a$ implies that
\bea
T_{\a (bc)} = H_{\a cd} =  H_{\a\bh\g} =
T_{\a\b c} + H_{\a\b c}=
T_{\a\bh c} - H_{\a\bh c}=0 
\label{plugII}
\eea
$$  T_{\a c}{}^\b + T_{\a\gh c} P^{\b\gh} =
 T_{\a c}{}^\bh - T_{\a\g c} P^{\g\bh} =
T_{\a\b}{}^\gh  -\half H_{\a \b \g} P^{\g\gh} = T_{\a\gh}{}^\b=0,$$
$$ C^{\g\bh}_\a + \nabla_\a P^{\g\bh} - T_{\a\rho}{}^\g P^{\rho\bh}=
\hat R_{c\a\bh}{}^\gh +T_{\a\rho c} \hat C_{\bh}^{\gh \rho}=
\hat R_{\a\d\bh}{}^\gh -\half H_{\a\d\rho} C_\bh^{\gh \rho} =0,$$
$$
S_{\a\gh}^{\rho\dh} + \hat R_{\a\bh\gh}{}^\dh P^{\rho\bh} +\nabla_\a 
\hat C^{\dh\rho}_\gh - T_{\a\b}{}^\rho \hat C_\gh^{\dh\b} =0, $$
$$ \l^\a\l^\b  (R_{c \a \b}{}^\g  + T_{\a\dh c}C_\b^{\g\dh})=
\l^\a\l^\b  R_{\dh \a \b}{}^\g = 0,$$
$$
\l^\a\l^\b  (\nabla_\a C^{ \d\gh}_\b -R_{ \a \k\b}{}^\d P^{\k\gh}) =
\l^\a\l^\b  (\nabla_\a S_{\b\gh}^{\rho\dh} - \hat 
R_{\a\hat\kappa\gh}{}^\dh C_\b^{\rho\hat\kappa} -
R_{\a\kappa\b}{}^\rho \hat C_\gh^{\dh\kappa})=0,$$
where $\Pi^A = E^A_M \p Z^M$, $\bar\Pi^A = E^A_M \bar\p Z^M$,
$T_{ABc} = \eta_{cd}
T_{AB}{}^d$,
and all
superspace derivatives acting on unhatted spinor indices
are covariantized using the $\O_{M\a}{}^{\b}$ connection while
all superspace derivatives acting on hatted spinor indices
are covariantized using the $\hat\O_{M\ah}{}^\bh$ connection.
Furthermore, the torsion $T_{AB}{}^\a$ and curvature $R_{AB \g}{}^{\d}$
are defined using the $\O_{M\a}{}^\b$ connection
whereas the torsion
$T_{AB}{}^\ah$ and curvature $\hat R_{AB\gh}{}^{\dh}$ are defined
using the $\hat\O_{M\ah}{}^\bh$ connection.
Note that $T_{Abc}$ appears only in the combination $T_{\a(bc)}$.
This combination is independent of
the spin connections 
since $\O_M^{(s)}$ and ${\hat{\Omega}}_M^{(s)}$
only act on spinor indices and since
$\O_M^{ab}$ and ${\hat{\Omega}}_M^{ab}$ are antisymmetric in their vector indices.
The last two lines 
of equations must be satisfied for any pure spinor
$\l^\a$.
Antiholomorphicity of $\lh^\ah {\hat d}_\ah$ implies the hatted
version of the above equations. The only subtle point is
that it implies $T_{\ah\bh c}-H_{\ah\bh c}= T_{\ah\b c} + H_{\ah\b c}=0$, 
which together
with the above equations implies that 
\bea
T_{\a\b c}+ H_{\a\b c}= T_{\ah\bh c}- H_{\ah\bh c} = 
T_{\a\bh c} = H_{\a\bh c}=0.
\label{addhol}
\eea

The constraints of (\ref{nilII}) and (\ref{plugII}) will now
be shown to imply the correct Type II supergravity 
equations of motion.

{\em Supergravity constraints}

At scaling dimension
$-\half$, the constraints of (\ref{nilII}) imply
that 
\bea
H_{\a\b\g}=H_{\a\b\gh}=H_{\a\bh\gh}=
H_{\ah\bh\gh}=0
\label{Hzero}
\eea
since there is no non-zero symmetric 
$H_{\a\b\g}$ and $H_{\ah\bh\gh}$ satisfying $\l^\a\l^\b H_{\a\b\g}=0$
and $\lh^\ah\lh^\bh H_{\ah\bh\gh}=0$. 

At dimension 0, the constraints $\l^\a\l^\b T_{\a\b}{}^c=\lh^\ah\lh^\bh
T_{\ah\bh}{}^c= 0$ imply that  
$T_{\a\b}{}^c = i(\g^d)_{\a\b} f_d^c$ and 
$T_{\ah\bh}{}^c =i (\g^d)_{\ah\bh} \hat f_d^c$ 
for
some $f_d^c$ and $\hat f^d_c$. 
Using the dimension zero H Bianchi identities and
the local Lorentz and scale transformations
of (\ref{localtwo}) for the unhatted and hatted spinor indices
independently, both $f_d^c$ and $\hat f_d^c$ can be gauge-fixed to $\d_d^c$.
After this gauge-fixing, the only remaining gauge invariance is a single local
Lorentz invariance which acts on all spinor and vector indices in the standard
fashion.
Combining with the other dimension 0 constraints of (\ref{nilII}) and
(\ref{plugII}), one has
\bea
T_{\a\b}{}^c = - \eta^{cd} H_{\a\b d} =i(\g^c)_{\a\b}, \quad 
T_{\ah\bh}{}^c = \eta^{cd} H_{\ah\bh d} =i(\g^c)_{\ah\bh}, \quad
T_{\a\bh}{}^c=H_{\a\bh c}=0.
\label{scalarsII}
\eea

At dimension $\half$, the constraints $\l^\a\l^\b T_{\a\b}{}^\g=0$
and
$\lh^\ah\lh^\bh T_{\ah\bh}{}^\gh=0$
imply that $T_{\a\b}{}^\g
= f_c^\g (\g^c)_{\a\b}$ and
$T_{\ah\bh}{}^\gh
= \hat f_c^\gh (\g^c)_{\ah\bh}$ 
for some $f_c^\g$ and $\hat f_c^\gh$. 
Using the shift symmetries
\bea\nonumber
\d \Omega_\a^{(s)}  = (\g_c)_{\a\b} h^{c\b},\quad \d\Omega_\a^{bc}=
2(\g^{[b})_{\a\b} h^{c]\b},\\ \d d_\a = -\d\O_{\a\b}{}^\g \l^\b
w_\g,\quad \d U_{I\a}{}^{\b}=W_I^\g \d \O_{\g\a}{}^\b, \label{wgauge}
\eea
both $f_c^\g$ and $\hat f_c^\gh$
can be gauge-fixed to zero so that
$T_{\a\b}{}^\g= T_{\ah\bh}{}^\gh =0$.

At dimension one, the constraint $T_{c \a}{}^\b= T_{c\ah}{}^\bh=0$ 
decomposes into
\bea
T_{c\a}{}^\b= T_c^{defg}(\g_{defg})_\a{}^\b+ T_c^{de}(\g_{de})_\a{}^\b
+ T_c \d_\a^\b=0,
\label{decompII}
\eea
$$T_{c\ah}{}^\bh= \hat T_c^{defg}(\g_{defg})_\ah{}^\bh+ \hat 
T_c^{de}(\g_{de})_\ah{}^\bh
+ \hat T_c \d_\ah^\bh=0.$$
The constraints
$T_c=\hat T_c=0$ and
$T_c^{de}=\hat T_c^{de}=0$ determine the vector components
of the spin connections $\O_c^{(s)}$, ${\hat{\Omega}}_c^{(s)}$,
$\O_c{}^{de}$ and ${\hat{\Omega}}_c{}^{de}$, whereas the
constraint
$T_c^{defg}=\hat T_c^{defg}=0$ is implied by the Bianchi identities
$(DH + TH)_{bc\alpha\gamma} (\g^{bdefg})^{\a\g}=0$
and
$(DH + TH)_{bc\ah\gh} (\g^{bdefg})^{\ah\gh}=0$.
The constraints $T_{\a c}{}^\bh = (\g_c)_{\a\g} P^{\g\bh}$ 
and
$T_{\ah c}{}^\b = (\g_c)_{\ah\gh} \hat P^{\b\gh}$ 
for some
$P^{\g\bh}$ and $\hat P^{\b\gh}$ are implied by the Bianchi identities
$(\nabla T +TT)_{\a\b\g}^\dh = 
(\nabla T +TT)_{\ah\bh\gh}^\d = 0$. And $P^{\g\bh}=
\hat P^{\g\bh}$ is implied by the Bianchi identity 
$(\nabla T + TT)_{\a\bh c}^c=0$.
Similarly, all other constraints in (\ref{nilII}) and (\ref{plugII}) are
either implied by Bianchi identities or define $C_\a^{\b\gh}$,
$\hat C_\ah^{\bh\g}$ and $S_{\b\dh}^{\a\gh}$ in terms of the supervielbein.

The above constraints imply that all background superfields appearing in
the action of (\ref{curvedaction}) can be expressed in terms of the
spinor supervielbein $E_\a^M$ and $E_\ah^M$. Furthermore,
the constraints 
\bea
T_{\a\b}{}^c=i(\g^c)_{\a\b},\quad T_{\ah\bh}{}^c= i (\g^c)_{\ah\bh},\quad
T_{\a\bh}{}^c=0
\label{scalarcons}
\eea
imply the on-shell equations of motion for $E_\a^M$ and $E_\ah^M$.
So the constraints of 
(\ref{nilII}) and (\ref{plugII}) imply the Type II supergravity
equations of motion. 

\subsubsection{Invertible R-R superfield}

Let us specialize to the case in which the vacuum expectation  value of the RR superfield $P^{\a\ah}$ is invertible and denote its inverse by $P_{\a\ah}$ such that $P_{\a\ah}P^{\a\bh}=\delta_\ah^\bh$, $P_{\a\ah}P^{\b\ah}=\delta_\a^\b$. 
The variables $d$ and $\hat d$ couple to the R-R field strength through the term $d_\a\hat d_\ah P^{\a\ah}$ in the action (\ref{curvedaction}). If $P$ is invertible we can integrate $d$ and $\hat d$ out upon their equations of motion 
\bea\label{dout}
d_\a=&P_{\a\ah}(\hat\Pi^\ah+\l^\rho w_\sigma C_\rho^{\sigma\ah}) \ ,\\
\hat d_\ah=&-P_{\a\ah}(\bar\Pi^\a+\hat\l^{\hat\rho}\hat w_{\hat\sigma}\tilde C_{\hat\rho}{}^{\hat\sigma\a}) \ .
\eea
Substituting (\ref{dout}) into the stress tensor (\ref{stressca}) we find
\bea\label{stresscu}
T= -\half\Pi^a\Pi^b\eta_{ab}-P_{\g\gh}(\hat\Pi^\gh+\l^\rho w_\b C_\rho^{\b\gh})\Pi^\g -w_\a(\nabla\l)^\a \ .
\eea
The proof that the stress tensor (\ref{stressca}) is separately  invariant under the BRST transformations generated by the left and right-moving BRST charges
\bea\label{stressinv}
\{Q,T\}=\{\hat Q, T\}=0 \ ,
\eea
involves the supergravity constraints that we just introduced.

\medskip
{\bf Guide to the literature}

The fact that pure spinors can be used to describe ten-dimensional on-shell super Yang-Mills theory and supergravity was noticed a long time ago  \cite{Siegel:1978yi,Witten:1985nt,Howe:1991mf,Howe:1991bx}. Building on this fact, the pure spinor formalism for the superstring was put forward by Berkovits in \cite{Berkovits:2000fe}. 
The tree level prescription for scattering amplitudes is given  in \cite{Berkovits:2000fe}, while the multiloop prescription for scattering amplitudes in flat space appears in  \cite{Berkovits:2004px}. The multiloop prescription of \cite{Berkovits:2004px} involves picture changing operators and has been reformulated in terms of the non-minimal pure spinor formalism in \cite{Berkovits:2005bt,Berkovits:2006vi}.
An exhaustive discussion of the anomalies of the pure spinor curved beta-gamma system and its $U(5)$ decomposition can be found in \cite{Nekrasov:2005wg}. 

The pure spinor formalism in a generic type II supergravity background is discussed in \cite{Berkovits:2001ue}, where the holomorphicity and nilpotency constraints are derived (for a review, see \cite{Guttenberg:2008ic}).  A proof of the equivalence of such constraints with the more familiar Howe and West formalism \cite{Howe:1983sra} is given, too. The relation between the constraints and worldsheet one-loop conformal invariance, which we have omitted, can be found in \cite{Bedoya:2006ic}. The case of a supergravity background with an invertible RR superfield is discussed in \cite{Berkovits:2010zz}, where the properties of the stress tensor and the $b$ antighost are studied.  A general procedure to construct the pure spinor action in type IIA supergravity backgrounds using free differential algebras appears in \cite{D'Auria:2008ny,Tonin:2010mm}.

Since we have in mind the application of the pure spinor formalism to $AdS_5\times S^5$ background, we have not discussed at all the issue of the origin of the pure spinor formalism, namely how to obtain such a quantum action and BRST charge by performing a gauge fixing of a classical action. Some papers that address this problem are \cite{Grassi:2001ug,Matone:2002ft,Aisaka:2005vn}. In the superparticle case, it is actually possible to obtain the pure spinor formalism from the classical Brink-Schwarz action \cite{Berkovits:2002zk}.
The pure spinor superstring is equivalent to the RNS and the GS formalisms. Although we will not address this issue in this review, the interested reader can find the proof of equivalence to the RNS string in \cite{Berkovits:2001us,Berkovits:2007wz} and to the light-cone GS string in \cite{Berkovits:2004tw,Gaona:2005yw}.

\newpage


\section{Supercoset sigma models}
\label{section:supercoset}

In this Section, we introduce non-linear sigma models on supercosets. We first derive the curved AdS superspace geometry using the supergravity constraints from the previous Section, then discuss the issues related to the choice of the light-cone gauge in the GS formalism, as a motivation to introduce the pure spinor formalism. After a brief introduction to supergroups and supercosets, we construct the GS action on a supercoset that admits a $\ZZ_4$ grading and prove its kappa-symmetry. We finally introduce the pure spinor action in $AdS_5\times S^5$ and briefly mention the generalization to other interesting supercosets.
In \ref{appendix:supergroups} the reader can find more details on supercosets and the structure constants of the $PSU(2,2|4)$ supergroup.

\subsection{AdS geometry}

In this section we will solve the supergravity constraints discussed in Section~\ref{section:constraints} and derive the value of the superfields in the particular $AdS_5\times S^5$. We will then plug such background superfields in the action (\ref{curvedaction}) and derive the worldsheet non-linear sigma model action.

The type IIB supergravity solution for $AdS_5\times S^5$ is given by the metric and five-form flux
\bea\label{adssugra}
 &ds^2=e^{2\phi/R}[-dt^2+(dx^i)^2]+d\phi^2+R^2d\Omega_5^2 \ ,\\
 &F_5=(1+*){\rm Vol}(S^5){N_c/R^5} \ ,
\eea
where $N_c$ are the units of RR flux through the five-sphere and $R^2=\a'(4\pi g_s N_c)^{1/ 2}$ is the radius of the sphere and of AdS and the AdS boundary is at $\phi\to\infty$.

The type IIB Ramond-Ramond fluxes couple to the worldsheet fields through the vertex operator 
$$
{\cal V}^{(0)}_{RR}=\int d^2z\, d_\alpha P^{\alpha\bh}\widehat d_\bh \ .
$$ 
The bispinor superfield
 \bea\label{bispi}
 P^{\alpha\bh}=g_s\left((\gamma^m)^{\alpha\bh}F_m+
{1\over 3!}(\gamma^{mnp})^{\alpha\bh}F_{mnp}+{1\over 5!}(\gamma^{mnpqr})^{\alpha\bh}F_{mnpqr}\right) \ ,
 \eea
contains the  one- three- and five-form fluxes $F$'s and in the background (\ref{adssugra}) takes the form
 \bea\label{bispifive}
 P^{\alpha\bh}={\eta^{\alpha\bh}\over (g_s N_c)^{1/4}} \ ,\qquad \eta^{\alpha\bh}=(\gamma^{01234})^{\alpha\bh} \ .
 \eea
The matrix $\eta^{\alpha\bh}$ has rank sixteen and is numerically equal to the identity
matrix, in particular it is invertible.

Consider the definition of the superfield three-form flux with curved indices $H=dB$. One can pass to flat indices by introducing the super-vielbeins and recalling the definition of the torsion $T^A=\nabla E^A=dE^A-E^B\wedge \Omega_B{}^A$, where $\Omega_B{}^A$ is the spin connection. We obtain then the flat index equation $H=\nabla B+TB$ and in particular the component $H_{a\alpha\beta}=T_{a\alpha}{}^\bh B_{\beta\bh}$.  The supergravity constraints fix $H_{a\alpha\beta}=-T_{\alpha\beta}^b\eta_{ab}=-(\gamma_a)_{\alpha\beta}$ and the Bianchi identity for the torsion implies that $T_{a\alpha}{}^\bh=-(\gamma_a)_{\alpha\beta}P^{\beta\bh}$. Hence we obtain a non-vanishing component of the super B-field
 \bea\label{bfield}
 B_{\alpha\bh}=\half P^{-1}_{\alpha \bh}=\half (g_s N_c)^{1/4}\eta_{\alpha\bh} \ .
 \eea

The pull-back of the spin connection is given by $d Z^M\Omega_{M\alpha}{}^\beta=\half J^{mn}(\gamma_{mn})_\alpha{}^\beta$, while the Riemann curvature is computed as follows. From the torsion Bianchi identity one finds $R_{abcd}={1\over 8} (\gamma_{cd})_\beta{}^\alpha T_{\alpha[a}{}^{\bh}T_{b]\ah}{}^\beta$, then by using $T_{a\alpha}^\bh=(\gamma_a)_{\alpha\beta}P^{\beta\bh}$ and
$T_{a\ah}^\beta=(\gamma_a)_{\ah\bh}P^{\beta\bh}$, together with the definitions of the gamma matrices in the Appendix and using the fact that $[\gamma_{ab},\gamma_{a'b'}]=0$, we find that
 \bea\label{riemannads}
 R_{abcd}={2\over (g_sN_c)^{1/2}}\eta_{a[b}\eta_{c]d} \ , \qquad
 R_{a'b'c'd'}=-{2\over (g_sN_c)^{1/2}}\eta_{a'[b'}\eta_{c']d'} \ .
 \eea
where the unprimed indices denote the $AdS$ directions and the primed indices denote the sphere directions.

\subsubsection{Green-Schwarz}

We write the sigma model by making use of the pull-backs of the supervielbeins $J^A=dZ^M E^A_M$, hence the sigma model fields will take values in the tangent superspace. The Green-Schwarz-Metsaev-Tseytlin sigma model on the $AdS_5\times S^5$ background depends only on $B_{\alpha\bh}$, which takes care of the Wess-Zumino term\footnote{Unlike in the pure spinor formalism, in which $\theta^\alpha$ and $\theta^\ah$ have independent Lorentz transformations, in the Green-Schwarz formulation they transform with the same Lorentz parameter. Hence, in the Green-Schwarz-Metsaev-Tseytlin action we could drop the hatted notation on the $\theta^\ah$. However, we keep it anyway because our goal here is to construct the pure spinor action.}
\bea\label{GSMT}
 S={R^2\over 2\pi}\int\,d^2\sigma \left( \sqrt{g}g^{ij}J^a_iJ^b_j\eta_{ab}+{\epsilon^{ij}\over 2}\eta_{\alpha\bh}J^\alpha_iJ^{\bh}_j\right) \ .
\eea
We have performed a rescaling of the currents $J^a\to (g_s N_c)^{1/4}J^a$, $J^\alpha\to(g_s N_c)^{1/8}J^\alpha$ and $J^{\ah}\to (g_sN_c)^{1/8} J^\ah$. The sigma model inverse coupling squared is usually set to be $R^2/\a'$. By the AdS/CFT correspondence, such quantity is equal to the 't Hooft coupling $\mL=g_{YM}^2 N_c$ of $\cN=4$ super Yang-Mills theory
\be
R^2/\alpha'=\sqrt{\mL} \ .
\ee
We will set $\alpha'=1$ in the following.

By performing a $\kappa$-symmetry transformation
 \bea\label{kappads}
 \delta_{\kappa} Z^M E_M^a=0 \ ,\quad \delta_\kappa Z^M E_M^\alpha=\rho^\alpha \ ,\quad\delta_\kappa Z^M E_M^\ah=\widehat\rho^\ah \ ,
 \eea
where $\rho^\alpha=\half (\gamma_a)_{\ah\bh}\eta^{\alpha\bh}J^a_i\kappa^{i\beta}$ and $\widehat\rho^\ah=-\half(\gamma_a)^{\alpha\beta} \eta^{\beta\ah} J^a_i\kappa^{i\alpha}$, the variation of the action is proportional to the Virasoro constraints, which can be canceled by an appropriate transformation of the worldsheet metric
 \bea\label{kappah}
 \delta_\kappa (\sqrt{g} g^{ij}) = 4 \sqrt{h}\, \eta_{\alpha\ah}
  (P_-^{ik} J_{k}^{\hat \alpha} \kappa^{j \alpha} - P_+^{i k}
  J_{k}^\alpha k^{j \hat \alpha}) \ ,
\eea
where $\kappa^{i\alpha}=P_-^{ij}\kappa^\alpha_j$, $\kappa^{i\ah}=P_+^{ij}\kappa^\ah_j$ and $P_\pm$ are the usual worldsheet projectors.

Note that while in flat space the WZ term transforms as a total derivative under supersymmetry, the action (\ref{GSMT}) is manifestly invariant under the target space isometry $PSU(2,2\vert 4)$.

{\em Light-cone gauge issues}

In flat space and in a plane wave background, the Green-Schwarz (GS) action can be quantized in the light-cone gauge, where the theory becomes free and the spectrum can be easily computed. In AdS background (\ref{adssugra}), however, the use of the GS sigma model is limited, due to the fact that it is hard to impose  the light-cone gauge when expanding around empty AdS. Therefore, it is hard to study the perturbative spectrum of the GS formulation around empty AdS.  On the other hand, the light-cone gauge can be imposed consistently if we expand around a classical solution of the GS equations of motion for string configurations with energies of order ${\cal O}(\l)$. This is the solitonic sector of string theory and it contains a variety of macroscopic strings. By the AdS/CFT correspondence, these strings are dual to gauge invariant local operators with an asymptotically large number of elementary fields on the super Yang-Mills side. Let us see why it is hard to fix the light-cone gauge around empty AdS background.

In flat space or in a plane wave background, the worldsheet symmetries can be fixed in the following way. As a first step, we choose the $\kappa$-symmetry light-cone gauge $\gamma^+ \theta^{I}=0$, where $\gamma^+=\gamma^3+\gamma^0$,  that drastically simplifies the equation of motion for $X^+$ to 
\bea\label{lighteo}
\partial_i(\sqrt{g}g^{ij}\partial_j X^+)=0 \ .
\eea
Then one chooses the conformal gauge $g^{ij}=\eta^{ij}$ on the worldsheet, which leaves the residual invariance $\{\sigma^+,\sigma^-\}\to\{\tilde \sigma^+(\sigma^+),\tilde \sigma^-(\sigma^-)\}$. This implies that $\tau(\sigma^+,\sigma^-)$ is a solution of $\partial_+\partial_-\tau=0$. In conformal gauge, the equation of motion (\ref{lighteo}) becomes $\partial_+\partial_- X^+=0$. We can thus fix the residual invariance by setting $\tau\sim X^+$ in a way consistent with the equations of motion for $X^+$.

The above procedure can be implemented whenever the background geometry is a direct product $\RR^{1,1}\times M_8$. The $AdS_5\times S^5$ metric (\ref{adssugra})
does not fall in this category. Let us try to apply the naive gauge fixing procedure outlined above and see what goes wrong. The $\kappa$-symmetry light-cone gauge $\gamma^+\theta^I=0$ simplifies the GS action, leaving in the sigma model a linear dependence on the $X^-$ coordinates. Upon integrating it out, we obtain the constraint
\bea\label{constraintxplus}
\partial_i(e^{2\phi}\sqrt{g}g^{ij}\partial_j X^+)=0 \ ,
\eea
which, unlike (\ref{lighteo}), depends explicitly on the radial coordinate. After choosing the conformal gauge $g^{ij}=\eta^{ij}$, the residual invariance, encoded in the equation $\partial_+\partial_-\tau=0$, is not compatible with the equation for $X^+$ anymore, because of the explicit appearance of the warp factor depending on the radial direction $e^{2\phi}$. We can rephrase this obstruction in a more geometrical language, by observing that in AdS there is no globally defined null Killing vector. The norm of a would-be null Killing vector is proportional to $e^{2\phi}$, which vanishes near the AdS horizon $\phi=-\infty$.

One might try a different approach and, after fixing the $\kappa$-symmetry, solve first the constraint (\ref{constraintxplus}) in terms of a generic function $f(\sigma,\tau)$ as
\bea\label{solvexplus}
e^{2\phi}\p_i X^+=g_{ij}{\epsilon^{jk}\over \sqrt{g}}\partial_k f \ .
\eea
This is just a field redefinition, since the GS action in the $\kappa$-symmetry light-cone gauge depends on $X^+$ only through the combination $e^{2\phi}\partial_i X^+$. Then one fixes the reparametrization invariance by choosing
\bea\label{fixrep}
\sqrt{g}g^{ij}={\rm diag}(-e^{-2\phi},e^{2\phi}) \ .
\eea
It turns out that the gauge (\ref{fixrep}) is compatible with the special choice $f=\sigma$ and $X^+=\tau$ in (\ref{solvexplus}), so that the kinetic term for the remaining fermions, schematically of the form $\bar \theta D\theta$, is non-degenerate, i.e. it does not depend on the AdS background variables.

The gauge choice (\ref{fixrep}), unlike the naive light-cone gauge, is consistent even in an AdS background, but it leads to two problems. Firstly, the resulting sigma model is defined on a {\it curved} worldsheet geometry, where the worldsheet metric is related to the radial profile in AdS and one looses the two dimensional Lorentz symmetry. Secondly, the resulting action is an interacting sigma model: unlike the light-cone action in flat space, it cannot be quantized in terms of free fields. It remains an open problem to find a change of variables that may lead to further progress in the quantization. 

Despite these difficulties, it has been recently shown that one can extrapolate the spectrum of perturbative strings by first quantizing the sigma model around a semi-classical string with energy of order ${\cal O}(\mL)$ and large spin and/or R-charge, and then take the limit in which spin and/or R-charge become of order one. This extrapolation to the perturbative string regime reproduces the results that we will discuss in Section~\ref{section:spectrum}, where we will compute the perturbative string spectrum using the pure spinor formalism. It is an open problem to understand why  this limit is valid.

Having seen that the GS sigma model is hard to quantize around empty AdS, we turn now to the study of the pure spinor sigma model.

\subsubsection{Pure spinor}

The pure spinor action in a general background is given by (\ref{curvedaction}), that we can rewrite schematically as
 \bea\label{actionss}
 S=S_{GS}+S_{\kappa}+S_{ghost} \ .
 \eea
By plugging the values of the background fields (\ref{bispifive}), (\ref{bfield}), (\ref{riemannads}), together with the super-vielbeins and the spin connections, we specialize the pure spinor action to the maximally supersymmetric $AdS_5\times S^5$. The first term is the Green-Schwarz action (\ref{GSMT}) in conformal gauge
 $$
 S_{GS}={R^2\over 2\pi}\int\,d^2z \left( \half J^a\bar J^b\eta_{ab}-{1\over 4}\eta_{\alpha\bh}J^\alpha\bar J^{\bh}+
{1\over 4}\eta_{\alpha\bh}\bar J^\alpha J^{\bh})\right) \ .
 $$
The fermionic part of the action, antisymmetric in the worldsheet index, is usually called the WZ term and it does not provide a kinetic term for the fermions. The second term is the coupling to the background super-vielbeins and the RR superfield $P^{\alpha\ah}$ of the conjugate momenta to the $\theta$'s, which breaks $\kappa$-symmetry of the GS action and introduces a kinetic term for the fermions
 $$ S_{\kappa}={R^2\over 2\pi}\int\,d^2z \left(
 d_\alpha \bar J^\alpha+\hat d_\ah J^\ah+d_\alpha \hat d_{\ah}\eta^{\alpha\ah}\right) \ .
 $$
The matter part of the action contains the coupling of the dilaton to the worldsheet curvature through the Fradkin-Tseytlin term, but since in the $AdS_5\times S^5$ background the dilaton is constant, we will not write it explicitly. The last term in (\ref{actionss}) contains the action for the ghosts and their coupling to the spin connections and the Riemann curvature\footnote{We rescaled the $d$'s by $(g_sN_c)^{3/8}$, the Lorents generators for the ghost by $(g_sN_c)^{1/2}$ and the matter currents as in (\ref{GSMT}).}
\bea\label{sghosts} S_{ghost}={R^2\over 2\pi}\int\,d^2z\left(
 -w_\alpha (\bar\nabla \lambda)^\alpha+\hat w_\ah(\nabla\hat\l)^\ah-{1\over2}( N^{ab}\hat N_{ab}-N^{a'b'}\hat N_{a'b'})\right) \ ,
 \eea
where
 \bea\label{covlambda}
 (\bar\nabla\l)^\alpha=&\bar\partial \l^\alpha+\half \bar J_{ab}(\gamma^{ab}\l)^\alpha +\half \bar J_{a'b'}(\gamma^{a'b'}\l)^\alpha \ ,
 \\
 (\nabla\hat\l)^\ah=&\partial \hat\l^\ah+\half J_{ab}(\gamma^{ab}\hat\l)^\ah +\half J_{a'b'}(\gamma^{a'b'}\hat\l)^\ah \ , \nonumber
 \eea
and $(N,\hat N)$ are the Lorentz generators in the ghost sector, defined in (\ref{lorentz}). Since the RR superfield is a rank sixteen matrix, the $d$'s are auxiliary fields and we can integrate them out upon their equations of motion
 $$
 d_\alpha= J^\ah\eta_{\alpha\ah} \ ,\quad \hat d_\ah=-\bar J^\alpha\eta_{\alpha\ah} \ ,
 $$
obtaining the final form of the pure spinor action
 \bea\label{actionads}
 S={R^2\over 2\pi}\int\,d^2z \left( \half J^a\bar J^b\eta_{ab}-{1\over 4}\eta_{\alpha\bh}J^\alpha\bar J^{\bh}-
{3\over 4}\eta_{\alpha\bh}\bar J^\alpha J^{\bh})\right)+S_{ghost} \ .
 \eea
The asymmetry in the factors $1/4$ and $3/4$ in front of the fermionic terms is easily understood. It is just the sum of the WZ term in the GS action, which carries a $1/4$ factor, with the kinetic term for the fermions $-\eta_{\a\bh}\bar J^\a J^\bh$, coming from integrating out the auxiliary variables $d_\a$ and $\hat d_\ah$.

The action (\ref{actionads}) is invariant under the BRST transformations generated by
 $$Q=\oint\, dz \lambda^\alpha d_\alpha+\oint\,d\bar z\hat\lambda^\ah \hat d_\ah = -\oint\, dz \lambda^\alpha J^\ah\eta_{\alpha\ah}+\oint\,d\bar z\hat\lambda^\ah \bar J^\alpha\eta_{\alpha\ah} \ ,
 $$
and under the local Lorentz symmetry $SO(1,4)\times SO(5)$ and the global $PSU(2,2|4)$ supergroup. We postpone the discussion of the action and its symmetries to the next section, in which we will arrive at (\ref{actionads}) from a supercoset construction, that will give us a deeper insight into the geometry.

\subsection{Supercosets}

In this section we will rederive the pure spinor action (\ref{actionads}) using a supercoset construction.
The maximally supersymmetric type IIB background $AdS_5\times S^5$ is described by the supercoset 
$$
{G\over H}={PSU(2,2|4)\over SO(1,4)\times SO(5)} \ .
$$
After reviewing the salient features of this supercoset, we will introduce the superstring sigma model and discuss its local and global symmetries. The crucial ingredient will be the presence of a $\ZZ_4$ automorphism of the $\mathfrak{psu}(n,n|2n)$ super Lie-algebra, whose invariant locus is the gauge symmetry group $H$. The same $\ZZ_4$ grading will be used later in Section~\ref{section:integrability} to prove the integrability of the sigma model.

The supergroup $U(m|m)$ can be represented in terms of the $m|m$ complex unitary supermatrices. Its bosonic subgroup is $U(m)\times U(m)$ and the fermionic generators transform in the $({\bf m}\otimes \bar{\bf m})\oplus (\bar{\bf m}\otimes {\bf m})$. Its Lie superalgebra ${\cal G}$ admits a representation in terms of supermatrices
\bea\label{superM}
M= \left(\ba{cc}
A&X\\
Y &B\ea\right)
\eea
where $A$ and $B$ are bosonic hermitian matrices and $X,Y$ are fermionic matrices. We can define the supertrace operation by
$$
\str M=\tr A-(-)^{deg(M)} \tr B \ ,
$$
where $deg(M)=0$ if $A,B$ are bosonic and one otherwise. The supergroup can be decomposed as $U(m|m)=PSU(m|m)\times U(1)_Y\times U(1)_D$, where the $2m^2-2$ generators of $PSU(m|m)$ are the traceless and supertraceless ones, that we denote $\bf T_A$, while the two remaining generators are
\bea
\label{remaining}
\bf Y=\left(\ba{cc}
\Id_m&0\\
0&-\Id_m\ea\right)\ ,\quad
I=\left(\ba{cc}
\Id_m&0\\
0&\Id_m\ea\right)\ ,
\eea
where $\bf Y$, usually denoted hypercharge, acts as an outer automorphism and $\bf I$, generator of the diagonal $U(1)_D$, is a central extension
\bea
 &\left[{\bf T_A}, {\bf T_B}\right\}=F^C_{AB} {\bf T_C}+d_{AB} {\bf I} \ ,\nonumber\\
 &\left[{\bf Y},{\bf T_A}\right]=C^B_A {\bf T_B} \ ,\label{extension}\\
 & [{\bf Y},{\bf I}]=[{\bf I},{\bf T_A}]=0 \ . \nonumber 
 \eea
Imposing the condition $\str M=0$ removes ${\bf Y}$ and gives the superalgebra of $SU(m|m)$. If we quotient the supergroup by the action of ${\bf I}$ we get the supergroup denoted by $PSU(m|m)$. One can still define the superalgebra $\mathfrak{psu}(m|m)$ by the supermatrix
(\ref{superM}), subject to the condition $\tr A=\tr B=0$. Since it is not possible to represent this last condition in a $PSU(m|m)$ invariant way (which would involve super-traces, super-determinants or super-matrix relations), this superalgebra does not admit a matrix representation; instead, we realize it by a coset construction as we just described.

The superalgebra $\mathfrak{psu}(2n|2n)$ admits a $\ZZ_4$ automorphism generated by
 $$
 \Omega(M)=\left(\ba{cc}
 JA^tJ &-JY^tJ\\
 JX^tJ& JB^tJ\ea\right)\ ,\qquad J=\left(\ba{cc}0 & -\Id_n\\ \Id_n &0 \ea\right)\ ,
 $$
which splits the superalgebra into its four eigenspaces
\bea\label{eigenfour}
 {\cal G}={\cal H}_1\oplus\cH_2\oplus \cH_3\oplus \cH_4 \ ,
 \eea
where $\Omega(\cH_k)=i^k \cH_k$. The fermionic generators belong to $\cH_1$ and $\cH_3$ and they are related by hermitian conjugation $(\cH_1)^\dagger=\cH_3$. This last fact will imply that the left and right moving pure spinors in the AdS background are the complex conjugate of each other and will bear important consequence for the scattering amplitude computations.

The invariant locus of this $\ZZ_4$ automorphism is given by the bosonic subalgebra $usp(2n)\times usp(2n)$. Since the $\ZZ_4$ grading is an automorphism of the Lie superalgebra, we have that
 $$
 [\cH_k,\cH_l\}\subset \cH_{k+l} \qquad {\rm mod}\,4 \ ,
 $$
and it is compatible with a supertrace operation
$$
 \str \cH_k\cH_l=0\quad {\rm unless} \quad k+l=0 \quad {\rm mod}\,4 \ .
 $$

Let us specialize to the supergroup $PSU(2,2|4)$, whose bosonic subgroup is $SO(2,4)\times SO(6)$. The invariant locus of the superalgebra under the $\ZZ_4$ automorphism is $\cH_0=SO(1,4)\times SO(5)$ and the bosonic part of the ten dimensional geometry of $AdS_5\times S^5$ is described by the coset
$$
{SO(2,4)\over SO(1,4)}\times {SO(6)\over SO(5)}.
$$
We divide the $\mathfrak{psu}(2,2|4)$ generators in the fundamental representation according to their grading. $\cH_0=\{{\bf T}_{[ab]}, {\bf T}_{[a'b']}\}$ are the generator of the Lorentz group, where $a,a'=1,\ldots,5$ are indices in the fundamental representation of $SO(1,4)$ and $SO(5)$ respectively. The other bosonic generators are the ten $\cH_2=\{{\bf T}_a,{\bf T}_{a'}\}$, which generate the translations along AdS and the five-sphere respectively. Finally, the thirty-two fermionic generators are $\cH_1=\{{\bf T}_\a\}$ and $\cH_3=\{{\bf T}_\ah\}$, where $\alpha,\ah=1,\ldots,16$ are ten dimensional Majorana-Weyl spinor indices (they both have the same chirality, as required by the type IIB superspace). The relation of this ten dimensional notation with the supermatrix representation in (\ref{superM}) is schematically ${\bf X}={\bf T}_\a+i{\bf T}_\ah$ and ${\bf Y}={\bf T}_\a-i{\bf T}_\ah$.
The non-vanishing supertraces of the generators are
 $$
 \str {\bf T}_a{\bf T}_b=\eta_{ab} \ ,\qquad \str {\bf T}_{a'}{\bf T}_{b'}=\delta_{a'b'} \ ,
 $$
 $$
 \str {\bf T}_{ab}{\bf T}_{cd}=\half\eta_{a[c}\eta_{d]b} \ ,\qquad
 \str {\bf T}_{a'b'}{\bf T}_{c'd'}=-\half\delta_{a'[c'}\delta_{d']b'} \ ,
 $$
 $$
 \str {\bf T}_\a {\bf T}_\ah=-\str {\bf T}_\ah {\bf T}_\a=\eta_{\a\ah} \ ,
 $$
where $\eta_{\a\ah}$ is numerically equal to the identity matrix and is the object we defined in (\ref{bispifive}). The supertrace of the generators in the fundamental representation $\str {\bf T}_A{\bf T}_B=\eta_{AB}$ defines the metric on the supergroup.
The non-vanishing structure constants of the $\mathfrak{psu}(2,2|4)$ superalgebra are given in Appendix B.

\subsubsection{Green-Schwarz}
\label{section:GS}

Let us construct the GS sigma model on the supercoset $G/H$, where the supergroup $G$ admits a $\ZZ_4$ automorphism, by gauging its invariant locus, whose superalgebra is $\cH_0$. Consider the maps 
\be
g(\sigma,\tau):\Sigma\rightarrow G/H \ ,
\ee 
from the string worldsheet to the supercoset. The coset element transforms as
\be
g(\sigma,\tau)\to g_0g(\sigma,\tau)h(\sigma,\tau) \ ,
\ee
under the global $g_0\in G$ and local $h(\sigma,\tau)\in H$ symmetry transformations. The pull-backs on the worldsheet of the target space supervielbeins are the Maurer-Cartan one-forms $J=g^{-1}dg$, which are invariant under the global symmetry that acts by left multiplication by a constant group element. The Maurer-Cartan currents take values in the Lie superalgebra of $G$ and can be decomposed according to their $\ZZ_4$ grading 
\be
J=J_0+J_1+J_2+J_3 \ .
\ee 
They satisfy the Maurer-Cartan equation
 \bea\label{maurercartan}
 dJ+J\wedge J=0 \ ,
 \eea
which split according to the $\ZZ_4$ grading of the generators and, in conformal gauge, read
\bea\label{eqMaurerCartan}
   \partial \bar J_0 -\bar\partial J_0+ [J_0, \bar J_0] + [J_1 ,\bar J_3]+[J_3,\bar J_1] + [J_2 ,\bar J_2]   =  0 \ , \nonumber\\
  \nabla \bar J_1 -\bar\nabla J_1+ [J_2,\bar J_3] + [J_3,
  \bar J_2]  =  0 \ ,\\
  \nabla \bar J_2+\bar \nabla J_2 + [J_1,\bar J_1] +[J_3,\bar
   J_3]  =  0 \ , \nonumber\\
  \nabla \bar J_3+\bar \nabla J_3 + [J_1 ,\bar J_2] + [J_2, \bar J_1] =
   0 \ , \nonumber
  \eea
  where $\nabla J_i=\partial J_i+[J_0,J_i]$.
We would like to gauge the part of the supergroup which is generated by $\cH_0$ under its right action $g(\sigma)\to g(\sigma)h(\sigma)$. While $J_1,J_2,J_3$ transform by conjugation $J\to h^{-1}Jh$, or infinitesimally
\bea\label{gaugeJ}
  \delta J = d \Lambda + [J, \Lambda] \ , \quad \Lambda \in
  \cH_0 \ ,
\eea
the grading zero current transforms inhomogeneously as $J_0\to h^{-1}J_0h+h^{-1}dh$. It is then clear that any gauge invariant lagrangian on the supercoset is given by a bilinear in the currents, whose total grading charge vanishes and does not contain $J_0$.

Using these ingredients, let us first construct the Green-Schwarz action on the supercoset. We have a kinetic term for the bosonic currents $J_2$, but we cannot allow a kinetic term for the fermionic currents, which would break $\kappa$-symmetry. The fermionic currents can only enter through a Wess-Zumino term and the sigma model action is given by
\bea\label{eqcosetGSaction}
  S_{GS} =& {1\over4} \int d^2\sigma\,\str \left(J_2 \wedge *J_2 + J_1
  \wedge J_3 \right)\\
  &= {1\over4} \int d^2 \sigma\, \str  \left(\sqrt{g}g^{ij}J_{2i}J_{2j}+\e^{ij}J_{1i}J_{3j}\right)\ . \nonumber
\eea
which is a particular form of (\ref{taction}). The coefficient of the WZ term is fixed by $\kappa$-symmetry, as we will shortly check. The slightly unusual feature is the form of the WZ term, that we will now discuss.

A sigma model on a supergroup manifold $G$ admits a WZ term constructed from a closed three-form, whose pull-back is written in terms of the Maurer-Cartan currents as
$$
 \Omega^{(3)}=\str J\wedge (J\wedge J)=\eta_{AB}f^B_{CD} J^A\wedge (J^C\wedge J^D) \ .
 $$
The three-form is closed because of the Maurer-Cartan equation (\ref{maurercartan}) and the Jacobi identity on the structure constants $f^B_{CD}$. The WZ term is the integral of $\Omega^{(3)}$ on a three dimensional manifold, whose boundary is the string worldsheet. Because the three-form is closed, it is possible to write it locally as a two form, then we can write the WZ term as a regular integral on the two-dimensional worldsheet as in (\ref{taction}).
In the case in which the target space is not a supergroup, but a supercoset equipped with a $\ZZ_4$ automorphism, we can readily write down a WZ term by using the fact that the gauge invariant operators have total grading zero and do not depend on $J_0$. The three-form $\Omega^{(3)}=\str [ (J_1\wedge J_1 -J_3\wedge J_3)\wedge J_2]$ takes values in the supercoset and it is exact due to the Maurer-Cartan equations (\ref{eqMaurerCartan}), hence $\Omega^{(3)}=d\Omega^{(2)}$, where $\Omega^{(2)}=\str J_1\wedge J_3$.

Another way to understand the two dimensional WZ term in the AdS background is the fact that in such background there is one tensor with the correct spinor indices, namely the inverse of the RR flux $\eta_{\a\ah}$, that yields a grading zero term in the action. Hence, the only possibility for a WZ term is the form $\eta_{\a\ah}\epsilon^{ij}J^\a_i J^\ah_j$ in (\ref{GSMT}).

The coefficient of the WZ term is fixed by requiring that the GS action (\ref{eqcosetGSaction}) is invariant under Siegel's $\kappa$-symmetry transformations. It is convenient to parameterize the
$\kappa$-transformation by
$$
  \delta_\kappa x_i \equiv \delta_\kappa X^M J_{i M} \;\;  \ ,
$$
where the index $M$ runs over the target superspace indices and
$X^M$ are the superspace coordinates, while $i = 1,
  \dots, 3$ denotes the $\ZZ_4$ grading. Since $J_i = d X^M
J_{i M}$ we obtain the following transformations of the currents
$$
 \delta_\kappa J_0  =  d \delta_\kappa x_0 + [J_0, \delta_\kappa
  x_0] + [J_1, \delta_\kappa x_3] + [J_2, \delta_\kappa x_2] + [J_3,
  \delta_\kappa x_1] \ ,
  $$
  $$
    \delta_\kappa J_1  =  d \delta_\kappa x_1 + [J_0, \delta_\kappa
  x_1] + [J_1, \delta_\kappa x_0] + [J_2, \delta_\kappa x_3] + [J_3,
  \delta_\kappa x_2] \ ,
  $$
  $$
  \delta_\kappa J_2  =  d \delta_\kappa x_2 + [J_0, \delta_\kappa x_2] +
  [J_2, \delta_\kappa x_0] + [J_1, \delta_\kappa x_1] + [J_3,
  \delta_\kappa x_3] \ ,
  $$
  $$
   \delta_\kappa J_3  =  d \delta_\kappa x_3 + [J_0, \delta_\kappa
  x_3] + [J_1, \delta_\kappa x_2] + [J_2, \delta_\kappa x_1] + [J_3,
  \delta_\kappa x_0] \ .
  $$
Using these transformations and taking into account the
Maurer-Cartan equations, the $\kappa$-transformation of the
action is
\bea\label{kappags}
  \delta_\kappa S_{GS} = & {1\over4}\int d^2 \sigma\,\str
  \Bigl( \epsilon^{i j} \partial_i (J_{3 j} \delta_\kappa x_1 -
  J_{1 j} \delta_\kappa x_3) + \delta_\kappa (\sqrt{g}
  g^{ij}) J_{2 i} J_{2 j} + \nonumber\\
  & {} +2 \sqrt{g} g^{i j} (J_{2 i}
  \partial_j \delta_\kappa x_2 + [J_{2 i}, J_{0 j}] \delta_\kappa x_2)
  + \epsilon^{ij} ([J_{1 i}, J_{1 j}] \nonumber \\
  & - [J_{3 i}, J_{3 j}])
  \delta_\kappa x_2 {} - 2 (\sqrt{g} g^{ij} + \epsilon^{ij}) [J_{1 j},
  J_{2 i}] \delta_\kappa x_1 + 2 (\sqrt{g} g^{ij} \nonumber \\
  & - \epsilon^{ij})
  [J_{2 i}, J_{3 j}] \delta_\kappa x_3 \Bigr) \ .
\eea
The $\kappa$-transformation is parameterized by
\bea\label{kappasym}
  \delta_\kappa x_2 = 0 \ , \quad \delta_\kappa x_1 = [J_{2 i},
  \kappa_3^i] \ , \quad \delta_\kappa x_3 = [J_{2 i}, \kappa_1^i] \ ,
\eea
where $\kappa_3^i \in \cH_3$ and $\kappa_1^i \in
\cH_1$. By substituting this and expressing the result in
terms of the structure constants and the Cartan metric $\eta$ one
finally has
\bea\label{deltagss}  \delta_\kappa S_{GS}  = & {1\over4} \int d^2 \sigma \,\str\Big[
  \epsilon^{ij} \partial_i (J_{3 j} \delta_\kappa x_1 - J_{1
  j} \delta_\kappa x_3) + \delta_\kappa (\sqrt{g} g^{ij})
  \eta_{a b} J_{2 i}^a J_{2 j}^b + \nonumber\\
  & {} + 4 \sqrt{g} (P_+^{ij} \eta_{\hat \beta \beta} f_{\alpha a}^{\hat
  \beta} f_{b \hat \alpha}^\beta J_{1 j}^\alpha \kappa_3^{k \hat
  \alpha} -  P_-^{ij} \eta_{\beta \hat \beta} f_{a \hat \alpha}^\beta
  f_{b \alpha}^{\hat \beta} J_{3 j}^{\hat \alpha} \kappa_1^{k \alpha})
  J_{2 i}^a J_{2 k}^b \Big] \ ,
\eea
where we have defined the projectors $P_\pm^{ij} =
\half (g^{ij} \pm {1\over \sqrt{g}} \epsilon^{ij})$. Since
$\delta_\kappa (\sqrt{g} g^{ij})$ must be symmetric and
traceless and not Lie-algebra valued, we have to require that
\bea\label{eqkappasymmetrycondition}
   \eta_{\beta \hat \beta} f_{\hat \alpha (a}^\beta
  f_{b) \alpha}^{\hat \beta} = c_{\alpha \hat \alpha} \eta_{a b}
\eea
for some matrix $c_{\alpha \hat \alpha}$. Then one obtains
$$
  \delta_\kappa (\sqrt{g} g^{ij}) = 4 \sqrt{g} c_{\alpha \hat \alpha}
  (P_-^{ik} J_{3 k}^{\hat \alpha} \kappa_1^{j \alpha} - P_+^{ik}
  J_{1 k}^\alpha k_3^{j \hat \alpha}) \ ,
$$
which is automatically symmetric in $a$ and $b$ if we require that
$$  \kappa_1^i = P_-^{ij} \kappa_{1 j} \ , \quad \kappa_3^i = P_+^{ij}
  \kappa_{3 j}
$$
since $P_\pm^{ik} P_\pm^{jk} = P_\pm^{jk} P_\pm^{ik}$. It is
also traceless because $P_-^{ji} \kappa_{1 j} = P_+^{ji}
\kappa_{3 j} = 0$.

The relation (\ref{eqkappasymmetrycondition}), required for
$\kappa$-symmetry, is a condition on the structure constants of the
supergroup, listed  in \ref{appendix:supergroups}, which can be easily checked to hold. By making use of the Bianchi identity for the torsion and the expression for the super B-field (\ref{bfield}), one can translate this condition to a constraint on the supergravity background
 $(\gamma_{(a})_{\ah\bh}(\gamma_{b)})_{\a\g} P^{\g\bh}=c_{\a\ah}\eta_{ab}$. It is straightforward to check that this holds in our supergravity background with $c_{\a\ah}=\eta_{\a\ah}$.

\subsubsection{Pure spinor}
\label{secpurespinorsigmamodel}

The worldsheet action in the pure spinor formulation of the
superstring consists of a matter and a ghost sector. The
worldsheet metric is in the conformal gauge and there are no
reparameterization ghosts. The matter fields are written in terms
of the same left-invariant currents $J = g^{-1}
d g$ that appear in the GS action (\ref{eqcosetGSaction}). We just need to rewrite the ghost variables in a way adapted to the supercoset construction. The Lie algebra-valued
pure spinor fields and their conjugate momenta are defined as \bea\label{purepara}
  \lambda = \lambda^\alpha {\bf T}_\alpha \ ,\quad
  w = w_\alpha \eta^{\alpha \hat \alpha} {\bf T}_{\hat \alpha} \ , \quad
  \hat \lambda = \hat\lambda^{\hat\alpha} {\bf T}_{\hat \alpha} \ , \quad
  \hat w = \hat w_{\hat \alpha} \eta^{\alpha \hat \alpha} {\bf T}_\alpha \
\eea
Just as for the matter variables, the spinor indices in the ghost sector are unhatted
for the left moving quantities and hatted for right moving
ones.  Using these conventions, the
pure spinor Lorentz generators $N^{ab}=\half (w\gamma^{ab}\l)$ and their hatted sibling take the form
$$
  N = - \{ w, \lambda \} \ , \quad \widehat N = - \{ \hat w, \hat \lambda
  \} \ .
$$
They generate the Lorentz
transformations on the pure spinor variables that correspond to multiplication by elements
of $H$. $N, \widehat N \in \cH_0$ so they act  as Lorentz transformations on the
tangent-space indices $\alpha$ and $\hat \alpha$ of the pure
spinor variables. The pure spinor
constraints read
 $$
 \{\l,\l\}=0,\qquad \{\hat\l,\hat\l\}=0 \ .
 $$

The sigma-model is invariant under the global
transformation $g \to g_0 g$, $g_0 \in G$. The
sigma-model is also invariant under the gauge
transformation (\ref{gaugeJ}) and the analogous ones acting on the ghosts
\bea\label{gaugepure}  \delta_\Lambda \lambda = [\lambda, \Lambda] \ &,\quad
  \delta_\Lambda w = [w, \Lambda] \ , \\
  \delta_\Lambda \hat\lambda  =  [\hat\lambda, \Lambda] \ &,
  \quad
  \delta_\Lambda \hat w = [\hat w, \Lambda] \ , \nonumber
\eea
where $\Lambda \in \cH_0$. Moreover, it is invariant under the gauge transformation on the antighosts (generated by the pure spinor constraint)
\bea\label{gaugegh}
\delta w=[\lambda, \Omega_2] , \qquad \delta \hat w=[\hat \lambda, \hat \Omega_2] \ ,
\eea
where $\Omega_2=\Omega_2^a T_a$ is an arbitrary function.

The BRST operator for the pure spinor sigma model on our supercoset is
  \bea\label{brstads} Q_B = -\oint dz\str\left(  \lambda J_3\right) + \oint d\bar z \str\left(\hat\lambda \bar J_1\right)
  \ .
\eea
It acts on the supercoset element as a derivation
$$
Q_B \,g=g(\l+\hat\l) \ ,
$$
and on the Maurer-Cartan currents as
\bea\label{brstrans}
\delta_B J_j  =& \delta_{j+3,0}\,d
(\epsilon\l)+[J_{j+3},\epsilon\l]+\delta_{j+1,0}\,d(\epsilon\hat\l)+[J_{j+1},\epsilon\hat\l],\\
\delta_B w = & -J_3\epsilon,
\qquad \delta_B \hat w=-\bar J_1\epsilon,\\
\delta_B N  = &
[J_3,\epsilon\l],\qquad \delta_B \hat N=[\bar J_1,\epsilon\hat
\l] \ ,
\eea
where we defined $\delta_B=[\epsilon Q_B,\cdot]$.

The coefficients of the various terms in the action
are determined by requiring that the action be BRST invariant (the
details can be found in
\ref{secpurespinorsigmamodelfromBRST}). The
BRST-invariant sigma-model thus obtained is
\bea\label{eqAdSpsaction}
  S = {R^2\over2\pi}\int d^2 z \str\Bigl( \half J_2 \bar J_2 + {1\over4} J_1
  \bar J_3 + {3\over 4}J_3 \bar J_1 + w \bar\nabla  \lambda + \hat w
  \nabla\hat\lambda -N \hat N
  \Bigr) \ ,
\eea
where the covariant derivatives are defined in (\ref{covlambda}).

It is worth explaining a subtlety related to the index contractions. The ghost part of the action reads explicitely
\bea\label{sghoste}
S_{gh}= S_{flat}+{R^2\over2\pi}\int(N_{ab}\bar J_0^{ab}+N_{a'b'} \bar J_0^{a'b'}+\hat N_{ab} J_0^{ab}+\hat N_{a'b'} J_0^{a'b'}\\ \nonumber +N_{ab}\hat N^{ab}-N_{a'b'}\hat N^{a'b'}) \ .
\eea
Note the relative sign in the contraction of the pairs of indices $[ab]$ and $[a'b']$ in the last two terms as opposed to the other terms. The reason is the following. The term $\Str N\bar J_0$ comes from the covariant derivative acting on the spinor $ \lambda$
$$
\Str N \bar J_0= \Str w [\bar J_0,\lambda]=w_\alpha [(T_{{ab}})^\alpha_\beta\bar\Omega^{{ab}}- (T_{{a'b'}})^\alpha_\beta\bar\Omega^{{a'b'}}] \lambda^\beta \ ,
$$ 
where $\Omega$ is the spin connection and its indices have been contracted using the supermetric $\eta_{[\underline{ab}][\underline{cd}]}$. The generator of the Lorentz group $SO(1,4)\times SO(5)$ in the adjoint representation is given by the structure constants, namely $({\bf T}_{{ab}})^\alpha_\beta=+\half (\gamma_{ab})_\beta{}^\alpha$ and $({\bf T}_{{a'b'}})^\alpha_\beta=-\half (\gamma_{a'b'})_\beta{}^\alpha$. The minus signs in the generators cancels the minus sign in the supermetric. The last two terms in (\ref{sghoste}), on the other hand, do not come from any covariant derivative, but are the coupling to the curvature, which is given just by the supermetric, so a minus sign appears.

The equations of motion of the currents $J_i$ are obtained by
considering the variation $\delta g = g X$ under which $\delta J =
\partial X + [J, X]$ and using the $\ZZ_4$ grading and the
Maurer-Cartan equations, so that we get
\bea\label{purespinoreoms}
  \nabla \bar J_3  = & - [J_1, \bar J_2] - [J_2, \bar J_1] + [N, \bar
  J_3] + [\hat N, J_3] \ , \nonumber\\
  \bar \nabla J_3  = & [N, \bar J_3] + [\hat N, J_3] \ , \nonumber\\
  \nabla \bar J_2  = & - [J_1, \bar J_1] + [N, \bar J_2] + [\hat N,
  J_2] \ ,\\
  \bar \nabla J_2  = & [J_3, \bar J_3] + [N, \bar J_2] + [\hat N,
  J_2] \ , \nonumber\\
  \nabla \bar J_1  = & [N, \bar J_1] + [\hat N, J_1] \ , \nonumber\\
  \bar \nabla J_1  = & [J_2, \bar J_3] + [J_3, \bar J_2] + [N, \bar
  J_1] + [\hat N, J_1] \ , \nonumber
\eea
where $\nabla J = \partial J + [J_0, J]$ and $\bar \nabla J = \bar
\partial J + [\bar J_0, J]$ are the gauge covariant derivatives.
The equations of motion of the pure spinors and their Lorentz currents are
\bea\label{ghosteoms}
  \bar \nabla \lambda = & [\hat N, \lambda] \ , \quad
  \nabla \hat \lambda = [N, \hat \lambda] \ , \nonumber\\
  \bar \nabla w = & [\hat N, w] \ , \quad
  \nabla \hat w = [N, \hat w] \ ,\\
  \bar \nabla N = & -[N, \hat N] \ , \quad \nabla \hat N = [N, \hat
  N]  \ . \nonumber
\eea

Let us briefly comment on the relation between the pure spinor
action (\ref{eqAdSpsaction}) and the GS action
(\ref{eqcosetGSaction}). The latter, when written in
conformal gauge, reads
$$  S_{GS} ={R^2\over2\pi} \int d^2 z \str\Bigl( \half J_2 \bar J_2 +
  {1\over4}J_1 \bar J_3 - {1\over4}J_3 \bar J_1 \Bigr) \ .
$$
To this one has to add a term which breaks $\kappa$-symmetry and
adds kinetic terms for the target-space fermions and coupling to
the RR-flux $P^{\a\ah}$
$$  S_\kappa = {R^2\over2\pi}\int d^2 z (d_\alpha \bar J_1^\alpha + \bar d_{\hat \alpha}
  J_3^{\hat \alpha} + P^{\alpha \hat \alpha} d_\alpha \bar d_{\hat
  \alpha}) = {R^2\over2\pi}\int d^2 z \str\Bigl( d \bar J_1 - \bar d J_3 + d \bar d
  \Bigr) \ ,
$$
where, in curved backgrounds, the $d$'s are the conjugate
variables to the superspace coordinates $\theta$'s. After
integrating out $d$ and $\bar d$ we get the complete matter part
\bea\label{matteraction}
 S_{matter}= S_{GS} + S_\kappa ={R^2\over2\pi} \int d^2 z \str\Bigl( \half J_2 \bar
  J_2 + {1\over4}J_1 \bar J_3 + {3\over4}J_3 \bar J_1 \Bigr) \ .
\eea
This sigma-model can be recognized as taking the same form as the
sigma-model for the
compactification of type II superstring on $AdS_2 \times S^2\times
CY_3$ in the hybrid formalism. It is a general fact that the
matter part of the hybrid and the pure spinor formalism is the
same. As usual this has to be supplemented with kinetic terms for
the pure spinors and their coupling to the background (\ref{sghoste}), in order to obtain the full superstring sigma-model
(\ref{eqAdSpsaction}) with action $S=S_{GS}+S_\kappa+S_{ghost}$.

\subsubsection{Lower dimensions}

There are many superstring theory backgrounds supported by RR flux that are described by covariant sigma models on supercosets. Some of them are critical ten dimensional backgrounds, for which there exists a weakly coupled supergravity regime, while others are genuine non-critical string theories, for which no supergravity regime exists, namely the spacetime curvature is of the order of the string scale (required to cancel the worldsheet Weyl anomaly). In the case in which the relevant supercoset admits a $\ZZ_4$ grading decomposition (\ref{eigenfour}), whose invariant locus is gauged, then we can use the formalism developed in the previous section to write down their sigma model action.

The first example is the AdS/CFT correspondence between three-dimensional $\cN=6$ Chern-Simons theory with bi-fundamental matter and type IIA superstring theory on $AdS_4\times CP^3$ with RR two- and four-form fluxes. The full type IIA worldsheet action is not a supercoset. However, one can partial fix the kappa-symmetry of the GS action and reduce it to a sector which is described by the supercoset 
$$
 AdS_4\times CP^3\oplus24\,\,\textrm{fermions}:\qquad {Osp(6|4)\over SO(1,3)\times
 U(3)}\ .
 $$
 
The following backgrounds are interpreted as the non-compact part
of a ten-dimensional type II background $AdS_p\times S^p\times
M_{5-p}$, where $M_{5-p}$ is a Ricci flat manifold of complex dimension $5-p$.
\begin{enumerate}
\item $AdS_2\times S^2$ with RR two-form flux, realized as
 $$
 AdS_2\times S^2\oplus8\,\,\textrm{fermions}:\qquad {PSU(1,1|2)\over U(1)\times
 U(1)}\ .
 $$
 \item $AdS_3\times S^3$ with RR three-form flux, realized as
$$
 AdS_3\times S^3\oplus16\,\,\textrm{fermions}:\qquad {PSU(1,1|2)^2\over SO(1,2)\times
 SO(3)}\ ,
$$
\end{enumerate}

\noindent The action for the matter part of these two sigma models is the same as the matter part of the action (\ref{eqAdSpsaction}), where the currents take values in the corresponding supercosets. For the ghost sector, one can add the relevant hybrid action, or a lower dimensional pure spinor action. In the latter case, the full action is formally the same as (\ref{eqAdSpsaction}), including the ghost sector, where the currents and the pure spinor variables take values in the appropriate supercoset.

The following backgrounds are interpreted as non-critical
superstrings.
The $AdS_{2n}$ backgrounds with spacefilling RR-flux,
realized as
$$
 AdS_2\oplus4\,\,\textrm{fermions}:\qquad{Osp(2|2)\over SO(1,1)\times SO(2)}
 $$
 $$
 AdS_4\oplus8\,\,\textrm{fermions}:\qquad {Osp(2|4)\over SO(1,3)\times SO(2)}
 $$
 $$
 AdS_6\oplus16\,\,\textrm{fermions}:\qquad {F(4)\over SO(1,5)\times SL(2)}
$$
describe type II non-critical superstrings in $2n$
dimensions. These backgrounds are not super Ricci flat, and a  non-vanishing string scale curvature is needed to cancel the Lioville central charge, which is non-zero when the target space dimension is sub-critical. 

{\bf Guide to the literature}

An introduction to supergroups and super Lie-algebras can be found in the comprehensive reviews \cite{Pais:1975hu,Frappat:1996pb}.

The Green-Schwarz action on the $AdS_5\times S^5$ background was introduced by Metsaev and Tseytlin in \cite{Metsaev:2000bj}, where the supercoset nature of the background was thoroughly exploited. The GS was found to be kappa-symmetric in \cite{Siegel:1985xj}. The issue of fixing the light-cone gauge in two-dimensional sigma models is discussed in general in \cite{Rudd:1994ss} and in the particular case of $AdS_5\times S^5$ in \cite{Metsaev:2000yf}. An exhaustive treatment of the GS action in AdS and its semi-classical quantization in the light-cone gauge can be found in the recent review \cite{Arutyunov:2009ga}, where its applications to the AdS/CFT correspondence are analyzed in great details.

The complete pure spinor action in  $AdS_5\times S^5$ was derived in \cite{Berkovits:2000fe,Berkovits:2000yr}. The matter part of such action is the same for any supercoset that admits a $\ZZ_4$ grading and coincides with the one for the hybrid formalis, which appeared in \cite{Berkovits:1999zq,Berkovits:1999im}, before the advent of the pure spinor formalism. The full GS action on the $AdS_4\times CP^3$ background is given in \cite{Gomis:2008jt} and its subsector, which is described by a supercoset, can be found in \cite{Arutyunov:2008if,Stefanski:2008ik} for the GS action and in \cite{Fre:2008qc} for the pure spinor action. Actions on lower dimensional supercoset backgrounds are constructed in \cite{Adam:2006bt,Adam:2007ws}. An alternative formulation of GS on $AdS_2$ can be found in \cite{Verlinde:2004gt}.

\clearpage

\section{Quantum effective action and conformal invariance}
\label{section:effectiveaction}

The pure spinor sigma model contains kinetic terms for both fermions and bosons and it is in the conformal gauge on the worldsheet. It is suitable for quantization, which we will now perform in order to evaluate the quantum effective action at one-loop. The salient features of the sigma model that we will extract from the effective action are the following:
\begin{enumerate}
\item Absence of divergent terms. This implies that the beta functions vanish and the theory is {\it conformally invariant} at one-loop. This is no surprise of course, since one-loop beta functions are as usual equivalent to the background supergravity fields be on-shell.
\item All the finite terms in the effective action are local. This means that they can be removed by adding local counter-terms. This fact has two important consequences. First, {\it gauge or BRST anomalies are absent}, that would arise as non-local terms that are not gauge invariant or BRST invariant. Secondly, {\it the AdS radius is not renormalized}, as opposed to what happens in bosonic WZW models.
\end{enumerate}

In the first part of this Section, we use the background field method to compute the one-loop 1PI effective action and prove these statements. The background field expansion is useful to perform perturbative computations in the sigma model, such as to compute the energy spectrum of string states.

In the rest of the Section, we derive some interesting results using this formalism. We show that the stress tensor has zero central charge at one-loop. Then we compute the algebra of OPE of the worldsheet left-invariant currents and find that it is not a chiral algebra, but left- and right-moving currents mix in the OPE. Finally, we extend the proof of conformal invariance and absence of BRST and gauge anomalies to all loops in the sigma model perturbation theory. In \ref{appendix:worldsheet} we derive several results pertaining this Section, that we omitted from the main discussion to ease the reading.

\subsection{Background field method}

Let us review the computation of effective actions
in the closed bosonic and RNS string. We will set the notations and show why the bosonic string renormalizes, while worldsheet supersymmetry protects the metric from $\alpha'$ corrections at one-loop in the RNS formalism.
We will then apply the background field method to the superstring sigma model in AdS.

\subsubsection{Bosonic string}

The bosonic string in a curved background is (we are
assuming that $B_{mn}=0$)
\bea\label{boso}S_{bos}=\int d^2z \left[ \p x^m \pb x^n G(x)_{mn}\right].\eea
In the covariant background field expansion we fix a classical solution of the
worldsheet equations of motion $x_0$ and expand around
it in quantum fluctuations $X$,
\bea\label{bo}S_{bos}= S_0+\int d^2z \left[\eta_{ab}\nabla X^a\nb X^b  + ... \right], \eea
where $...$ are terms depending on the curvature,
$\nabla X^a=\p X^a +  A^{ab}X_b$, $A^{ab}=\p x_0^m \omega_m^{ab}$ and
$\omega_m^{ab}$ is the spin connection. When one uses the normal coordinate expansion
within the background field method, local Lorentz invariance is used to
fix the spin connection to zero. In this case the resulting effective action
will not be manifestly invariant under this symmetry. We then have to
check whether the effective action is not anomalous under this symmetry.

\begin{figure}[tbp]
\begin{center}
\centerline{\includegraphics[scale=1]{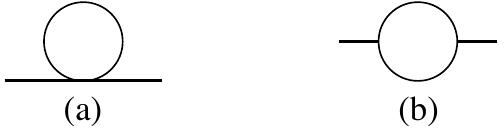}}
\end{center}
\begin{center}
\caption{Diagrams $(a)$ and $(b)$ correspond respectively to the first and second term in (\ref{effact}).} 
\label{holon}
\end{center}
\end{figure}

The classical worldsheet action is conformally invariant. This classical symmetry might be broken at the quantum level, by a conformal anomaly arising from the regularization of the divergent diagrams. To check that conformal invariance persists in the quantum theory, we need to evaluate the effective action and, in particular, its divergent part. The computation of the effective action at one-loop order proceeds as follows. First we observe that, by simple power counting arguments, the only UV divergent terms will be the ones with two external background currents $A^{ab}$ or $\bar A^{ab}$. The propagators are $1/p^2$ while the insertion of a current carries at most one derivative, so the terms with two external currents will be log divergent in the momentum cutoff. There are two kind of terms that we need to compute, schematically
\bea\label{effact}
S_{eff}=\int d^2z\langle {\cal L}_{int}(z)\rangle-\half \int d^2z\int d^2w\langle {\cal L}_{int}(z){\cal L}_{int}(w)\rangle \ ,
\eea
where ${\cal L}_{int}$ denotes the interaction part of the lagrangian inside the square brackets in (\ref{bo}), and $\langle\cdot\rangle$ denotes functional integration over the fluctuating fields. The first term $\langle {\cal L}_{int}(z)\rangle$ corresponds to the normal ordering of the composite operators in the lagrangian: it is just given by the one-loop self energy of the fluctuations at the same point, in operators with two external currents inserted at the same point (diagram $(a)$ in Figure~\ref{holon}). The second term $\langle {\cal L}_{int}(z){\cal L}_{int}(w)\rangle $ corresponds to the one-loop fish diagram generated by the contraction of the operators with one external current (diagram $(b)$ in Figure~\ref{holon}).

The effective action is
\bea\label{effboso} S_{eff}= \half \int d^2z\,d^2w\,[ {1\over 4}A^{ab}(z)A_{ab}(w)
( {{\ln|z-w|^2}\over{(\zb-\wb)^2}} + {1\over(\zb-\wb)^2}) + \eea
$${1\over 4}\bar A^{ab}(z)\bar A_{ab}(w)
({\ln|z-w|^2\over (z-w)^2} + {1\over (z-w)^2} ) +$$
$$ - {1\over 2}A^{ab}(z)\bar A_{ab}(w)({1\over|z-w|^2} -
\delta(z-w)\ln|z-w|^2)] -\half\int d^2z A^{ab}(z)\bar A_{ab}(z)\ln(0) .$$
It is not clear upon inspection which terms above are finite or divergent. Also, we have
the usual complications due to infrared divergencies. To clarify the interpretation, we will transform
the above two point functions into loop integrals in the momentum space. We
perform the loop integral using dimensional regularization adding a small
mass $m$ to the $X^a$ fields in order to regularize IR divergencies.
All the UV divergences cancel, and the dependence on the
dimensional regularization mass scale $\mu$ is an infrared effect, so
we can identify the mass regulator $m$ with $\mu$.
In order to
simplify the calculation, we build a dictionary between the above two point
functions and the corresponding result of the integration over the
momenta\footnote{We can ignore terms like $\ln( {|p|^2\over \mu^2})$.
These terms are an infrared
effect. They reflect the fact that $X^a$ is not a primary fields. It is expected
that after summing up the entire perturbation series these terms vanish \cite{deWit:1993qv}.}
\bea\label{dictio}{1\over{(z-w)^2}} &\quad&\leftrightarrow \quad -{ p \over{\bar p} }\ , \nonumber\\
{1\over{(\bar z-\bar w)^2}} &\quad&\leftrightarrow\quad -{\bar p \over p} \ , \nonumber\\
{\ln|z-w|^2\over{(z-w)^2}} &\quad&\lfa \quad -{ \bar p \over p}(1+
2\ln( {|p|^2\over \mu^2} ))\ , \nonumber\\
{\ln|z-w|^2\over{(\bar z-\bar w)^2}} &\quad&\lfa \quad -
{ p \over{\bar p}}(1 + 2\ln( {|p|^2\over \mu^2} ))\ ,\\
{1\over|z-w|^2}&\quad&\lfa\quad 1+{1\over \epsilon} -2\ln( {|p|^2\over \mu^2} ) \ , \nonumber\\
-\ln|z-w|^2\delta(z-w) &\quad&\lfa\quad 1+{1\over \epsilon} \ , \nonumber\\
-\ln(0) &\quad&\lfa\quad {1\over \epsilon}\ . \nonumber
\eea

In momentum space the effective action is
\bea S_{eff}=
 \half \int d^2k[ {1\over 2}A^{ab}(-k)A_{ab}(k)
{\bar k\over k}
+
{1\over 2}\bar A^{ab}(-k)\bar A_{ab}(k){k\over \bar k}
- {1\over 2}A^{ab}(-k)\bar A_{ab}(k)]\ .\nonumber\\ \label{effbosomom}\eea
As anticipated above, UV divergent terms, which would be proportional to $1/\epsilon$, are absent. The one-loop beta function is proportional to such terms and, hence, it vanishes. As explained in the footnote above, we ignored all IR divergent terms proportional to $\lnk$. Even though UV divergent terms are absent, there are finite renormalization terms, which depend on the pull back of the spin connection $A^{ab}=d x_0^m \omega_m^{ab}$. Due to the explicit dependence of these finite terms on $A^{ab}$, they could in principle affect the gauge invariance of the effective action.  The first two terms in  (\ref{effbosomom}) are gauge invariant, but the last term is not. Since this last term is local it can be removed, as usual, by adding a local counter-term, that will restore gauge invariance. We conclude that there is no anomaly in the gauge invariance. The local counter-term
that restores gauge invariance is just a redefinition of the target space metric
$$G(x_0)_{mn}\to \widetilde G(x_0)_{mn}= G(x_0)_{mn} +
\alpha'{1\over 4}\omega_m^{ab}\omega_{n ab},$$
and the new metric now has a gauge transformation
$$\delta \widetilde G(x_0)_{mn} = \alpha'{1\over 4} \p_m \L_{ab}\omega_n^{ab}+
\alpha'{1\over 4} \p_n \L_{ab}\omega_m^{ab}.$$
Note that the new one-loop counter-term explicitly contains a factor of $\a'$. The new gauge transformation of the metric compensates the gauge transformations of the finite renormalization terms, restoring the gauge invariance of the full effective action. The anomaly is trivial, this is the reason why we can fix the connection
to be zero when using normal coordinates. To
compute higher $\alpha'$ corrections in this scheme, e.g. to compute
scattering amplitudes or beta functions, we have to take into
account this $\alpha'$ correction to the classical action.
The conclusion is that once we choose a scheme of computation, in this case
without fixing the connections,  we cannot
ignore the finite local counter-terms in the classical action, even when
the anomaly is trivial.

\subsubsection{RNS string}

Let us see what happens in the RNS formalism. Its action in a curved
background is
\bea\label{crns}S_{RNS}=\int d^2z\left[ (\p x^m \pb x^n +\half \psi^m\pb \psi^n +
\half \bar\psi^m\p\bar\psi^n + \right. \eea
$$\left. \half \psi^m \Gamma^n_{qp}(x)\pb x^q\psi^p +
\half \bar\psi^m \Gamma^n_{qp}(x)\p x^q\bar\psi^p)G(x)_{mn} +
{1\over 4}R_{mnqp}\psi^m\psi^n\bar\psi^q\bar\psi^p \right].$$
Again, we fix a classical solution of the
worldsheet equations of motion $(x_0,\psi_0,\bar\psi_0)$ and expand around
it in the corresponding quantum fluctuations $(X,\Psi,\bar\Psi)$,
\bea\label{rns}S_{RNS}= S_0+\int d^2z \left[\eta_{ab}\nabla X^a\nb X^b  +
\eta_{ab}\half\Psi^a \nb \Psi^b + \eta_{ab}\half\bar\Psi^a \nabla \bar\Psi^b
+...\right],\nonumber \eea
where $...$ are terms depending on the curvature. We can compute the effective action just as in (\ref{effact}): we evaluate diagrams with two background fields $A^{ab}$, by integrating over the quantum fluctuations. The effective action is just
$$ {1\over 2}\int d^2k[-A^{ab}(-k)\bar A_{ab}(k) ]\ ,
$$
where again we ignored the IR divergent terms (such terms would be gauge invariant anyway).
Non-local terms cancel due to worldsheet supersymmetry. There are no divergent terms either, i.e. the beta function vanishes. 
We have to add a local counter-term to cancel
the anomalous variation of this term.
We see that in the case of RNS superstring, even without gauge fixing
the connection, there are no finite counter-terms in the effective action.

In the case case of Type I or Heterotic string we would have the usual
local Lorentz anomaly that appears because we have only left moving fermions,
which can be canceled by a variation of the $B$ field.
In coset models, like the superstring in $AdS_5\times S^5$ space, it is
useful to keep the connection unfixed since this
simplifies significantly the background field expansion.

\subsubsection{$AdS_5\times S^5$ sigma model}

The first step in the quantization is to expand the coset element in fluctuations around a classical configuration.  Let us pick an element $g=\tilde g e^{X/R}$, where $\tilde g$ represents a point in the coset and the $X$'s are the fluctuations around that point. We will expand the sigma model around the point $\tilde g$: since the sigma model coupling is the inverse of the AdS radius $R$, we are expanding locally around flat space. The functional integration is performed over the fluctuating fields. In this subsection we consider the matter part, which is the same for the pure spinor as well as the hybrid formalism. In the next subsection we will discuss the ghost part of the effective action.

Let us consider first the matter part of the action. By using a gauge transformation $g\to g e^{h_0}$, with $h\in {\cal H}_0$, we can choose a gauge such that the fluctuations do not depend on the grading zero part, i.e. $X\in {\cal G}\backslash {\cal H}_0$. The Maurer-Cartan form can be expanded around the classical configuration $\tilde g$ 
\bea\label{maurex}
 J=&e^{-X/R}\tilde J e^{X/R}+e^{-X/R}d e^{X/R} \ ,
\eea
where $\tilde J=\tilde g^{-1}d \tilde g$ is the background current, and its components can be isolated according to their grading
\bea\label{gradingex}
J|_i=&\tilde J|_i+{1\over R}\left(dX+[\tilde J,X]\right)|_i+{1\over 2R^2}\left[dX+[\tilde J,X],X\right]|_i +{\cal O}(R^{-3}) \ ,
 \eea 
 where $i=0,\ldots,3$ and we choose the gauge $X_0=0$.

When we plug the expressions (\ref{gradingex}) into the matter part of the action (\ref{matteraction}) we get a certain number of terms, that we expand in powers of the fluctuating fields. The terms which do not depend on the fluctuations correspond just to the classical action, written in terms of the background currents $\tilde J_i$. The terms linear in the fluctuations do not contribute to the effective action, so we are only interested in the terms quadratic in the fluctuations. Upon performing the functional integration, these terms will produce functional determinants depending on the background currents, that we can express in an appropriate form to represent the effective action.

The terms in the action which are quadratic in the fluctuations separate into kinetic terms for the fluctuations and terms that depend on the background currents. The kinetic terms are
\bea\label{kinaction}
S_{\rm kin}=\int d^2z\str\left( \half \nabla X_2\bar \nabla X_2+{1\over4}\nabla X_1\bar \nabla X_3+{3\over 4}\nabla X_3\bar \nabla X_1\right) \ ,
\eea
where the covariant derivative $\nabla X_i=\partial X_i+[J_0,X_i]$ depends on the background gauge current. In the following, to avoid cluttering the equations we remove the tilde from the background currents. We can read off the free field OPE's of the fluctuations from (\ref{kinaction})
\bea\label{freeope}
X_i^A(z)X_j^B(0)\sim&-\eta^{BA}\log |z|^2 \ ,
\eea

For convenience, we will collect the terms that depend on the background currents into three parts. We will put in $S_I$ all the terms that contain either $J_2$ or $\bar J_2$ or both
\bea\label{sone}
S_I=&\int d^2z\str\Bigr( \half J_2[X_1,\bar\nabla X_1]+\half \bar J_2[X_3,\nabla X_3]\\&
+{1\over4}J_2\left[[\bar J_2,X_1],X_3\right]-{1\over4}J_2\left[[\bar J_2,X_3],X_1\right] \Bigr)\ . \nonumber
\eea
We will put into $S_{II}$ all the terms that depend on $J_1$ or $\bar J_3$ or both
\bea
S_{II}=&\int d^2z\str\Bigr( {1\over 8}J_1(3[X_1,\bar \nabla X_2]+{5}[X_2,\bar \nabla X_1])+{1\over 8}\bar J_3(3[X_3, \nabla X_2] \nonumber\\
&+{5}[X_2,\nabla X_3]) -\half J_1\left[[\bar J_3,X_2],X_2\right] \label{stwo}\\
&+{1\over4}J_1\left[[\bar J_3,X_1],X_3\right]-{1\over 4}J_1\left[[\bar J_3,X_3],X_1\right] \Bigr) \ . \nonumber
\eea
Finally, we collect in $S_{III}$ the terms that depend on $J_3$ or $\bar J_1$ or both
\bea
S_{III}=&\int d^2z\str\Bigl( {1\over8}\bar J_1([X_1,\nabla X_2]-[X_2,\nabla X_1])+{1\over 8}J_3([X_3,\bar\nabla X_2] \nonumber\\
&-[X_2,\bar\nabla X_3]) +\half\bar J_1\left[[J_3,X_2],X_2\right]\label{sthree}\\
&+{3\over4}\bar J_1\left[[ J_3,X_1],X_3\right]+{1\over 4}\bar J_1\left[[J_3,X_3],X_1\right] \Bigr) \ . \nonumber
\eea


\subsection{Effective action and conformal invariance}
\label{section:conformalinv}

Everything is in place now to compute the one-loop effective action. There are two kinds of terms that we need to compute, looking back at (\ref{effact}). The terms contributing to $\langle {\cal L}(z)\rangle $ are the second lines in (\ref{sone}), (\ref{stwo}) and (\ref{sthree}). The terms contributing to $\langle {\cal L}(z){\cal L}(w)\rangle $ are the first lines in (\ref{sone}), (\ref{stwo}) and (\ref{sthree}).

In this section, we will compute the effective action, extract
the divergent terms and show that they vanish. This proves that the theory is conformally invariant. In the next section we will discuss the leftover finite renormalization.

The philosophy of the computation is the following: once we have fixed the coefficients of the operators in the action by classical BRST invariance, then the matter part and the ghost part of the action will be separately conformal invariant. For the matter part, the divergent contributions to the 1PI are proportional to the classical action itself, times the second Casimir of the supergroup $c_2(G)$, which vanishes for the $PSU(2,2|4)$ supergroup
 \bea\label{matterdiv}
 S_{1\rm PI}|_{\rm matter}=[1-{c_2(G)\over4\pi}\log(\Lambda/\mu)]S_{cl}|_{\rm matter}=S_{cl}|_{\rm matter} \ .
 \eea
The ghost part of the action again has a 1PI action proportional to the second Casimir as well, so it is not renormalized
 \bea\label{ghostdiv}
 S_{1\rm PI}|_{\rm gh}=S_{cl}|_{\rm gh}-{c_2(G)\over4\pi}\log(\Lambda/\mu)\int d^2z \str\left( N_{cl}\hat N_{cl}\right) =S_{cl}|_{\rm gh} \ .
 \eea
Even though all the divergent contributions cancel, there will be finite terms generated. We will discuss that at the end of this section.
\smallskip

\noindent{\it Matter}

\noindent Let us work this out explicitly for the part of the effective action generated by (\ref{sone}). The structure constants $f^A_{BC}$ of the $\mathfrak{psu}(2,2|4)$ superalgebra can be found in \ref{appendix:psu}. We will summarize at the end the contributions of the other parts (\ref{stwo}) and (\ref{sthree}). Working out the supertrace in (\ref{sone}) we get the lagrangian
\bea\label{supersone}
{\cal L}_I=&\half J_2^a\bar J_2^b\left(X_2^cX_2^df_{bc}^{[ef]}f_{a[ef]d}-\half X_1^\a X_3^\ah(f_{b\a}^\bh f_{a\bh\ah}-f_{b\ah}^\b f_{a\b\a})\right)\\&+J_2\bar\partial X_1^\a X_1^\b f_{a\a\b}+\bar J_2^b \partial X_3^\ah X_3^\bh f_{a\ah\bh}\ . \nonumber
\eea
The self energy contribution is
\bea\label{selfen}
\langle {\cal L}_I\rangle =&\half J_2^a\bar J_2^b\left(\langle X_2^cX_2^d\rangle f_{bc}^{[ef]}f_{a[ef]d}-\half\langle X_1^\a X_3^\ah\rangle(f_{b\a}^\bh f_{a\bh\ah}-f_{b\ah}^\b f_{a\b\a})\right) \nonumber\\
=&\half J_2^a\bar J_2^b\left[-\ln(0)\right]\left(f_{a[ef]}^cf_{bc}^{[ef]}-\half f_{b\a}^\bh f_{a\bh}^\a+\half f_{a\b}^\ah f_{b\ah}^\b\right) \ .
\eea
The second line in (\ref{supersone}) contributes by a fish diagram as
$$
-\half \langle {\cal L}_I(z){\cal L}_I(w)\rangle=-{1\over4}J_2^a(z)\bar J_2^b(w)f_{a\a\b}f_{b\ah\bh}\langle \bar \partial X_1^\a X_1^\b(z)\partial X_3^\ah X_3^\bh(w)\rangle $$
\bea\label{fishen}
={1\over 4} J_2^a(z)\bar J_2^b(w)f_{a\a}^\bh f_{b\bh}^\a\left(-{1\over |z-w|^2}+\delta^{(2)}(z-w)\ln|z-w|^2\right) \ . 
\eea

Collecting (\ref{selfen}) and (\ref{fishen}) and using (\ref{dictio}) we find
$$
S_{eff,I}=\half\int d^2z J_2^a\bar J_2^b\left[-\ln(0)\right]\left(f_{a[ef]}^cf_{bc}^{[ef]}-\half f_{b\a}^\bh f_{a\bh}^\a+\half f_{a\b}^\ah f_{b\ah}^\b\right)
$$
\bea\label{seffi}
+{1\over 4}\int d^2z \int d^2w J_2^a(z)\bar J_2^b(w)f_{a\a}^\bh f_{b\bh}^\a\left(-{1\over |z-w|^2}+\delta^{(2)}(z-w)\ln|z-w|^2\right) \ . 
\eea

By using the map (\ref{dictio}) from position to momentum space, we can read in (\ref{seffi}) the divergent contribution to the effective action that originates from the terms in (\ref{supersone})
\bea\label{divI}
S_{div,I}=-{c_2(G)\over 4\pi}\ln(\Lambda/\mu)\int d^2z\, \str\,\half J_2\bar J_2 \ .
\eea
We have introduced some group theory notations that we are going to explain now (for more details see Appendix B). Consider a supergroup $G$ with a subgroup $H$ and use the following letter assignment to denote the generators: $\{M,N,\ldots\}\in {\cal G}$, $\{A,B,\ldots\}\in {\cal G}\backslash {\cal H}$ and $\{I,J,\ldots\}\in {\cal H}$. The super Ricci tensor of the supergroup $G$ is defined as
\bea\label{superricci}
R_{MN}(G)=-{1\over4}f_{MQ}^Pf_{NP}^Q(-)^{\rm deg(P)}={c_2(G)\over 4}\eta_{MN} \ ,
\eea
and it is proportional to the second casimir of the supergroup.
In the presence of the subgroup $H$, we can express it as
\bea\label{superri}
R(G)_{AB}=-{1\over4}f_{AC}^Df_{BD}^C(-)^{\rm deg(D)}-\half f_{AC}^If_{BI}^C(-)^{\rm deg(I)} \ ,
\eea
In our particular case $G/H=PSU(2,2|4)/SO(1,4)\times SO(5)$, for the expression (\ref{divI}) we find
\bea\label{riccipsu}
R_{ab}(G)=&{1\over4}f_{a\a}^\ah f_{b\ah}^\a+{1\over4}f_{a\ah}^\a f_{b\a}^\ah -\half f_{ac}^{[ef]}f_{b[ef]}^c ={c_2(G)\over 4}\eta_{ab}\ ,
\eea
and we obtain the effective action (\ref{divI}), using the fact that $f_{ac}^{[ef]}f_{b[ef]}^c=f_{a[ef]}^cf_{bc}^{[ef]}$. It is proportional to the classical action and the coefficient is given by the second Casimir of the supergroup $PSU(2,2|4)$.

Let us look at the other contributions to the effective action. The operators in (\ref{stwo}) give the following contribution to the one-loop effective action
\bea\label{leffII}
{S}_{eff,II}=&\int d^2z\,J_1^\a \bar J_3^\ah (z)[-\ln(0)]\left({3\over4}f_{\ah a}^\b f_{\a\b}^a+{1\over4}f_{\ah \b}^{[ef]}f_{\a[ef]}^\b\right)
\eea
$$
+\int d^2z\int d^2w\,J_1^\a(z)\bar J_3^\ah(w)f_{\ah a}^\b f_{\a\b}^a\left({34\over 64}\delta^{(2)}(z-w)\ln|z-w|^2-{30\over 64}{1\over|z-w|^2}\right) \ , 
$$
where we used the identity $f_{\ah a}^\b f_{\a\b}^a=-f_{\ah\bh}^a f_{\a a}^\bh$. We extract the divergent contribution
$$
S_{div,II}=-{c_2(G)\over 4\pi}\ln(\Lambda/\mu)\int d^2z\, \str\,{1\over4} J_1\bar J_3 \ . 
$$

The operators in (\ref{sthree}) give the following contribution to the one-loop effective action
\bea\label{leffIII}
{S}_{eff,III}=&\int d^2z\,\bar J_1^\a J_3^\ah (z)[-\ln(0)]\left({3\over4}f_{\ah \bh}^a f_{\a \b}^\bh+{3\over4}f_{\ah \b}^{[ef]}f_{\a[ef]}^\b\right)\eea
$$
+\int d^2z\int d^2w\,\bar J_1^\a(z)J_3^\ah(w){1\over 32}f_{\ah a}^\b f_{\a\b}^a\left(\delta^{(2)}(z-w)\ln|z-w|^2+{1\over|z-w|^2}\right) \ , 
$$
from which we extract the divergent contribution
\bea\label{divIII}
S_{div,III}=-{c_2(G)\over 4\pi}\ln(\Lambda/\mu)\int d^2z\, \str\,{3\over4} \bar J_1 J_3 \ .
\eea

Collecting all the divergent contributions to the matter part of the effective action, we find that it is proportional to the classical action
\bea\label{divaction}
S_{div,matter}=-{c_2(G)\over 4\pi}\ln(\Lambda/\mu) S_{cl} \ ,
\eea
where the coefficient is the second Casimir of the supergroup. Because of the crucial fact that
\bea\label{casimirpsu}
c_2\left(PSU(2,2|4)\right)=0 \ ,
\eea
the divergent part of the effective action vanishes. The matter part of the action is separately conformally invariant at one-loop.

Finally, the operators $\str (J_i J_j)$ for any $i,j$ and $\str (J_i\bar J_j)$ for $i+j\neq0$ are not generated at one-loop. Consider, for example, the term $\str J_1 J_1$. The possible one-loop contributions to this operator come from evaluating 
$$
-\half \int d^2z\int d^2w\langle {\cal L}_{int}(z){\cal L}_{int}(w)\rangle \ ,
$$
where the relevant ${\cal L}_{int}$ is given by the first two terms in (\ref{stwo}), proportional to the background current $J_1$. It is immediate to see that, while the propagator $\langle X_2(z)X_2(w)\rangle$ is non-zero, there is no propagator $\langle X_1(z)X_1(w)\rangle$. Hence, this contribution vanishes. Most of such operators are not generated, due to the same mechanism. For other operators, where all propagators are non-vanishing, one finds a remarkable cancellation of diagrams thanks to the vanishing of the quadratic Casimir. 
\smallskip

\noindent{\it Ghosts}

\noindent Let us consider the ghost part of the one-loop effective action. We expand the left and right moving ghosts into upper case background fields and lower case fluctuations
\bea\label{ghostexp}
(w,\lambda)\to(W+w,L+\lambda)\ ,\qquad (\hat w,\hat \lambda)\to(\hat W+\hat w,\hat L+\hat \lambda)\ .
\eea
The ghost Lorentz currents are expanded as
\bea
N\to N_{(0)}+{1\over R}N_{(1)}+{1\over R^2}N_{(2)} ,\qquad
\hat N\to \hat N_{(0)}+{1\over R}\hat N_{(1)}+{1\over R^2}\hat N_{(2)} \ ,
\eea
where $(N_{(0)},\hat N_{(0)})$ denote the background currents while
\bea\label{fluctuan}
N_{(1)}=-\{W,\l\}-\{w,L\}\ ,\qquad &N_{(2)}=-\{w,\l\}\ ,\\
\hat N_{(1)}=-\{\hat W,\hat \l\}-\{\hat w,\hat L\}\ ,\qquad & N_{(2)}=-\{\hat w,\hat\l\} \ . \nonumber
\eea
We expand the classical ghost action (\ref{sghoste}) according to (\ref{fluctuan}) and collect the terms quadratic in the fluctuations
\bea
S=&\half\int d^2z\,\str\,\Bigl\{
N_{(0)}\left([\bar\nabla X_3,X_1]+[\bar \nabla X_2,X_2]+[\bar\nabla X_1,X_3]\right) \nonumber\\&+\hat N_{(0)}\left([\nabla X_3,X_1]+[\nabla X_2,X_2]+[\nabla X_1,X_3]\right)\label{ghostactio}\\&-N_{(1)} \hat N_{(1)}+N_{(2)}(\bar J_0-\hat N_{(0)})+(J_0-N_{(0)})\hat N_{(2)}\Bigr\} \ . \nonumber
\eea
In addition to (\ref{ghostactio}) there are several other terms, that do not contribute to the effective action. We list them in~\ref{extraterms}.

The terms with the partial derivatives in the first two lines in (\ref{ghostactio}) give rise to divergent terms in the effective action through a fish diagram
\bea\label{NNfirst}
{1\over 4}\int d^2z\int d^2w\,N_{(0)}^{[ef]}\hat N_{(0)}^{[lm]}\left(2f_{[ef]\a}^\b f_{[lm]\b}^\a-f_{[ef]b}^cf_{[lm]c}^b\right)\times\\
\times\left(\delta^{(2)}(z-w)\ln|z-w|^2-{1\over |z-w|^2}\right) \ . \nonumber
\eea
The combination of structure constants in (\ref{NNfirst}) amounts to
\bea\label{structurecom}
2f_{[ef]\a}^\b f_{[lm]\b}^\a-f_{[ef]b}^cf_{[lm]c}^b=4R_{[ef][lm]}(G)-4R_{[ef][lm]}(H)\\=(c_2(G)-c_2(H))\eta_{[ef][lm]}
\ , \nonumber
\eea
and (\ref{NNfirst}) can be recast in the form
\bea
{(c_2(G)-c_2(H))\over 4}\int d^2z\int d^2w\,\str N_{(0)}\hat N_{(0)}\left(\delta^{(2)}(z-w)\ln|z-w|^2-{1\over |z-w|^2}\right) \ ,\nonumber \\ \label{NNfirsttwo}
\eea
and we recall that $H=SO(1,4)\times SO(5)$ and $c_2(H)=3$.
The terms in the last line in (\ref{ghostactio}) contribute to the effective action through a fish diagram. To evaluate it, we need the OPE's between the fluctuations of the ghosts. At one-loop order, we can use the free field OPE's for the fluctuations of the ghost
\bea\label{freeghost}
\lambda^\b(z)w_\a(w)={\delta^\b_\a\over z-w} \ ,\qquad
\hat\lambda^\bh(z)\hat w_\ah(w)=-{\delta^\bh_\ah\over \bar z-\bar w} \ , \eea
since the pure spinor nature of the ghosts is relevant only at the next order of perturbation theory. To see this, recall the OPE's between the ghost Lorentz currents
\bea\label{lorentzope}
N^{[ef]}(z)N^{[pq]}(w)={\eta^{e[p}N^{f]q}(w)-\eta^{f[p}N^{q]e}(w)\over z-w}+3{\eta^{[ef][pq]}(w)\over (z-w)^2}\ ,\\
\hat N^{[ef]}(z)\hat N^{[pq]}(w)={\eta^{e[p}\hat N^{f]q}( w)-\eta^{f[p}\hat N^{q]e}(w)\over \bar z-\bar w}+3{\eta^{[ef][pq]}(w)\over (\bar z-\bar w)^2}\ ,\nonumber
\eea
The first term on the right hand side of (\ref{lorentzope}) is obtained by applying the naive OPE's (\ref{freeghost}), and the pure spinor nature of the curved beta-gamma system only affects the second terms in (\ref{lorentzope}). On the other hand, the divergent part of the effective action that we will compute below gets contribution only from the first term, so we can safely use the free field OPE's for the ghosts for computations at one-loop order.
The fish diagram contribution to the effective action coming from the last line in (\ref{ghostactio}) is given by
\bea\label{NNsecond}
\int d^2z\int d^2w\,{1\over |z-w|^2} \Bigl(W_\a\hat W_\gh(z)L^\rho \hat L^{\hat\rho}(w) f_{[ef]\hat\delta}^\gh  f_{[pq]\hat\rho}^{\hat \delta}\\
-W_\a \hat L^{\hat \rho}(z)\hat W_\gh L^\rho(w) f_{[ef]\hat\rho}^{\hat\delta}f_{[pq]\hat\delta}^\gh\Bigr)f_{[lm]\b}^\a f_{[rs]\rho}^\b\eta^{[ef][lm]}\eta^{[pq][rs]}\ .\nonumber
\eea
We can simplify the expression involving the structure constants by using the identities
\bea\label{lorentzstru}
f_{[ef]\hat\delta}^\gh f_{[pq]\hat\rho}^{\hat\delta}-f_{[pq]\hat\delta}^\gh f_{[ef]\hat\rho}^{\hat\delta}=&f_{[ef][pq]}^{[uv]}f_{[uv]\hat\rho}^\gh \ ,\\
f_{[ef]\delta}^\gamma f_{[pq]\rho}^{\delta}-f_{[pq]\delta}^\gamma f_{[ef]\rho}^{\delta}=&f_{[ef][pq]}^{[uv]}f_{[uv]\rho}^\gamma \ ,\nonumber
\eea
and by recalling the definition of the Ricci tensor (\ref{superricci})
\bea\label{riccih}
R_{[gh][rs]}(H)=-{1\over4}f_{[gh][ij]}^{[lm]}f_{[rs][lm]}^{[ij]}={c_2(H)\over4}\eta_{[gh][rs]}  \ ,
\eea
we can further simplify (\ref{NNsecond}) and extract its divergent part in the form
\bea\label{NNsecfin}
-{c_2(H)\over4\pi}\ln(\Lambda/\mu)\int d^2z \str\left\{N_{(0)}\hat N_{(0)}\right\} \ .
\eea

We collect the two contribution to the effective action (\ref{NNfirsttwo}) and (\ref{NNsecfin}) coming from (\ref{ghostactio}) and we find the divergent contribution
\bea\label{divghost}
S_{div,gh}=-{c_2(G)\over 4\pi}\ln(\Lambda/\mu)\int d^2z\str N_{(0)}\hat N_{(0)} \ .
\eea
Once again, (\ref{divghost}) vanishes due to the fact that $c_2(G)=0$.

There are a few more terms that might give contributions to the divergent part of the effective action in the ghost sector. Those are the terms that generate the operators $\str N_{(0)}\bar J_0$, $\str \hat N_{(0)}J_0$, and $\str J_0\bar J_0$, which would break gauge invariance. We need to check that these terms are absent.

The term $\str N_{(0)}\bar J_0$ is
generated by two different diagrams. The first is the self-energy diagrams involving the terms proportional to $\bar J_0$ in the first line of (\ref{ghostactio}). The second  contribution to the term $\str N_{(0)}\bar J_0$ involves a fish diagram with the contraction of terms with partial derivatives in the first line of (\ref{ghostactio}) together with the covariant derivatives, proportional to one power of $\bar J_0$, in the kinetic terms for the fluctuations in (\ref{kinaction}). Collecting the two contributions we find
\bea\label{njezero}
S_{eff,V}=&-{c_2(G)-c_2(H)\over2}\Bigl\{\int d^2z [-\ln(0)]\str N_{(0)}\bar J_0 \eea
$$
+\half\int d^2z\int d^2w \left({1\over |z-w|^2}-\delta^{(2)}(z-w)\ln|z-w|^2\right)\str N_{(0)}(z)\bar J_0(w) \Bigr\}\ . $$
By summing up the two terms we see that the divergent terms exactly cancel. The same story applies to the effective action for $\str \hat N_{(0)} J_0$.

Finally, we need to check the terms proportional to $\str J_0\bar J_0$. Those are generated by the self-energy diagrams and the fish diagrams coming from the terms with $J_0$ and $\bar J_0$ in the covariant derivatives in the kinetic terms (\ref{kinaction}), that sum up to give
\bea\label{effjzero}
S_{eff,VI}=&\int d^2z [-\ln(0)]J_0^{[ef]}\bar J_0^{[pq]}\left(-\half f_{[ef]a}^cf_{[pq]c}^a+f_{[ef]\a}^\b f_{[pq]\b}^\a\right) 
\eea
$$
+\int d^2z \int d^2w J_0^{[ef]}(z)\bar J_0^{[pq]}(w)\Bigl[{1\over4}f_{[ef]a}^b f_{[pq]b}^a\left({1\over|z-w|^2}-\delta^{(2)}(z-w)\ln|z-w|^2\right)
$$
$$+f_{[ef]\a}^\b f_{[pq]\b}^\a\left({5\over 8}\delta^{(2)}(z-w)\ln|z-w|^2-{3\over8}{1\over|z-w|^2}\right)\Bigr] \ \ , $$
where we used the identity $f_{[ef]\ah}^\bh f_{[pq]\bh}^\ah=f_{[ef]\a}^\b f_{[pq]\b}^\a$. It is immediate to check that the divergent part of (\ref{effjzero}) is identically zero.

\subsubsection{Non-renormalization of the radius}

We have compute the quantum effective action and checked that all divergent terms vanish. However, there are leftover finite terms. We will first compute them and then discuss their significance. Collecting them from (\ref{seffi}), (\ref{leffII}), (\ref{leffIII}), (\ref{NNfirsttwo}), (\ref{NNsecond}), (\ref{njezero}), (\ref{effjzero}) we obtain the total one-loop effective action
\bea\label{effectiveaction}
S_{eff}=&\int d^2z\,\str \Bigl(a_1J_2\bar J_2+a_2 J_1\bar J_3+a_3 J_3\bar J_1 \ \\
&+\half c_2(H)(J_0\bar J_0-N\bar J_0-\hat N J_0) \Bigr)\ . \nonumber
\eea
where $a_1=8$, $a_2=-10$, $a_3=5/4$, and $c_2(H)=3$ is the quadratic Casimir of the group $H=SO(1,4)\times SO(5)$. We did not include the IR singular terms proportional to $\ln |p|^2/\mu^2$, which can be removed once the full perturbative series is included \cite{deWit:1993qv}. The expression (\ref{effectiveaction}) is local, hence it can be removed by adding a local counter-term
\bea\label{counterterm}
S_{c.t.}=-S_{eff}\ ,
\eea
to the action.

Because there are no non-local finite terms, we proved that there are {\it no gauge nor BRST anomalies} at one-loop. But there is a further result we can infer from this fact.

In bosonic WZW models, describing the string propagation on the coset $G/H$, the level of the current algebra gets shifted at one-loop from $k$ to $k+\half c_2(G)$, where $c_2(G)$ is the quadratic Casimir of the group $G$. In the sigma model interpretation of WZW theory, the level is related to the radius of the target space
manifold, which is the inverse of the sigma model coupling constant. Therefore, the classical relation $R^2/\alpha'=k$ gets modified at one-loop to $R^2/\alpha'=k+\half c_G$ and in the full quantum theory there is a minimal value for the radius of the manifold, set by the quadratic Casimir of the group.
The situation is different for gauged WZW models with worldsheet supersymmetry. In that case, the fermionic and bosonic determinants cancel out and the relation between the radius and the level is not renormalized. These sigma models describe bosonic or RNS string theory on backgrounds supported by NS-NS flux. What happens  with Ramond-Ramond flux?

In the case of $AdS_5\times S^5$, we have a sigma model on a supercoset. The AdS radius is again equal to the inverse of the sigma model coupling constant and is related to the 't Hooft coupling $\mL$ of the dual ${\cal N}=4$ super Yang-Mills theory through the dictionary
$$
R^2/\alpha'=f(\mL) \ , \qquad f(\mL)\sim_{\mL\to\infty}\sqrt{\mL} + a+{\cal O}(1/\sqrt{\mL})\ ,
$$
The leading term in the large 't Hooft coupling expansion corresponds to the usual AdS/CFT dictionary, but in principle subleading terms are allowed and $a$ would arise at one-loop in the sigma model perturbation theory. This would be the analogue of the finite shift by $\half c_2(G)$ in the level of the current algebra in bosonic WZW models.

To study the renormalization of the radius, we need to consider the sigma model quantum effective action
$$
S_{eff}=S_{div}+S_{finite} \ .
$$
We have shown that the divergent part of the effective action vanishes, which implies that the sigma model is conformally invariant and the radius does not run. However, one still needs to evaluate the finite part of the effective action, which may consist of local as well as non-local terms. The local terms can be reabsorbed by local counter-terms and play no role. On the other hand, the presence of finite non-local contributions to the effective action would not be removed and generate a non-zero shift $a$.
But we have just concluded that all non-local contributions vanish. Hence, we have proven that there is no departure from
the classical AdS/CFT dictionary at one-loop
$$
a=0 \ .
$$

\subsubsection{Central charge}

We can use the background field method to compute the one-loop correction to the central charge of the supercoset sigma model. Consider the left and right moving components of the stress tensor
\bea\label{stresste}
T=&-{\rm Str}\left( \half J_2 J_2+J_1J_3+w\nabla \lambda \right)\ , \\
\bar T=&-{\rm Str}\left( \half \bar J_2\bar J_2+\bar J_1\bar J_3+\hat w\bar \nabla \hat
\lambda \right)\ .
\eea
We want to compute the one-loop correction to the central charge. It corresponds to the quartic pole in the OPE
$$
\langle T(z) T(0) \rangle= {c/2 \over z^4}+\ldots ,
$$
where $\langle\cdot\rangle$ denotes functional integration.
We expand $T$ according to (\ref{gradingex}) and we compute the contractions of the fluctuations. The terms coming from the action do not contribute to the central charge. We will find a leading tree level contribution, proportional to $1/R^4$, where $R$ is the radius. The one-loop correction is proportional to $1/R^6$. To compute terms of order $1/R^8$ we need to expand (\ref{gradingex}) up to ${\cal O}(R^{-3})$, so they will be neglected and we will stop at one-loop. We find
\bea
\langle \half {\rm Str} J_2J_2(z)\half {\rm Str} J_2J_2(0)\rangle={1\over R^4}{1\over z^4}\left({10\over2}-{1\over 2R^2} [1+\ln|z-w|^2]\eta^{lm} f_{l\alpha}^{\hat \delta}f_{m\hat\delta}^\alpha\right) \ ,\nonumber\\ \label{twotwo}
\eea
where $\eta^{lm} f_{l\alpha}^{\hat \delta}f_{m\hat\delta}^\alpha={1\over4}{\rm Tr}\gamma^a\gamma_a=40$.
The first term arises from the double contraction at tree level, while the second comes from the triple contraction at one-loop.
The second contribution is
\bea
\langle {\rm Str} J_1J_3(z){\rm Str} J_1J_3(0)\rangle={1\over R^4}{1\over z^4}\left({32\over2}-{1\over 2R^2} [1+\ln|z-w|^2]\eta^{lm} f_{l\alpha}^{\hat \delta}f_{m\hat\delta}^\alpha\right) \ .\nonumber\\ \label{onethree}
\eea
The mixed term is
\bea\label{mixed}
\langle {\rm Str}  J_2J_2(z){\rm Str} J_1J_3(0)\rangle={1\over R^6}{1\over z^4} [1+\ln|z-w|^2]\eta^{lm} f_{l\alpha}^{\hat \delta}f_{m\hat\delta}^\alpha\ .
\eea
By summing up (\ref{twotwo}), (\ref{onethree}) and (\ref{mixed}) we get the total contribution of the matter part. The one-loop correction cancels out exactly, leaving only the tree level part, which is the same as in flat space
\bea\label{matter}
\langle T_{matter}(z) T_{matter}(0)\rangle = -{1\over R^4}{22\over z^4} \ .
\eea

Let us look at the ghost part. The tree level contribution involves a trace on the ghost spinor indices and is equal to the analogous flat space contraction. In the gauge $X_0=0$ the ghost sector does not give any one-loop correction and it starts contributing only at two loops (the leading term in $w [J_0,\lambda]$ with no external fields is ${\cal O}(R^{-4})$), so we find
\bea\label{ghostt}
\langle T_{gh}(z)T_{gh}(0)\rangle ={1\over R^4}{22\over z^4}\ ,
\eea
and by adding (\ref{ghostt}) and (\ref{matter}) we proved that the total central charge vanishes at one-loop:
\be
\label{czero}
c=0 \ .
\ee
Since the effective action does not receive any finite corrections at one-loop, there is no correction to the stress tensor either.

\subsection{OPE algebra}
\label{section:OPE}

Let us consider the algebra of the Operator Product Expansion of the left invariant currents in the pure spinor sigma model. In the case of bosonic sigma models on a group manifold, for instance WZW models, the OPE's of the currents realize a Kac-Moody algebra, where the structure constants take values in the Lie algebra of the group. Such sigma models are parity invariant, hence the Kac-Moody algebra comes in two copies: holomorphic and anti-holomorphic.

In the $AdS_5\times S^5$ sigma model, the RR flux couples left and right movers, so the OPE's of the left invariant currents form a non-chiral algebra, that we will now compute at the tree-level. Note that logarithms are present in the OPE's.

The tree level OPE's are proportional to one structure constant and the sigma model coupling constant $R^{-2}$. Only the $(1/R)$ term in (\ref{gradingex}) contributes, together with the first lines in (\ref{sone}), (\ref{stwo}) and (\ref{sthree}) and the first two lines in (\ref{ghostactio}). Schematically, the OPE's are given by
 \bea\label{opescheme}
 R^2 \langle J^A(z) J^B(0)\rangle=  \langle \partial X^A (z) (\partial X^B(0)+ [J,X]^B (0)) \rangle + \langle [J,X]^A(z) \partial  X^B(0) \rangle \ ,
\nonumber
 \eea
On the left hand side we put the full sigma model current, while in the right hand side we expanded it using the background field method (\ref{gradingex}) and we keep only the leading term. Note that, for ease of notations, we dropped the tilde's from the currents in the r.h.s., even if they are understood as background currents. Since these expressions are evaluated inside a correlator, the terms on the r.h.s. pick up a contribution from the couplings in the action, namely
\bea\label{examples}
\langle \partial X^A (z) \partial X^B(0) \rangle= \partial X^A (z) \partial X^B(0) -\partial X^A (z) \partial X^B(0) \cdot S \ ,
\eea
where $S=S_I+S_{II}+S_{III}$ are the terms in the classical action, expanded at quadratic order in the quantum fluctuations and given in (\ref{sone}), (\ref{stwo}), (\ref{sthree}) and (\ref{ghostactio}). We compute the r.h.s. of (\ref{examples}) with the propagators in (\ref{freeope}), some details are collected in \ref{appendix:OPE}.

We find\footnote{We omit the overall factor $R^{-2}$ in the r.h.s.}
\bea\label{opes}
J_2^a(z)J_2^b(0)\to&-{\eta^{ab}\over z^2}-{f_{[ef]}^{ab}\over z}[N^{[ef]}-{\bar z\over z}\hat N^{[ef]}]\ ,\eea
$$
J_2^a(z)\bar J_2^b(0) \to 2\pi\delta^{(2)}(z)\eta^{ab}-f_{[ef]}^{ab}[{N^{[ef]}\over \bar z}+{\hat N^{[ef]}\over z}+\half \ln|z|^2 (\partial \hat N^{[ef]}-\bar \partial N^{[ef]})]\ ,\hfill
$$
$$
J_2^a(z) J_1^\alpha(0) \to  -{f_{\ah}^{a\alpha}\over z}\left(2J_3^\ah+{\bar z\over z} \bar J_3^\ah\right)\ ,\hfill
$$
$$
J_2^a(z) \bar J_1^\alpha(0) \to  -f^{a\alpha}_\ah\left({J_3^\ah\over \bar z}+{1\over 8}\ln|z|^2(\partial \bar J_3^\ah+\bar \partial J_3^\ah)\right)\ ,\hfill
$$
$$
J_2^a(z) J_3^\ah (0) \to  -{f_{\alpha}^{a\ah}\over z}J_1^\alpha\ ,\hfill
$$
$$
J_2^a(z) \bar J_3^\ah(0) \to  -f_\alpha^{a\ah}\left({\bar J_1^\alpha\over z}+{1\over4}\ln|z|^2(\bar \partial J_1^\alpha+\partial\bar J_1^\alpha)\right)\ ,\hfill
$$
$$
J_1^\alpha(z) J_1^\beta(0)\to -{f_a^{\alpha\beta}\over z}\left(2J_2^a+{\bar z\over z}\bar J_2^a\right)\ ,\hfill
$$
$$
J_1^\alpha(z)\bar J_1^\beta(0) \to -f^{\alpha\beta}_a\left({J_2^a\over \bar z}-{1\over4} \ln|z|^2( \partial \bar J_2^a+\bar \partial J_2^a\right)\ ,\hfill
$$
$$
J_1^\alpha(z) J_3^\ah(0) \to  {\eta^{\alpha\ah}\over z^2}-{f_{[ef]}^{\alpha\ah}\over z}\left(N^{[ef]}-{\bar z\over z}\hat N^{[ef]}\right) \ ,\hfill
$$
$$
J_1^\alpha(z) \bar J_3^\ah(0) \to -2\pi\delta^{(2)}(z)\eta^{\alpha\ah}-f_{[ef]}^{\alpha\ah}\left({N^{[ef]}\over \bar z}+{\hat N^{[ef]}\over z}-\half \ln|z|^2(\bar\partial \hat N^{[ef]}-\partial \hat N^{[ef]})\right)  \ ,\hfill
$$
$$
\bar J_1^\alpha(z) J_3^\bh(0) \to -2\pi\delta^{(2)}(z)\eta^{\alpha\ah}-f_{[ef]}^{\alpha\ah}\left({N^{[ef]}\over \bar z}+{\hat N^{[ef]}\over z}+\half\ln|z|^2(\partial \hat N^{[ef]}-\bar\partial N^{[ef]})\right)\ ,\hfill
$$
$$
J_3^\ah(z) J_3^\bh(0) \to f^{\ah\bh}_a{J_2^a\over z}\ ,\hfill
$$
$$
J_3^\ah(z) \bar J_3^\bh(0) \to -f_a^{\ah\bh}\left({\bar J_2^a\over z}+{1\over4}\ln|z|^2(\partial \bar J_2^a+\bar \partial J_2^a\right)\ ,\hfill
$$
while all the OPE's with $J_0$ and $(N,\hat N)$ vanish at tree level. On the right hand sides, we omitted terms proportional to the product of two currents, whose singularity is logarithmic. By use of the Maurer-Cartan identities, that are Ward identities for the supergroup symmetry, these terms can be reshuffled with the terms proportional to the derivative of the currents, that we have included. We refer to the literature for a more detailed discussion of this point.

We do not need to compute all the OPE's, since the pure spinor sigma model exhibits a symmetry under the simultaneous exchange of $z\leftrightarrow \bar z$ and grading one with grading three. The remaining OPE's may be obtained from (\ref{opes}) by applying this symmetry.

Using the background field method we can push the computation at any order in the perturbative expansion. For example, we can compute the one-loop corrections to the tree level OPE algebra. This can be used to study the worldsheet anomalous dimensions $\gamma_{ws}$ of the left-invariant currents $J_i$, which are given by the loop corrections to the double pole in their OPE with the stress tensor
$$
T(z)J_i(0)\to {1+\gamma_{ws}\over z^2}J_i(0)+\ldots \ ,
$$
where we omitted the single pole terms in the OPE. One finds that at one loop the anomalous dimensions of $J_2$ and $J_0$ vanishes, but the fermionic currents have
\bea
\gamma_{ws}(J_1)=&-{1\over R^2}{5\over 16} \ ,\\
\gamma_{ws}(J_3)=&+{1\over R^2}{5\over 16} \ .\nonumber
\eea
Since the anomalous dimensions of $J_1$ and $J_3$ have opposite sign and only their product appears in the stress tensor, we just proved that the stress tensor does not receive any anomalous dimension at one-loop and thus it is still conserved.

\subsection{All-loop quantum consistency}
\label{sectionquantum}

It turns out that the pure spinor sigma model is gauge
invariant and BRST invariant to all orders in the sigma-model
perturbation theory. Let us extend the one-loop results discussed above to all-loops.

\subsubsection{Gauge invariance}

The pure spinor action (\ref{eqAdSpsaction}) is classically gauge invariant
under the right multiplication $g\to gh$, where $h\in H$. We will
prove that we can always add a local counterterm such that the
quantum effective action remains gauge invariant at the quantum
level. Quantum gauge invariance will then be used to prove
BRST invariance.

An anomaly in the $H$ gauge invariance would show up as a
nonvanishing gauge variation of the effective action
$\delta_\Lambda S_{eff}$ in the form of a local operator. Since
there is no anomaly in the global $H$ invariance, the variation
must vanish when the gauge parameter is constant and, moreover, it
must have grading zero. Looking at the list of our worldsheet
operators, we find that the most general form of the variation is
 \bea\label{lambdas}
 \delta S_{eff}=&\int d^2z\langle c_1 N\bar\partial \Lambda+\bar c_1\hat
 N\partial \Lambda+2c_2J_0\bar\partial\Lambda+2\bar c_2\bar
 J_0\partial\Lambda\rangle,
 \eea
where $\Lambda=T_{[ab]}\Lambda^{[ab]}(z,\bar z)$ is the local
gauge parameter and $(c_1,\bar c_1,c_2,\bar c_2)$ are arbitrary
coefficients. By adding the counterterm
 $$
 S_c=-\int d^2z\langle c_1 N\bar J_0+\bar c_1\hat NJ_0+(c_2+\bar
 c_2)J_0\bar J_0\rangle,
 $$
we find that the total variation becomes
 $$
 \delta_\Lambda (S_{eff}+S_c)=(c_2-\bar c_2)\int d^2z\langle
 J_0\bar\partial\Lambda-\bar J_0\partial\Lambda\rangle.
 $$
On the other hand, the consistency condition on the gauge anomaly
requires that
 $$
( \delta_\Lambda\delta_{\Lambda'}-\delta_{\Lambda'}\delta_\Lambda)
S_{eff}=\delta_{[\Lambda,\Lambda']}S_{eff},
 $$
which fixes the coefficients $c_2=\bar c_2$.
Therefore the action is gauge invariant quantum mechanically.

\subsubsection{BRST invariance}
\label{quantumbrst}

In order to prove the BRST invariance of the superstring at all
orders in perturbation theory we will first show that the classical BRST charge is nilpotent. We will
then prove that the effective action can be made classically BRST
invariant by adding a local counterterm, using triviality of a
classical cohomology class. Then we will prove that order by order
in perturbation theory no anomaly in the BRST invariance can
appear.

As we have shown in the previous section, the action
(\ref{eqAdSpsaction}) in the pure spinor formalism is
classically BRST invariant. It is easy to prove that the pure spinor BRST charge is classically nilpotent
on the pure spinor constraint, up to gauge invariance and the
ghost equations of motion. The second variation of the ghost
currents reads indeed
 $$
 Q^2(N)=-[N,\Lambda]-\{\l,\nabla\hat\l-[N,\hat\l]\},
 $$
 $$
 Q^2(\hat N)=-[\hat N,\Lambda]-\{ \hat \l,\bar\nabla\l-[\hat N,\l]\},
 $$
for the particular gauge transformation parameterized by
$\Lambda=\{\l,\hat \l\}$ and the equations of motion (\ref{ghosteoms}). Therefore the classical BRST charge is well defined.

Consider now the quantum effective action $S_{eff}$. After the
addition of a suitable counterterm, it is gauge invariant to all
orders. Moreover, the classical BRST transformations of
(\ref{brstrans}) commute with the gauge transformations, since the
BRST charge is gauge invariant. Therefore, the anomaly in the
variation of the effective action, which is a local operator, must
be a gauge invariant integrated vertex operator of ghost number
one
 $$
  \delta_{BRST}S_{eff}=\int d^2z\langle {\Omega}_{z\bar z}^{(1)}\rangle.
$$
In \ref{empti} we show that the cohomology of such
operators is empty, namely that we can add a local counterterm to
cancel the BRST variation of the action. A crucial step in the
proof is that the symmetric bispinor, constructed with the product
of two pure spinors, is proportional to the middle dimensional
form. Schematically, this means that in $d=2n$ dimensions we can
decompose
$$ \l^\a\l^\b\sim\gamma_{m_1\ldots
m_n}^{\a\b}(\l\gamma^{m_1\ldots m_n}\l).
 $$

Since there are no conserved currents of ghost number two in the
cohomology, that could deform $Q^2$, the quantum modifications to
the BRST charge can be chosen such that its nilpotence is
preserved. In this case, we can
use algebraic methods to extend the BRST invariance of the
effective action by induction to all orders in perturbation
theory. Suppose the effective action is invariant to order
$h^{n-1}$. This means that
 $$\tilde Q S_{eff}=h^n\int d^2z\langle {\Omega}^{(1)}_{z\bar
z}\rangle+{\cal O}(h^{n+1}).
$$
The quantum modified BRST operator $\tilde Q=Q+Q_q$ is still
nilpotent up to the equations of motion and the gauge invariance.
This implies that $Q\int\,d^2z\langle\Omega^{(1)}_{z\bar
z}\rangle=0$. But the cohomology of ghost number one integrated
vertex operators is empty, so $\Omega^{(1)}_{z\bar z}=Q
\Sigma^{(0)}_{z\bar z}$, which implies
$$ \tilde Q\left(S_{eff}- h^n\int\,d^2z\langle\Sigma^{(0)}_{z\bar
 z}\rangle\right)= {\cal O}(h^{n+1}).
 $$
Therefore, order by order in perturbation theory it is possible to
add a counterterm that restores BRST invariance.

\subsubsection{Conformal invariance}
\label{allloopconf}

In Section~\ref{section:conformalinv} we gave an explicit diagrammatic proof of the one-loop conformal invariance of the pure spinor sigma model in $AdS_5\times S^5$. This result is of course expected on general grounds, since in a generic curved background classical BRST invariance implies one-loop conformal invariance. But we can actually do much better and prove conformal invariance at all orders in perturbation theory.

Let us extend the $PSU(2,2|4)$ supergroup to $U(2,2|4)$ by adding the hypercharge ${\bf Y}$ and the central extension ${\bf I}$, as discussed in (\ref{remaining}). The new generators have grading two and act on the $PSU(2,2|4)$ ones as
\bea\label{newutwo}
[{\bf Y},{\bf T}_{1}]={\bf T}_{3}\ , \qquad
[{\bf Y},{\bf T}_{3}]={\bf T}_{1} \\
\{{\bf T}_1,{\bf T}_1\}=\{{\bf T}_3,{\bf T}_3\}= {\bf T}_2+{\bf I} \ .\nonumber
\eea
We may parameterize the element of the $U(2,2|4)/SO(1,4)\times SO(5)$ coset as
\bea\label{cosetu}
h=e^{u{\bf I}+v{\bf Y}}g(x,\theta,\hat\theta) \ ,
\eea
where $g(x,\theta,\hat\theta)$ is the element of the original $PSU(2,2|4)/SO(1,4)\times SO(5)$ coset. The new Maurer-Cartan currents read
\bea\label{maureru}
h^{-1}dh=K_I+K_Y+K_0+K_1+K_2+K_3 \ ,
\eea
where
$$
K_I=\left(d u+[g^{-1}(dv {\bf Y})g]^I+[g^{-1}dg]^I\right){\bf I} \ ,
$$
$$
K_Y=dv{\bf Y} \ ,
$$
$$
K_0+K_1+K_2+K_3=\left([g^{-1}(d v{\bf L})g]^A+[g^{-1}dg]^A\right) {\bf T}_A \ .
$$
In particular, note that $u$ only appears as a free boson, since ${\bf I}$ commutes with all the generators, and $v$ only appears linearly, because the hypercharge ${\bf Y}$ never appears on the r.h.s. of the commutation relations (\ref{newutwo}).

Under the new BRST transformation $\epsilon Q'(h)=h(\epsilon \lambda+\epsilon\hat\lambda)$, the currents transform as
\bea\label{brstcu}&
\epsilon Q'(K_I)=[K_3,\epsilon\hat\lambda]+[K_1,\epsilon\lambda] \ ,\qquad \epsilon Q'(K_Y)=0 \ ,\\&
\epsilon Q'(K_j)=\delta_{j+3,0}(\epsilon\partial \lambda+[K_Y,\epsilon\hat\lambda])+[K_{j+3},\epsilon\lambda]\nonumber\\
&+\delta_{j+1,0}(\epsilon\partial \hat\lambda+[K_Y,\epsilon\lambda])+[K_{j+1},\epsilon\hat\lambda] \ .\nonumber
\eea

Let us consider the action for the extended supercoset $U(2,2|4)/SO(1,4)\times SO(5)$. Since $\str {\bf I}^2=\str {\bf Y}^2=0$ and the only non-zero supertrace involving the extra generators is $\str {\bf I}{\bf Y}\neq 0$, we find the action
\bea
S\,'=S\,'_0+\int d^2z\left(\partial u\bar\partial v+\bar \partial u\partial v+j_z(x,\theta,\hat\theta)\bar\partial v+j_{\bar z}(x,\theta\,\hat\theta)\partial v+f(x,\theta,\hat\theta )\partial v\bar \partial v\right) \ ,\nonumber\\ \label{actionu}
\eea
where $S\,'_0$ is the original pure spinor action (\ref{eqAdSpsaction}) with the original currents $J$'s replaced by the new $K$'s. By using the Maurer-Cartan equations for the currents $K$ one can check that the action (\ref{actionu}) is invariant under the BRST transformations (\ref{brstcu}). One can go through the proof of quantum gauge and BRST invariance of the previous sections and repeat them for the action (\ref{actionu}). In particular, one can add a local counterterm to remove the BRST variation of the quantum effective action $S\,'_{eff}$.

By inverting the kinetic terms in the action (\ref{actionu}), we find a propagator for $\langle uu\rangle$ and $\langle uv\rangle$ but no propagator for $\langle v v\rangle$. Moreover, we see that only $v$ couples to the original $PSU(2,2|4)$ variables. Therefore, we readily conclude that in the perturbative expansion of the theory (\ref{actionu}), all the diagrams with external $PSU(2,2|4)$ lines are exactly the same as the corresponding diagrams computed in the original theory with action $S_0$. If the effective action $S\,'_{eff}$ computed from (\ref{actionu}) is conformally invariant, then it follows that the effective action for the original $PSU(2,2|4)$ supercoset is also conformally invariant.

We are now going to prove that $S\,'_{eff}$ is conformally invariant. In order to do this, we will at first assume that quantum conformal transformations commute with the quantum BRST transformation
\bea\label{confbrst}
[\delta_{c},Q']=0 \ ,
\eea
and we postpone the proof of this statement to the end of this section. From (\ref{confbrst}) it follows that $Q'( \delta_{c} S\,'_{eff})=0$ and so $\delta_{c} S\,'_{eff}$ is a BRST invariant and local expression. But there is only one local and BRST invariant object of ghost number zero and weight two and that is precisely the classical action $S\,'$. This fact is proven in~\ref{empti} for the original $PSU(2,2|4)$ invariant action, and the same argument holds for the present action. Hence
\bea\label{confcla}
\delta_{c}S\,'_{eff}=a S\,' \ ,
\eea
where $a$ is a numerical coefficient.

Now $S\,'$ contains the term $\int \partial u\bar \partial v$, but this term cannot get any correction because there is no interaction that contains $u$, so this term cannot appear in the 1PI effective action. We deduce that the coefficient $a$ in (\ref{confcla}) must be zero and so the effective action is conformally invariant
\bea\label{confinv}
\delta_{c}S\,'_{eff}=0 \ .
\eea

Let us finally prove (\ref{confbrst}) by cohomology arguments. We denote the quantum BRST charge by $\tilde Q=Q^{cl}+\hbar Q^q$, where the last term is the possible $(n-1)$-loop correction to the classical charge and $\hbar$ is our loop counting parameter. We further denote by $\delta_{c}$ the quantum conformal transformation at $(n-1)$-loops. Suppose now that (\ref{confbrst}) does not hold, namely that $[\tilde Q_,\delta_c]=\hbar^n\delta'+{\cal O}(\hbar^{n+1})$, where $\delta'$ is a local transformation of ghost number one. Since $\{\tilde Q,\delta'\}=0$ we have in particular $\{Q^{cl},\delta'\}=0$. It turns out that there is no such local charge at ghost number one in the cohomology, which we prove in Appendix C. We conclude that $\delta'$ must be BRST trivial, namely $\delta'=-[Q^{cl},\delta_q]$, for some ghost number zero charge $\delta_q$. The charge $\delta_c+\hbar^n\delta_q$ now commutes with the BRST charge
\bea\label{proofconf}
[\tilde Q,\delta_c+\hbar^n\delta_q]=\hbar^n\delta'-\hbar^n\delta'+{\cal O}(\hbar^{n+1}),
\eea
so we conclude that conformal transformations commute with the BRST charge at the quantum level.

{\bf Guide to the literature}

A textbook introduction to the background field method as a means to study two-dimensional non-linear sigma models can be  found in Polyakov's book \cite{Polyakov:1987ez}. The method we use to evaluate the one-loop effective action was introduced in \cite{deWit:1993qv} and further discussed in \cite{DeBoer:1995hv,deBoer:1995cb,deBoer:1996kt}, where the map between the position space and momentum space integrands is explained.

The divergent part of the one-loop effective action for  string sigma models on supercosets with RR flux was computed for the matter part in \cite{Berkovits:1999zq} and for the full pure spinor superstring in \cite{Vallilo:2002mh}. The finite part of the effective action and the non-renormalization of the radius was computed in \cite{Mazzucato:2009fv}, where the vanishing of the central charge at one-loop was also proven. The reader may find a discussion of the finite shift in the radius of WZW models in \cite{Knizhnik:1984nr}, where in the bosonic case the radius gets shifted, and in \cite{Tseytlin:1993my}, where in the supersymmetric RNS case the radius is not shited. The authors of \cite{Berenstein:2009qd} argue that the classical relation between the radius and the 't Hooft coupling is exact in the planar limit, based on completely different considerations of S-duality of the type IIB superstring and of the dual ${\cal N}=4$ super Yang-Mills theory.

The algebra of OPE for the pure spinor superstring was derived at the tree level in \cite{Bianchi:2006im,Puletti:2006vb,Mikhailov:2007mr}. The one-loop corrected OPE algebra can be found in \cite{Bedoya:2010av}.

The quantum consistency of the pure spinor sigma model in AdS at all-loops was proven in \cite{Berkovits:2004xu}.

\clearpage

\section{String spectrum}
\label{section:spectrum}

In this section, we will compute the perturbative spectrum of superstrings propagating in $AdS_5\times S^5$, with its application to the AdS/CFT correspondence in mind. We set the sigma model coupling equal to the inverse of the $\cN=4$ SYM 't Hooft coupling
$R^2/\alpha'=\sqrt{\mL}$,
which is more convenient for the AdS/CFT analysis.

We will give different examples of physical string states. We start with a short review of some aspects of the AdS/CFT correspondence, related to the spectrum of anomalous dimensions of local gauge invariant operators on the gauge theory. We first discuss the massless vertex operators, corresponding to type IIB supergravity compactified on the five-sphere and dual to the half BPS sector of $\cN=4$ SYM theory. As an example of a massive string state, we derive the energy of a vertex operator at the first massive string level, that is dual to a member of the Konishi multiplet, a long super-multiplet of $\cN=4$ SYM. Since this operator is not protected, its anomalous dimension is non-vanishing. We compute the one-loop corrections to the Virasoro constraint, whose solution determines the energy, which in turn gives the anomalous dimension. The result confirms its earlier conjectured value, obtained using integrability and the Y-system. Finally, we briefly discuss the $n$-point correlation functions of gauge invariant local operators in $\cN=4$ SYM, dual to the $n$-point function of vertex operators on the string worldsheet. We construct the zero mode measure for the worldsheet variables and show that there is a well-defined higher genus amplitude prescription, which computes $1/N_c$ corrections to the planar limit of the AdS/CFT system. In the course of the discussion we introduce the $b$ antighost in the AdS, whose construction is significantly easier than its flat space cousin, due to the RR flux present in the AdS background.

\subsection{AdS/CFT correspondence}

Let us recall some general facts about the AdS/CFT correspondence. Consider maximally supersymmetric Yang-Mills theory in four dimensions ($\cN=4$ SYM). We will restrict for the moment to the planar sector of the theory, in which the number of colors $N_c\to\infty$ at fixed 't Hooft coupling $\mL$. $\cN=4$ SYM is a conformal field theory with global symmetry group $PSU(2,2|4)$, in particular its global symmetry contains the dilatation operator $D$. The observables we will focus on in this Section are the anomalous dimensions of gauge invariant local operators $\cO(x)$. The eigenvalues of the dilatation operator are the conformal dimensions $\Delta$, defined as
\be
[D,\cO_\Delta(x)]=i\Delta\cO_\D(x) \ .
\ee
If we know the conformal dimensions of all operators we have solved the two-point function sector of $\cN=4$ SYM, since
\be
\langle \cO^\dagger_{\D_I}(x)\cO_{\Delta_J}(y)\rangle={\delta^I_J\over|x-y|^{2\D_I} 
}\ .
\ee
What do we know about the conformal dimensions of operators in $\cN=4$  SYM? In the classical theory at zero coupling, conformal dimensions are equal to the "engineering dimensions" $\D_0$, i.e. the free field values that can be read off the classical Lagrangian. As we turn on the coupling $\mL$, quantum corrections generate anomalous dimensions
\be
\gamma=\D-\D_0 \ .
\ee
At weak coupling, i.e. when $\mL<<1$, the anomalous dimensions can be computed in a perturbative expansion loop by loop
\be
\gamma=c_1\mL+c_2\mL^2+\ldots 
\ee
where $c_i$ is a numerical coefficient obtained by an $i$-th loop computation. An exception is given by the BPS operators, namely short multiplets in $\cN=4$ SYM whose anomalous dimensions vanish due to supersymmetry. An example of a BPS operator is the chiral primary $\Tr \phi^{\{i}\phi^{j\}}$, where $\phi^i$ are the six scalars of $\cN=4$ SYM and their $SO(6)$ vector indices are in a symmetric traceless combination.

In the strong coupling regime, $\mL>>1$, perturbation theory breaks down and we need new tools to study the spectrum of anomalous dimensions. Thanks to the AdS/CFT correspondence, type IIB superstring theory on $AdS_5\times S^5$ comes to our rescue.

According to the AdS/CFT dictionary, conformal dimensions on the CFT side are equal to energies of string states on the AdS side
\be
\label{deltae}
\Delta=E \ .
\ee
In the strong coupling regime $\mL>>1$, the dual AdS background has a very large radius in units of the string length, hence its curvature is weak and we are in the supergravity approximation. In particular, the string sigma model coupling, proportional to the inverse radius, is small and we are in the perturbative string theory regime. As a result, the perturbative worldsheet computation of energy of string states in AdS gives us the strong coupling expansion of conformal dimensions of local operators on the gauge theory side. 

At strong coupling, we can match conformal dimensions of local operators with energies of strings as in (\ref{deltae}), and we can classify the various strings moving in AdS in three different sectors according to the different scaling of their energy as follows
$$
\begin{array}{ccc|cc}
		\textrm{SYM side} &&& \textrm{String side}&\\
\textrm{BPS:}    && \gamma=0       & \textrm{supergravity:}&E\sim \cO(1) \\
\textrm{``short" non-BPS:} && \gamma\sim \sqrt[4]{\mL} & \textrm{massive:}& E\sim\sqrt[4]{\mL}  \\
\textrm{``long" non-BPS:} && \gamma\sim \sqrt{\mL} & \textrm{semi-classical:} & E\sim\sqrt{\mL}
\end{array}
$$
The first sector contains the gauge theory BPS operators, whose anomalous dimensions vanish. This is dual to the massless sector of the superstring, which describes ten-dimensional type IIB supergravity compactified on the five-sphere. This is described on the worldsheet by physical vertex operator of weight zero on the worldsheet. 

In the second sector we put non-BPS operator that are made of a small number of elementary $\cN=4$ SYM fields. The prototypical example is the Konishi multiplet, whose highest weight state is classically $\Tr \phi^i\phi^i$. This kind of operators are dual to the massive superstrings. In flat space, the massive string states have a mass squared that scales as $M^2\sim1/\a'$, so in the large radius limit we see that their energy scales as $E\sim \sqrt[4]{\mL}$. The Konishi multiplet is dual to strings at the first massive level, described by physical worldsheet vertex operators of classical weight $(1,1)$.

The last sector contains non-BPS operators made out of an asymptotically large number of elementary fields. They correspond to semi-classical macroscopic strings with very large quantum numbers, whose energy scales as $E\sim \sqrt{\mL}$.

In this section, we will first consider the massless sector and show that it correctly describes type IIB supergravity on $AdS_5\times S^5$. We will also provide two examples of zero-momentum deformations of the action. We will then compute the energy of a string state at the first massive level, that will give us the anomalous dimension of the Konishi multiplet in the strong coupling regime. Finally, we will consider the BMN sector, which is a particular example of semi-classical strings with very large quantum numbers.

\subsection{Supergravity sector}
\label{section:sugra}

As we recalled in Section~\ref{section:closed}, the massless sector of the closed superstring is given in terms of vertex operators of ghost number two and weight zero.
On-shell fluctuations
around the $AdS_5\times S^5$ background are therefore described
by vertex operators of the form
\be
\cU^{(2)}=\l^a\lh^\ah A_{\a\ah}(x,\t,\th)
\ee
satisfying (\ref{cohomd}) where $Q$ and $\hat Q$ are the BRST charges defined in (\ref{brstads}). We will show that the cohomology condition for this vertex operator reproduces the supergravity equations of motion in AdS.

Consider the action of the BRST charge on a ghost-number $(M,N)$ vertex operator
$\Phi=
\l^{\a_1} ...\l^{\a_M} \lh^{\bh_1} ... \lh^{\bh_N}
A_{\a_1 ... \a_M \bh_1 ... \bh_N}(x,\t,\th),$
\bea\label{ansatz}Q \Phi
= \l^{\kappa}
\l^{\a_1} ...\l^{\a_M} \lh^{\bh_1} ... \lh^{\bh_N} \nabla_{\kappa}
A_{\a_1 ... \a_M \bh_1 ... \bh_N} {\rm ~~~ and ~~~}\eea
$$\hat Q \Phi
=\l^{\a_1} ...\l^{\a_M}\lh^{\widehat\kappa} \lh^{\bh_1} ... \lh^{\bh_N}
\nabla_{\widehat\kappa}
A_{\a_1 ... \a_M \bh_1 ... \bh_N} $$
where
$\nabla_\a=
E_\a^M(\p_M + \omega_M)$ and $\nabla_\ah= E_\ah^M(\p_M+\omega_M) $ are
the covariant supersymmetric derivatives in the $AdS_5\times S^5$
background, and $E_B^M$ and $\omega_M$ are the super-vierbein
and spin connection in the $AdS_5\times S^5$
background with $B$ ranging over the tangent-superspace
indices $(\underline{c},\a,\ah)$
and $M$
ranging over the curved superspace indices
$(m,\mu,\widehat\mu)$.
One can express $E_B^M$ and $\omega_M$ in terms of the coset
elements $g(x,\t,\th)$ by defining
$E_B^M =(E_M^B)^{-1}$
and $\omega_M^{[\underline{cd}]} = E_M^{[\underline{cd}]}$ where
$(g^{-1}dg)^B = E_M^B dX^M$ for
$X^M=(x^m,\t^\mu,\th^{\widehat\mu})$.

To justify (\ref{ansatz}), one can check that it is the unique definition
which preserves all $AdS$ isometries and reduces
to $Q\Phi = \l^\a D_\a\Phi$ and
$\hat Q\Phi = \lh^\ah  D_\ah\Phi$ in the flat limit.
Furthermore,
(\ref{ansatz}) is consistent with the nilpotency conditions
$Q^2 = \hat Q^2 = \{Q, \hat Q\}=0$.
The conditions $Q^2\Phi = \hat Q^2\Phi=0$ follow
from the fact that
$\g_{mnpqr}^{\a\b}\{\nabla_\a,\nabla_\b\}=
\g_{mnpqr}^{\ah\bh}
\{\nabla_\ah,\nabla_\bh\}=0$.
Although
$\{\nabla_\a,\nabla_\bh\}$
is non-vanishing, its symmetrical structure allows
$\{Q, \hat Q\}\Phi $ to vanish since
\bea\label{nilp} \{Q, \hat Q\} \Phi
= \l^{\kappa} \lh^{\widehat\tau}
\l^{\a_1} ...\l^{\a_M} \lh^{\bh_1} ... \lh^{\bh_N}
\{\nabla_{\kappa},\nabla_{\widehat\tau}\}
A_{\a_1 ... \a_M \bh_1 ... \bh_N} \eea
\bea\label{nilq}= \l^{\kappa} \lh^{\widehat\tau}
\l^{\a_1} ...\l^{\a_M} \lh^{\bh_1} ... \lh^{\bh_N}
((\g^{cd})_{\kappa}{}^{\sigma} \d_{\sigma\widehat\tau} \nabla_{[cd]}-
(\g^{c'd'})_\kappa{}^\sigma \d_{\sigma\widehat\tau} \nabla_{[c'd']})
A_{\a_1 ... \a_M \bh_1 ... \bh_N}, \nonumber\eea
where $\nabla_{[\underline{cd}]}$ acts as a Lorentz rotation in the
$[\underline{cd}]$ direction on the
$M+N$ spinor indices of
$A_{\a_1 ... \a_M \bh_1 ... \bh_N}$, i.e.
\bea\nabla_{[\underline{cd}]}A_{\a_1 ... \a_M \bh_1 ... \bh_N}=&
\half[(\g_{\underline{cd}})_{\a_1}{}^\g
A_{\g\a_2 ...  \bh_N}
+(\g_{\underline{cd}})_{\a_2}{}^\g
A_{\a_1 \g \a_3 ... \bh_N}
+ \ldots \nonumber\\&\ldots+ (\g_{\underline{cd}})_{\bh_N}{}^\gh
A_{\a_1 ... \bh_{N-1}\gh} ].\label{iea}\eea
But since all indices
of $A_{\a_1 ... \a_M \bh_1 ... \bh_N}$
are contracted with either $\l^\a$ or $\lh^\ah$,
one can use
(\ref{using})
to argue that
all terms in (\ref{nilq}) are proportional to
\bea\label{identitygamma}
\eta^{[\underline{a}\underline{b}][\underline{c}\underline{d}]}
\gamma_{\underline{a}\underline{b}}\gamma_{mnpqr}
\gamma_{\underline{c}\underline{d}}=\gamma^{ab}\gamma_{mnpqr}\gamma_{ab}-
\gamma^{a'b'}\gamma_{mnpqr}\gamma_{a'b'}=0 \ ,
\eea
which identically
vanishes for any choice of five-form directions $mnpqr$. This follows from the Jacobi identity for the $\mathfrak{psu}(2,2|4)$ structure constants as explained in \ref{appendix:gammahat}.
So $\{Q, \hat Q\}\Phi=0 $ as desired.

Using (\ref{ansatz}),
(\ref{cohomd}) implies that the bispinor superfield $A_{\a\bh}(x,\t,\th)$
in the physical vertex operator $\cU^{(2)}=\l^\a\lh^\bh A_{\a\bh}$
must satisfy the equations of motion and gauge invariances
\bea\label{eomtwo}\g_{mnpqr}^{\a\b} \nabla_\a A_{\b\gh} =
\g_{mnpqr}^{\ah\gh} \nabla_\ah A_{\b\gh} =
0,\eea
$$
\d A_{\b\gh} = \nabla_\b \widehat\Omega_\gh +
\nabla_\gh \Omega_\b {\rm ~~~with~~~}
\g_{mnpqr}^{\a\b} \nabla_\a \Omega_\b =
\g_{mnpqr}^{\ah\gh} \nabla_\ah \widehat\Omega_{\gh} = 0.$$
Although one could do a component analysis to check
that (\ref{eomtwo}) correctly describes
the on-shell fluctuations around the $AdS_5\times S^5$ background,
this is guaranteed to work since (\ref{eomtwo})
are the unique equations of motion and gauge invariances
which are invariant under the $AdS$ isometries and which reduce to the
massless Type IIB supergravity equations of (\ref{efms}) in the flat limit.

Sometimes it is useful to consider vertex operators that are also worldsheet primary fields, whose double pole with the stress tensor vanishes. In flat space, this is given by (\ref{gf}) and (\ref{residf}). In the $AdS_5\times S^5$ background, the left and right-moving stress
tensors
associated with the action of (\ref{eqAdSpsaction}) are
(\ref{stresste}).
As was done earlier with $Q$ and $\hat Q$, instead of
directly computing the OPE's of $T$ and $\overline T$, it will be simpler
to deduce them from the requirements that they preserve the $AdS$
isometries and reduce correctly in the flat limit. It turns out that, when acting on the physical
vertex operator $\cU^{(2)}=\l^\a\lh^\bh
A_{\a\bh}(x,\t,\th)$,
the condition of no double poles with $T$ or $\overline T$
implies that
\bea\label{imppl}\nabla_B\nabla^B
A_{\a\bh}(x,\t,\th)=0\eea
where
$\nabla_B\nabla^B = \eta^{BC}\nabla_B\nabla_C $, $B=
(\underline{c},[\underline{cd}],\a,\ah)$ ranges over all tangent space
indices of $\mathfrak{psu}(2,2|4)$,
$\nabla_B = E_B^M (\p_M + \omega_M)$ are the covariant derivatives
in the $AdS_5\times S^5$ background when $B=(c,c',\a,\ah)$,
and
$\nabla_{[\underline{cd}]}$
acts as a Lorentz rotation in the
$\underline{cd}$ direction on all tangent
space indices.
Note that $[\nabla_A,\nabla_B\} = f_{AB}^C\nabla_C $ where
$f_{AB}^C$ are the $PSU(2,2|4)$ structure constants and $[\nabla_B\nabla^B, \nabla_C]=0$ for all $C$.
In the flat limit,
$\nabla_B\nabla^B$ reduces to $\p_n\p^n$ since $\eta^{\underline{cd}}$
is the only surviving component of $\eta^{BC}$.

To find the $AdS_5\times S^5$ analog of
(\ref{gf}), it will be convenient to define $A_{B\gh}$ by
\bea 
\nonumber A_{\underline{c}\gh} = {1\over {16}}\g_{\underline{c}}^{\a\b}
\nabla_\a A_{\b\gh},\\
A_{\ah\gh} = {1\over {5}}\eta^{\b\dh}[
\g^c_{\ah\dh} (\nabla_c A_{\b\gh} -\nabla_\b A_{c\gh})
+\g^{c'}_{\ah\dh} (\nabla_{c'} A_{\b\gh} -\nabla_\b A_{c'\gh})], \nonumber\\
A_{[cd]\gh} = \nabla_{c} A_{d\gh}
-\nabla_{d} A_{c\gh}, \quad
A_{[c'd']\gh} = -\nabla_{c'} A_{d'\gh}
+\nabla_{d'} A_{c'\gh}.\label{defin}
\eea
Under the gauge transformation $\d A_{\a\bh}=\nabla_\a\widehat\Omega_\bh
+\nabla_\bh\Omega_\a$, one can check that $A_{B \bh}$ transforms as
\bea\label{gtr}\d A_{B\bh}= \nabla_B \widehat\Omega_\bh -(-1)^{(B)} \nabla_\bh\Omega_B
+(-1)^{(E)} f_{C \bh}^D \eta^{CE} \eta_{BD} \Omega_E\eea
where $(B)=0$ if $B$ is a bosonic index, $(B)=1$ if $B$ is
a fermionic index,
and
\bea\label{ome}\Omega_{\underline c} =
{1\over {16}}\g_{\underline{c}}^{\a\b}
\nabla_\a \Omega_{\b},
\qquad\Omega_{[\underline{cd}]}=0. \nonumber\\\Omega_{\gh} =
{1\over {5}}\eta^{\b\dh}[
\g^c_{\ah\dh} (\nabla_c \Omega_{\b} -\nabla_\b \Omega_{c})
+\g^{c'}_{\ah\dh} (\nabla_{c'} \Omega_{\b} -\nabla_\b \Omega_{c'})],
\eea
One can similarly define $A_{\a B}$.
Then the unique conditions on $A_{\a\bh}$ which preserve
the $AdS$ isometries and which reduce to (\ref{gf}) in the flat limit are
\bea\label{gfads}
\nabla^B\nabla_B A_{\a \bh} = 0,\quad
\nabla^B A_{B \bh}=
\nabla^B A_{\a B}= 0.\eea
Furthermore, using the gauge transformations of (\ref{gtr}) 
one can check that these conditions are invariant under the residual
gauge transformations
\bea\d A_{\b\gh} = \nabla_\b \widehat\Omega_\gh +
\nabla_\gh \Omega_\b
{\rm ~~~with~~~}\label{residfads}\\
\nabla_B\nabla^B \Omega_\a =
\nabla_B\nabla^B \widehat\Omega_\ah =
\nabla^B \Omega_B =
\nabla^B\widehat\Omega_B = 0,\nonumber\eea
which reduce to (\ref{residf}) in the flat limit.
So the conditions of (\ref{gfads}) and (\ref{residfads}) for the primary vertex
operator
describing on-shell fluctuations of the $AdS_5\times S^5$
background closely resemble the conditions of (\ref{gf}) and (\ref{residf})
for the primary
vertex operator describing the massless Type IIB supergravity fields in
a flat background.

\subsubsection{Radius deformation}
\label{secradius}

The first example of a vertex operator is the operator corresponding to a change in the radius of AdS. The pure spinor action is multiplied by the inverse sigma model coupling, which corresponds to the radius of AdS, in units of $\alpha'$.  As we have shown previously, the radius does not renormalize and is not shifted at one-loop, so it is a free parameter in the model. In the usual parlance of conformal field theories, such operator describes a line of fixed points and it is called an exactly marginal deformation. By the AdS/CFT dictionary, it is dual to the gauge theory 't Hooft coupling. Its insertion in an expectation value probes the response to an infinitesimal change in the radius of AdS. If we consider the action as an integrated vertex operator $\int\,d^2z\cV^{(0)}_{z\bar z}$, where $\cV^{(0)}_{z\bar z}$ is a ghost number zero worldsheet two-form, corresponding to the radius of AdS, by applying the standard descent procedure  we will obtain the unintegrated vertex operator (weight zero and ghost number two) corresponding to the radius of AdS.

The lagrangian (\ref{eqAdSpsaction}) is invariant under the BRST transformations generated by (\ref{brstads}) up to a total derivative. If we denote the worldsheet lagrangian $L_{AdS}\equiv \cV^{(0)}_{z\bar z}$ with
$$
\cV^{(0)}_{z\bar z}=\str\bigl[\half (J_2\bar J_2+J_1\bar J_3+J_3\bar J_1)+{1\over 4}(J_3\bar J_1-J_1\bar J_3)+(w\bar\nabla \lambda+\hat w\nabla\hat\lambda-N\hat N)\bigr] \ ,
$$
then the BRST variation of the first term is
$$
\half \str(\bar J_3\nabla \lambda +J_3\bar\nabla \l+J_1\bar\nabla \hat\lambda+\bar J_1\nabla \hat \l) \ .
$$
By using the Maurer-Cartan equations (\ref{eqMaurerCartan}), the variation of the second term is
$$
\half\str(\bar J_1\nabla \hat\l-J_1\bar\nabla \hat\l+J_3\bar\nabla\l -\bar J_3\nabla \l)
$$
$$
+{1\over 4}\str[\partial(\l\bar J_3-\hat\l \bar J_1)+\bar\partial (\hat\l J_1-\l J_3)] \ ,
$$
while the variation of the last term is
$$
-\str(\bar J_1\nabla\hat\l+J_3\bar\nabla \l) \ .
$$
By summing everything everything up, we conclude that the BRST variation of the action is equal to the total derivative
\bea\label{brstlagra}
Q \cV^{(0)}_{z\bar z}=\partial f^{(1)}_{\bar z}-\bar \partial f^{(1)}_z \ ,
\eea
where
$$
f^{(1)}_z={1\over4}\str(\l J_3-\hat\l J_1)\ ,\qquad
f^{(1)}_{\zb}={1\over4}\str(\l\bar J_3-\hat \l \bar J_1) \ .
$$
The last step in the descent gives the unintegrated vertex operator $\cU^{(2)}$
$$
Q f^{(1)}_z=\partial \cU^{(2)}\ ,\qquad Q f^{(1)}_\zb=\bar\partial \cU^{(2)} \ ,
$$
where
\bea\label{vertexradius}
\cU^{(2)}=-{1\over 4}\str \l\hat\l=-{1\over4}\eta_{\a\ah}\l^\a\hat\l^\ah\ .
\eea
We conclude that $(\eta \l\hat\l)$ belongs to the cohomology and corresponds to the radius deformation, that is the dilaton at zero momentum. The same procedure in the case of the flat space action gives the vertex operator $(\l\gamma^m\theta)(\hat\l\gamma_m\hat \theta)$, which is the unintegrated vertex operator for the trace of the graviton at zero momentum. 

\subsubsection{Beta deformation}

Another interesting physical vertex operator of the $AdS_5\times S^5$ superstring corresponds to the beta deformation at zero momentum. Let us introduce the following operator in the adjoint representation of $PSU(2,2|4)$
$$\Lambda^A=[g^{-1}(\hat\l-\l)g]^A \ ,
$$
and consider the ghost number two and weight zero vertex operator
$$
\cU^{[AB]}=\Lambda^{[A}\Lambda^{B]}\ .
$$
This operator is in the antisymmetric product of two adjoint representations. The part of $\cU^{[AB]}$ corresponding to the adjoint of $PSU(2,2|4)$ is BRST trivial
$$
f_{AB}^C\cU^{[AB]}=\{\Lambda,\Lambda\}^C
= Q[g^{-1}(\hat\l+\l)g]^C\ ,
$$
where but the remaining part is a physical deformation, to which we apply the inverse descent procedure to obtain the corresponding integrated vertex operator, by which we may deform the action.
The operator $\Lambda^A$ is the ghost number one cocycle associated with the Noether currents $j^A$ of the $PSU(2,2|4)$ global symmetry, namely $Q (j^A)= d\Lambda^A$, so that the descent procedure gives
$$
d\cU^{[AB]}=2Q(j^{[A}\Lambda^{B]})\ ,\qquad d(j^{[A}\Lambda^{B]})=
-\half j_{[A}\wedge j_{B]} \ .
$$
For any constant antisymmetric matrix $B_{[AB]}$ (where $[AB]$ is the product of two adjoint representations of $PSU(2,2|4)$ from which we remove the adjoint part) we have the following infinitesimal deformation of the action
\bea\label{integratedv}
\int d^2z\,B_{[AB]}j^A\wedge j^B\ .
\eea
Consider now the embedding of the five-sphere into $\RR^6$ through the embedding coordinates $Y^\mu$ such that $Y^\mu Y^\mu=1$, and take $B_{AB}$ in the direction of $S^5$
\bea\label{betadef}
V_{beta}=
B^{[\mu\nu][\rho\sigma]}\int d^2zY_{[\mu}dY_{\nu]}\wedge Y_{[\rho}dY_{\sigma]}+\ldots\ ,
\eea
where $\ldots$ include the terms with $\theta$'s present in the Noether currents. The vertex operator (\ref{betadef}) has the quantum numbers of the so-called {\it beta deformation} of ${\cal N}=4$ super Yang-Mills.

\subsection{Massive spectrum: Konishi multiplet}
\label{massiveK}

As we discussed above, short non-BPS operators in $\cN=4$ SYM at strong coupling are dual to massive perturbative string states, described in terms of worldsheet vertex operators. The conformal dimension of non-BPS operators, such as operators in the Konishi multiplet, receive quantum corrections, whose evaluation at strong coupling we are now going to perform.

By imposing that the worldsheet vertex operators satisfy the superstring physical state condition at the loop level, we derive an equation for the energy of the corresponding string state, which gives in turn the strong coupling expansion of the anomalous dimension of the dual gauge theory operator. We apply our method to the simplest non-BPS operator, a member of the Konishi multiplet. Our worldsheet computation of the anomalous dimension of a particular member of the multiplet  gives 
\be
\label{konishi}
\gamma=2\sqrt[4]{\mL}-4+{2\over\sqrt[4]{\mL}}+{\cal O}(1/\sqrt{\mL}) \ .
\ee
The classical dimension $\Delta_0$ of different members of the Konishi multiplet may take different integer or half-integer values, but the anomalous dimension (\ref{konishi}) is the same for all of them. In supersymmetric theories, a supermultiplet is obtained by applying the supercharges to operator corresponding to the highest weight state. The commutator of the supercharges with the dilation operator is not renormalized (because of the supersymmetry algebra), hence the anomalous dimension is the same for all elements of the supermultiplet. In light of this fact, we can pick our favourite member of the multiplet to perform the calculation. The leading term $2\sqrt[4]{\mL}$ reproduces the expected leading behaviour for an operator dual to a string state at the first massive level. The last term $2/\sqrt[4]{\mL}$ is the one-loop correction to the anomalous dimension of the Konishi multiplet at strong coupling. 

The method we are going to explain can be used to solve for the whole energy spectrum of massive states of type IIB superstring in $AdS_5\times S^5$ and can be expanded to any loop order at strong coupling. This will give an expansion of the anomalous dimensions of short operators in super Yang-Mills theory in inverse powers of $\sqrt[4]{\mL}$.

The way the massive string spectrum is computed in the pure spinor formalism in flat space proceeds usually as follows. Out of the worldsheet currents $J_i,\bar J_i, N,\hat N, J_\l,\bar J_\lh$, one writes down the most general vertex operator of ghost number two that is a two-form on the worldsheet.  Then one imposes that the vertex operator is in the BRST cohomology and finds some superspace equations of motion and gauge invariances, from which one identifies the spectrum content of the superfields. In principle, one could attempt to replicate this procedure in AdS as well. However, it is not going to be easy, as the superspace formulation of AdS, unlike flat space, has not been worked out in details. Another way would be to use the worldsheet current algebra that we derived in the Section~\ref{section:OPE} and solve for the primary operators, as it is usually done in the context of WZW models. We will use a different and simpler method instead.

\subsubsection{Classical configuration}

In order to study the massive string spectrum, we will expand the sigma model around a classical string configuration, describing a point-like string sitting at the center of AdS. Using the metric of AdS in Lorentzian global coordinates 
\be
ds^2=-\cosh^2\rho\,dt^2+d\rho^2+\sinh^2\rho\, dS_3^2 \ ,
\ee 
our string configuration sits at $\rho\sim0$ and evolves in time as $e^{iEt}$. In the static gauge, this is described by the coset element 
\be
\label{ground}
\tilde g(\sigma,\tau)=\exp[-\tau E {\bf T}/ \sqrt{\mL}] \ ,
\ee
that solves the worldsheet equations of motion (\ref{purespinoreoms}), where ${\bf T}$ is the anti-hermitian $PSU(2,2|4)$ generator corresponding to the AdS time translations and $\tau$ is the worldsheet time. In this Section we are using a different form of the $\mathfrak{psu}(2,2|4)$ algebra, that the reader can find in \ref{appendix:alternative}. The only non-vanishing left-invariant current in this background is 
\be
\tilde J_\tau=\tilde g^{-1}\partial_\tau \tilde g=-E{\bf T}/ \sqrt{\mL}\ .
\ee Hence, such classical configuration has vanishing BRST charge.

The Noether currents for the global $PSU(2,2|4)$ symmetry of the string sigma model are given by
\bea
j=&{\sqrt{\mL}\over2\pi}g\left(J_2+\half J_1+{3\over2}J_3+N\right)g^{-1} \ ,\nonumber\\
\bar j=&{\sqrt{\mL}\over2\pi}g\left(\bar J_2+\half \bar J_3+{3\over2}\bar J_1+\hat N\right)g^{-1} \ ,
\eea
and the Noether charges is 
\be
\label{noetherg}
G_{PSU}=\oint d\sigma\, j_\tau \ , 
\ee
where 
\be
j_\tau={\sqrt{\mL}\over 2\pi}g[J_1+J_2+J_3+N+\bar N]_\tau g^{-1} \ .
\ee
In particular, the $AdS$ energy operator $\textsf{E}$ evaluated on the string configuration (\ref{ground}) gives 
\be
\label{energyop}
\textsf{E}=\oint d\sigma\,\textrm{Str} {\bf T}\tilde j_\tau=E \ ,
\ee
where we used $\textrm{Str}({\bf T}{\bf T})=-1$. Since we consider positive energy configurations, we can take $E$ to be positive in the following.

The classical Virasoro constraint for such configuration reads
\be\label{classvir}
T+\bar T={\sqrt{\mL}\over2}\textrm{Str}\, \tilde J_\tau \tilde J_\tau=-{E^2\over2\sqrt{\lambda}} \ .
\ee
The classical Virasoro constraint (\ref{classvir}) will be modified by quantum effects, which are going to allow for a non-zero solution for $E$. Since for massive string states, such as the one we are considering, the energy scales as $E\sim\sqrt[4]{\mL}$, the classical contribution to (\ref{classvir}) is of order one and may be canceled against quantum effects. Note that the classical configuration (\ref{ground}) is completely analogous to the tachyon vacuum in bosonic string theory, whose profile is in fact $e^{ikX}$. In the bosonic string, the vacuum $|k\rangle$ is a tachyon with momentum $k$ and the classical part of the Virasoro constraint is $k^2=-M^2$, while in our case the vacuum $|E\rangle$ has energy $E$ and classical Virasoro constraint  (\ref{classvir}).

\subsubsection{Quantization}

Let us quantize the pure spinor action (\ref{eqAdSpsaction}) around the classical configuration (\ref{ground}) using the background field method. We parameterize the coset element by $g=\tilde g e^X$, where 
\be
\label{fluctux}
X=t {\bf T}+X^A {\bf P_A}+\Phi {\bf J}+X^I {\bf P_I}+\Theta^a {\bf Q_a}+\Theta^\aad {\bf Q_\aad}+\widehat \Theta^a {\bf \widehat Q_a}+\widehat \Theta^\aad {\bf \widehat Q_\aad} \ ,
\ee
are the quantum fluctuations and $\tilde g$ is given in (\ref{ground}). The new notations in eq. (\ref{fluctux}) are explained in~\ref{appendix:alternative}. As in the previous Section, we chose a coset gauge in which the grading zero part of the fluctuations vanishes. The left invariant currents are given by
\be
J_\tau= e^{-X}(\partial_\tau-E{\bf T}/ \sqrt{\mL}) e^X, \qquad J_\sigma=e^{-X}\partial_\sigma e^X \ .
\ee
By expanding the action (\ref{eqAdSpsaction}) up to quadratic order as in (\ref{sone}), (\ref{stwo}) and (\ref{sthree}), and using the $\mathfrak{psu}(2,2|4)$ structure constants, we can read the spectrum of fluctuations around the background (\ref{ground}). The quadratic part of the action for the fluctuations is 
$$
S={\sqrt{\lambda}\over2\pi}\int \Bigl[-\partial t\bar\partial t+\partial J\bar\partial J+\delta_{IJ}\partial X^I\bar\partial X^J\nonumber+\delta_{AB}\left(\partial X^A\bar\partial X^B+\left({E\over 2\sqrt{\lambda}}\right)^2X^AX^B\right) $$
$$
+\Pi_{ab}\partial\widehat \Theta^a\bar\partial \Theta^b+\Pi_{{\dot a}{\dot b}}\partial\widehat\Theta^{\dot a}\bar\partial \Theta^{\dot b}\nonumber-{E\over4\sqrt{\lambda}}\left[\delta_{ab}(\Theta^a\bar\partial\Theta^b+\widehat\Theta^a\partial\widehat\Theta^b)+\delta_{{\dot a}{\dot b}}(\Theta^{\dot a}\bar\partial\Theta^{\dot b}+\widehat\Theta^{\dot a}\partial\widehat\Theta^{\dot b})\right]$$
$$
+w_a\bar\partial l^a+w_{\dot a}\bar\partial l^{\dot a}+\hat w_a\partial\hat l^a+\hat w_{\dot a}\partial\hat l^{\dot a}\Bigr]
$$
The $AdS_5$ time direction as well as the five sphere directions remain massless, while the remaining four bosonic directions of $AdS_5$ acquire a mass squared  $m^2_X=(E/\sqrt{\mL})^2$. The fermionic spectrum consists of sixteen massless fermions and sixteen massive fermions with mass squared $m^2_\Theta=(E/2\sqrt{\mL})^2$. The ghosts remain massless. There is no relation between the bosonic and fermionic spectrum, reflecting the fact that this background is not BPS.

We can canonically quantize the theory imposing the usual equal time commutation relations for the coordinates and their conjugate momenta. The equations of motion of some of the fluctuations are
\bea
\partial\bar\partial X^A+m_X^2X^A=&0 \ ,\nonumber\\
\partial\bar\partial\Theta+m_\Theta\Pi\partial\widehat\Theta=&0 \ ,\nonumber\\
\partial\bar\partial\widehat\Theta-m_\Theta\Pi\bar\partial\Theta=&0 \ .
\eea
Their mode expansion is 
\bea
\Theta=&\Theta_0\sin m_\Theta\tau-\Pi\widehat\Theta_0\cos m_\Theta\tau\nonumber\\&+
\sum_{n\neq0}c_{\Theta,n}\left({im_\Theta\over\omega_{\Theta,n}+k_n}\varphi^1_{\Theta,n}\Theta_n-\varphi^2_{\Theta,n}\Pi\widehat\Theta_n\right)+ \sum_n \vartheta_n e^{-in(\tau-\sigma)}\ ,\nonumber\\
\widehat \Theta=&\widehat\Theta_0\sin m_\Theta\tau+\Pi\Theta_0\cos m_\Theta\tau\nonumber\\&+
\sum_{n\neq0}c_{\Theta,n}\left({im_\Theta\over\omega_{\Theta,n}+k_n}\varphi^2_{\Theta,n}\widehat\Theta_n+\varphi^1_{\Theta,n}\Pi\Theta_n\right)+ \sum_n \widehat\vartheta_n e^{-in(\tau+\sigma)}\ ,\nonumber\\
X^A=&x_0^A\cos m_X\tau+p_0^Am_X^{-1}\sin m_X\tau+i\sum_{n\neq0}{1\over\omega_{X,n}}\left(\varphi_{X,n}^1\alpha_n^{1A}+\varphi^2_{X,n}\alpha_n^{2A}\right) \ ,\nonumber
\\
\label{modeexpansion}\eea
where, for $s=(\Theta,X)$, we defined
\bea
\varphi^1_{s,n}=\exp[-i(\omega_{s,n}\tau-k_n\s)]\ , &\qquad&
\varphi^2_{s,n}=\exp[-i(\omega_{s,n}\tau+k_n\s)]\ ,\nonumber\\
\omega_{s,n}=\sqrt{m_s^2+k^2_n}\,\quad n>0\ ; &\qquad&
\omega_{s,n}=-\sqrt{m_s^2+k^2_n}\,\quad n<0\ ,\nonumber\\
k_n=2\pi n\ , &\qquad& c_{s,n}=\left(1+(\omega_{s,n}-k_n)^2/m_s^2\right)^{-1/2} \ ,\nonumber\\
\eea
We can now compute the conjugate momenta
\be\label{momentacon}
P_\Theta={\delta S\over\d\p_\tau \Theta}\ ,\quad P_X={\delta S\over\d\p_\tau X} \ ,
\ee
and impose equal time commutation relations between coordinates and momenta
\be
[P_X(\s),X(\s')]=-i\delta(\s-\s')\ ,\qquad
\{P_\Theta(\s),\Theta(\s')\}=-i\delta(\s-\s')\ ,
\label{commutatorspx}
\ee
by which we derive the commutation relations of the modes. In particular, the zero modes of the fermions have commutation relations
\bea
\{\Theta_0,\Theta_0\}=&-{\Id\over m_\Theta}&=\{\widehat\Theta_0,\widehat\Theta_0\} \ ,\nonumber\\
\{\Theta_0,\widehat\Theta_0\}=&0\ .& \label{commutheta}
\eea
Our vacuum state $|E\rangle$ is a scalar and is annihilated 
by all positive modes, including the zero modes of $w,\bar w$. This last requirement ensures that the Lorentz generators for the ghosts $N$ and $\bar N$ annihilate the vacuum.
We can choose sixteen fermionic zero modes as creation operators. Since we are using global AdS coordinates, these are linear combinations of supercharges and superconformal transformations. 
In the rest of the paper, we will evaluate the leading quantum contributions to the physical state condition $T+\bar T$, applied to a particular vertex operator to be introduced below. 

The first quantum correction to (\ref{classvir}) comes from the central charge. Even if for $E=0$, namely in empty AdS, the central charge vanishes (\ref{czero}), there is a normal ordering contribution coming from the quadratic part of the stress tensor when $E\neq0$. This is given as usual by the sum of the energies of the oscillator modes
\bea\label{zeropoint}
{2E\over \sqrt{\mL}}+\half\sum_{n=1}^{\infty}\Bigl(6\sqrt{n^2}+4\sqrt{n^2+(E/\sqrt{\mL})^2}
\\-16\sqrt{n^2}-16\sqrt{n^2+(E/2\sqrt{\mL})^2}+22\sqrt{n^2}\Bigr) \ .\nonumber
\eea
The first term is the contribution from the zero modes of the bosons; due to the commutation relations (\ref{commutheta}), there is no contribution from the fermionic zero modes, just as in the pp-wave Hamiltonian. Inside the sum, the first two terms come from the bosonic oscillators, the second two terms from the fermionic ones and the last term from the ghosts. We have not computed the precise value of $E$ yet, but we want it to correspond to a stringy state, whose energy scales as $E\sim\sqrt[4]{\mL}$, which gives $E/\sqrt{\mL}\sim 1/\sqrt[4]{\mL}\ll 1$. In this limit we obtain from (\ref{zeropoint}) the total contribution $2{E\over\sqrt{\mL}}-{3\over 16}\zeta(3)({E\over\sqrt{\mL}})^4$. We can drop the second term as it does not contribute to the energy 
at the order we are considering, so we are left with the contribution 
\be
\label{groundzero}
2{E/ \sqrt{\mL}} \ .
\ee 
Note that at order $({E\over\sqrt{\mL}})^2$ the four massive bosonic modes cancel with the massive sixteen fermionic modes. The contribution $2{E\over \sqrt{\mL}}$, which would vanish in a BPS background, will affect the one-loop correction to the energy of the string, contributing to the last term in (\ref{konishi}).

\subsubsection{Massive vertex operator}

Let us consider now a specific worldsheet vertex operator. All of the members of the Konishi multiplet have the same anomalous dimension and they are in one to one correspondence with the string states at the first massive level. Thus we will choose a particularly simple state in the first massive string level, that will simplify the computation. Physical states are given by unintegrated vertex operators of ghost number two.\footnote{Since we are using both the BRST and the Virasoro conditions on physical states, let us comment on their respective relevance. As we saw when discussing the massless vertex operators in Section~\ref{section:sugra}, the BRST condition applied to the most general massless vertex operator gives the covariant equations of motion and gauge invariances in superspace, that relate the members of the supermultiplet to each other. This allows to recognize the onshell degrees of freedom contained in the supermultiplet, e.g. the IIB supergravity multiplet in (\ref{eomtwo}). On the other hand, the Virasoro condition gives a polynomial expression in the $PSU(2,2|4)$ quantum numbers of the state. When acting on the vertex operator for the half-BPS supermultiplet in (\ref{imppl}), this polynomial expression turns out to be the quadratic Casimir of the $PSU(2,2|4)$ super-isometry group (realized as the super-Laplacian), which vanishes for such multiplet. We also showed that the Virasoro condition partially fixes some of the gauge redundancy in the superspace description of the supermultiplets in (\ref{gfads}), choosing the gauge in which the vertex operator is a worldsheet primary field. In the present discussion, we are interested in computing the eigenvalue of the energy operator for the Konishi multiplet, which is one of the generators of $PSU(2,2|4)$, appearing in the Virasoro condition. This condition is a polynomial expression in such eigenvalue, which we must set to zero and solve for the energy.} The simplest one is $\textrm{Str}\,\l\lh$ and it corresponds to the radius modulus at zero momentum (\ref{vertexradius}). We will denote the corresponding  state as 
\be
|\l\lh\rangle\equiv\textrm{Str}(\l\lh)|E\rangle \ .
\ee

We choose the simple state
\be\label{vertex}
|V\rangle=x_{-1}^+\bar x_{-1}^+|\l\lh\rangle \ ,
\ee
where $x_{-1}^+$ and $\bar x_{-1}^+$ are the first left- and right-moving oscillators coming from the fluctuations of the $AdS$ ``space-cone'' coordinate $x^+=X^1+iX^2$. We can interpret this state as being created by non-zero modes of the global symmetry right invariant currents. We should emphasize that this is {\em not} a global $PSU(2,2|4)$ transformation. We identify the vertex operator (\ref{vertex}) to be dual to a particular member of the  Konishi multiplet with classical dimension $\Delta_0=6$, Lorentz spin two and singlet of $SU(4)$. The operator $|\l\lh\rangle$, corresponding to the radius changing operator, is dual to the Yang-Mills lagrangian and it has $\Delta_0=4$, so it is natural to expect that (\ref{vertex}) has two unites more of classical dimension. This state is a two-magnon impurity of mass $E/\sqrt{\mL}$, whose contribution to the worldsheet energy is 
\be
\label{magnon}
2\sqrt{1+({E/\sqrt{\mL}})^2} \ .
\ee

\subsubsection{Quartic corrections}

Other possible contributions to the physical state condition at this order may come from the terms in the stress tensor $T+\bar T$, expanded to quartic order in fluctuations around the classical background (\ref{ground}) and acting on the specific vertex operator (\ref{vertex}). Let us analyze the possible terms.

The factor of $2$ in front of the square root in (\ref{magnon}) may get corrected by quartic terms in $T+\bar T$ of the form $(\partial X)^2 X^2$, due to normal ordering. However, there is no normal ordering due to this term. This comes from the fact that any correction to this term has to be proportional to the one-loop beta function, which vanishes. There are other corrections that are not protected by the beta function argument, but they are of the form $(E/\sqrt{\mL})\partial X X^3$ and $(E/\sqrt{\mL})^2 X^4$. In any case, they give higher order contributions to the energy and we can safely neglect them. 

The last possible contribution comes from the fact that the operator (\ref{vertex}) might mix with other operators due to quartic terms in $T+\bar T$. In order to study the mixing, we have to compute the momenta (\ref{momentacon}) conjugate to the fields up to quartic terms in the action, then plug these back into the stress tensor. The momenta are of the form
$$
P_i={\delta S\over \delta\partial_\tau X_i}=\partial_\tau X_i+\ldots \ ,
$$
where $\ldots$ are higher order terms in the fluctuation, which must be included up to the third order.
In this way we eliminate all the time derivatives in the stress tensor. For the particular vertex operator (\ref{vertex}) we may only consider the terms with four bosonic or two bosonic and two fermionic fields. The relevant terms in the hamiltonian are the following:
$$
\str \Bigl(
-{5\over16}[\p_\s X_1,X_2][P_2,X_3]+{1\over12}[\p_\s X_1,X_3][P_2,X_2]+{1\over8}[\p_\s X_2,X_1][P_2,X_3]
$$
$$
-{1\over 6}[\p_\s X_2,X_2][\p_\s X_2,X_2]
-{1\over3}[\p_\s X_2,X_2][X_3,X_1]-{1\over12}[\p_\s X_2,X_2][P_1,X_3]
$$
$$
+{1\over12}[\p_\s X_2,X_2][P_3,X_1]-{5\over24}[\p_\s X_2,X_3][P_1,X_2]
-{1\over12}[\p_\s X_3,X_1][P_2,X_2]
$$
$$
-{7\over24}[\p_\s X_3,X_2][P_1,X_2]-{5\over24}[\p_\s X_3,X_2][P_2,X_1]
+{1\over12}[P_1,X_2][P_3,X_2]
$$
$$
+{1\over3}[P_1,X_3][P_2,X_2]+{1\over12}[P_2,X_1][P_2,X_3]+{1\over12}[P_2,X_1][P_3,X_2]
$$
$$
{1\over6}[P_2,X_2][P_2,X_2]+{1\over3}[P_2,X_2][P_3,X_1]\Bigr) \ .
$$
Terms quartic in bosons can both introduce mixing and also correct the energy of our state. Terms with 
two fermions and two bosons will only give mixing. For the particular choice of the ``space-cone'' polarization in (\ref{vertex}), it is easy to see that there will be no mixing with other bosonic states, nor mixing with states created by two fermions, since the stress tensor will only have commutators and products of gamma matrices, which vanish for this particular choice of polarization. In fact, this simplification is what led us to choose that particular member of the Konishi supermultiplet in the first place. However, there is a non-vanishing correction to the energy, proportional to the state itself, coming from the term ${1\over6}(\textrm{Str}\,[P_2,X_2]^2-\textrm{Str}\,[\partial_\sigma X_2, X_2]^2   )$ in the stress tensor. At this point, we can
expand the Hamiltonian into modes as in (\ref{modeexpansion}) and use the canonical commutation relations in (\ref{commutatorspx}) when applying the Hamiltonian to the state (\ref{vertex}).  We obtain the following correction to the physical state condition:
\be
\label{quartic}
-2/\sqrt{\lambda_t} \ .
\ee

\subsubsection{Conformal dimension}

Summing up the contributions (\ref{classvir}), (\ref{groundzero}), (\ref{magnon}), and (\ref{quartic}) to the Virasoro constraint, we find that the physical state condition
\be
(T+\bar T) |V\rangle =0 \ ,
\ee
gives
\be
\label{marginality}
-{E(E-4)\over2\sqrt{\lambda_t}}+2\sqrt{1+({E\over\sqrt{\lambda_t}})^2}-{2\over\sqrt{\lambda_t}} =0 \ .
\ee
The positive energy solution of this equation gives the energy of our string state (\ref{vertex}). According to the AdS/CFT dictionary, the energy operator on the string side of the correspondence is mapped to the dilatation operator on the field theory side, whose eigenvalues are the conformal dimensions of operators. Above, we identified the field theory dual to the string state (\ref{vertex}) as a member of the Konishi multiplet with classical dimension $\Delta_0=6$, Lorentz spin two and singlet of $SU(4)$. Its conformal dimension at strong coupling is therefore
\be 
\Delta =E= 2\sqrt[4]{\lambda_t}+2+{2\over\sqrt[4]{\lambda_t}}+{\cal O}(\lambda_t^{-1/2}) .
\ee
Hence we derived (\ref{konishi}) with $\Delta_0=6$.

\subsection{String correlation functions}
\label{section:amplitudes}

After solving for the string spectrum, which corresponds to evaluating the two-point functions, the next step is to compute string $n$-point correlation functions on the worldsheet. These give the $1/N_c$ corrections to the planar limit of $\cN=4$ SYM. 
Correlation functions on the CFT side are dual to string scattering amplitudes on the AdS side, where each vertex operator insertion is dual to a specific gauge theory operator insertion, schematically\footnote{We consider here a sphere amplitude, so on the string side we only integrate over $n-3$ positions on the worldsheet and fix the remaining three using the $SL(2)$ symmetry.}
$$
\langle \prod_{i=1}^n\cO(x_i)\rangle_{CFT}=\langle \prod_{i=1}^3\cU^{(2)}_i(z_i;x_i)\prod_{i=4}^n\int d^2z_i\cV^{(0)}_i(z_i;x_i) \rangle _{AdS} \ .
$$
The right hand side is a tree level string amplitude. The vertex operators depend both on the insertion point $z_i$ on the worldsheet and on a parameter $x_i$, that represents the operator insertion on the AdS boundary. Here we will not evaluate any correlation function explicitly, but we will show how to integrate the zero modes of the worldsheet variables and define the higher genus amplitude prescription, which involves the construction of the $b$ antighost.

The pure spinor sigma model in AdS, albeit interacting, is in a sense simpler than the flat space one. In the $G/H$ supercoset realization with the $\ZZ_4$ grading, the ${\cal H}_1$ and the ${\cal H}_3$ fermionic subalgebras are complex conjugate of each other. As a consequence, the left and right moving pure spinors are complex conjugates
$$
(\l^\a)^\dagger=\eta_{\a\ah}\hat \l^\ah\ .
$$
This crucial fact implies that, unlike in flat space, we do not need to add any non-minimal pure spinor variables in the AdS sigma model. The $b$ antighost and the scattering amplitudes are easily defined just using the minimal variables. Moreover, the Riemann curvature coupling to the pure spinor Lorentz generators provides a natural gaussian regulator for the zero mode integral on the pure spinor cone. These facts imply that the ``minimal'' pure spinor formalism we have been using so far is suited for the amplitude computation.

The tree level prescription for string scattering amplitudes 
differs from the flat space one by the rule for the integration of the zero modes. The closed string $n$-point function
\be
{\cal A}_n(x_1,\ldots,x_n) = \langle \prod_{i=1}^3\cU^{(2)}_i(z_i;x_i)\prod_{i=4}^n\int d^2z_i\cV^{(0)}_i(z_i;x_i) \rangle \ ,
\ee
where the $\cU^{(2)}$'s are ghost number two unintegrated vertex operators and the $\cV^{(0)}$'s are ghost number zero integrated ones, is defined in flat space using the zero mode integration rule (\ref{flatzero}), that is
\bea
&\langle f(X,\t,\th,\l,\lh)\rangle_{flat}=\int d^{10}x\int d^5\t_{\a_1\ldots\a_5}\int d^5\th_{\ah_1\ldots\ah_5}\nonumber\\
&(\g^m\partial_\l)^{\a_1}(\g^n\partial_\l)^{\a_2}(\g^p\partial_\l)^{\a_3}(\g_{mnp})^{\a_4\a_5}
(\g^q\partial_\lh)^{\ah_1}(\g^r\partial_\lh)^{\ah_2}(\g^s\partial_\lh)^{\ah_3}(\g_{qrs})^{\ah_4\ah_5}\nonumber\\
&f(X,\t,\th,\l,\lh)|_{\t=\th=0} \ .
\eea
Note that the function $f(X,\t,\th,\l,\lh)$ in the previous equation was called $\l^\a\l^\b\l^\g$ $f_{\a\b\g}(X,\t,\th)$ in (\ref{flatzero}). 
In the $AdS_5\times S^5$ background, on the other hand, the zero mode saturation rule 
\bea
\langle f(X,\t,\th,\l,\lh)\rangle_{AdS}=&\\
=\int d^{10}x\int d^{16}\t\int d^{16}\th\,&\textrm{sdet}E^A_M\int d^{11}\l d^{11}\lh\, f(X,\t,\th,\l,\lh)|_{\t=\th=0} \ ,\
\nonumber\eea
where $E^A_M$ are the $AdS_5\times S^5$ super-vielbein, has a clear geometrical meaning as the Haar measure ${\cal D}g$ on the $G/H$ coset 
\be
{\cal D}g=\int d^{10}x\int d^{16}\t\int d^{16}\th\,\textrm{sdet}E^A_M \ .
\ee

\subsubsection{Amplitude prescription}

Since $(b,c)$ and $(\hat b,\hat c)$
reparameterization ghosts are not fundamental worldsheet
variables in the
pure spinor formalism, $g$--loop scattering amplitudes ${\cal A}_g$ are defined as
in topological string theory where the left-moving
$b$ antighost and right-moving $\hat b$ antighost are composite fields
constructed to satisfy 
$$
\{Q,b\}=T, \quad \{\hat Q,\hat b\}=\hat T \ ,
$$ 
where $Q$ and $\hat Q$ are the left and right-moving BRST operators and
$T$ and $\hat T$ are the left and right-moving stress tensors. As in topological
string theory, the integration measure
is then defined by contracting $(3g-3)$ composite $b$ antighosts with the
Beltrami differentials $\mu$ corresponding to the $(3g-3)$ 
Teichmuller moduli $\tau$ of
the genus $g$ Riemann surface
\bea\label{ampli}
{\cal A}_n=\int d^{3g-3}\tau \int d^{3g-3}\bar\tau\langle 
(\int \mu b)^{3g-3}(\int \bar\mu \hat b)^{3g-3}\prod_{i=1}^n 
\int d^2 z_i \cV^{(0)}_i(z_i) \rangle \ .
\eea
Note that in flat background, the integration over the $g$ zero modes of the worldsheet one-forms $w,\hat w$ requires the introduction of a special BRST exact regulator, depending on the ``non-minimal'' pure spinor variables. In the $AdS_5\times S^5$ case, however, the coupling of the Riemann curvature to the ghost Lorentz currents $\exp(-\str N\hat N)$, which is already present in the action, provides naturally such regulator. 

In a flat background, the construction of the $b$ antighost 
satisfying $\{Q,b\}=T$
is complicated and requires the introduction of non-minimal worldsheet
variables. However, in curved backgrounds where the R-R superfield (\ref{rrsuperfield}) is invertible, we do not need to introduce non-minimal variables. 

Instead of introducing new non-minimal variables $({\overline\l}_\a,
\hat{\overline\l}_\ah)$ to play the role of the complex conjugates of 
$(\l^\a,\lh^\ah)$, one can simply define 
\bea\label{relnmg}\overline\l_\a \equiv \eta_{\a\ah}
\lh^\ah, \quad \hat{\overline\l}_\ah \equiv \eta_{\a\ah} \l^\a.\eea
So after multiplying by the inverse RR superfield $P^{-1}=\eta$, the original 
left and right-moving pure spinor
ghosts can be interpreted as complex conjugates of each other. In a flat
background, this interpretation
is not possible since
$\l^\a \overline\l_\a = \eta_{\a\ah}\l^\a\lh^\ah$ is BRST-trivial, 
so it cannot be interpreted
as a positive-definite quantity. 
As we showed in the Section~\ref{secradius}, the vertex operator
$$
V=\eta_{\a\ah}\l^\a\hat\l^\ah \equiv (\l\hat\l)\ ,
$$
which is in the cohomology, is obtained by the descent relation from the action. So it is consistent to interpret
$\eta_{\a\ah}\l^\a\lh^\ah$ as a non-vanishing quantity
since it cannot be gauged away. After interpreting the complex conjugate
of the pure spinor variables as in (\ref{relnmg}),
the construction of the $b$ antighost is
straightforward.

However, unlike the left-moving $b$ antighost in a flat background, the 
$b$ antighost in a curved background is not holomorphic, i.e.
it does not satisfy $\bar\partial b=0$. Instead it satisfies 
\bea\label{nothol}\bar\partial b = [\hat Q,{\cal O}]\eea
where ${\cal O}$ is defined
by taking the antiholomorphic contour integral 
of $\hat b$ around $b$.  One similarly finds that
the $\hat b$ antighost 
is not antiholomorphic  
and instead satisfies
\bea\label{notantihol}\partial \hat b = [ Q,\hat{\cal O}]\eea
where $\hat {\cal O}$ is defined by taking the holomorphic contour
integral of $b$ around $\hat b$.

To prove (\ref{nothol}), one uses the properties
\bea\label{proper}\{Q,b\}=T,\quad \{\hat Q, b\}=0,\quad \{\hat Q,\hat b\}=\hat T,
\quad\{Q,\hat b\}=0\eea
to show that 
\bea\label{showt}\bar\partial b = [\hat T_{-1}, b] = [\{\hat Q, \hat b_{-1}\}, b] 
= [\hat Q, {\cal O}]\eea
where $[\hat T_{-1}, X]$ and 
$\{\hat b_{-1}, X\}$ denote the antiholomorphic
contour integral of $\hat T$ and $\hat b$ 
around $X$, and ${\cal O} \equiv \{\hat b_{-1}, b\}$. 
One can similarly use (\ref{proper}) to prove (\ref{notantihol}) where
$\hat{\cal O} \equiv \{ b_{-1},\hat b\}$ and $\{b_{-1},X\}$ denotes
the holomorphic contour integral of $b$ around $X$.

Although this non-holomorphic structure of the $b$ and $\hat b$
antighosts is unusual, (\ref{nothol})
and (\ref{notantihol})
should be enough
for consistency of the amplitude prescription of (\ref{ampli}).
In order that $\int \mu b$ in (\ref{ampli})
is invariant under the shift of the Beltrami differential $\mu\to\mu+\bar\partial\nu$ for any $\nu$, 
one usually requires that $\bar\partial b=0$. However, if one can ignore
surface terms coming from the boundary of Teichmuller moduli space,
it is sufficient to require the milder condition 
\bea\label{milder} \bar\partial b=[\hat Q,{\cal O}] \ .  \eea This can be shown
by pulling $\hat Q$ off of 
${\cal O}$ and using $[\hat Q,V]=\{\hat Q,b\}=0$ and $\{\hat Q,\hat b\}=\hat
T$ to obtain terms
which are total derivatives in the Teichmuller moduli. If one can ignore
surface terms from the boundary of moduli space, these 
total derivatives do not contribute.


\subsubsection{$b$ antighost}
\label{section:antighost}

It will be convenient to redefine the hatted worldsheet quantities by introducing a factor of the constant Ramond-Ramond superfield $\eta^{\a\ah}=(\g_{01234})^{\a\ah}$ and its inverse $\eta_{\a\ah}
=(\g_{01234})_{\a\ah}$ 
\bea\label{redefinition}
\hat\lambda_\a\equiv \eta_{\a\hat\a}\hat\l^{\hat\a} \ , \quad \hat w^\a\equiv \eta^{\a\ah}\hat w_\ah  \ ,\quad
(J_3)_\a\equiv \eta_{\a\hat\a}J_3^{\hat\a}  .
\eea

In terms of the $PSU(2,2|4)$ super Lie algebra, the grading one and the grading three subspaces are related by hermitian conjugation which implies
$$
(\l^\a)^\dagger=\hat\l_{\a} 
 \ .
 $$

After the redefinition (\ref{redefinition}), the stress tensor of the worldsheet theory reads 
\bea\label{stress}
T
=&-\half J_2^a J_2^b\eta_{ab}+J_1^\a J_{3\a}-w_\a\nabla\l^\a \ ,
\eea
and it is easy to check that it satisfies $\{Q,T\}=\{\hat Q,T\}=0$. 

Before we consider the $b$ antighost, let us introduce the projection operator
\bea\label{projector}
K^\a_\b={1\over 2(\l\hat\l)}(\gamma^a\hat\l)^\a(\l\gamma_a)_\beta ={1\over 2(\l\hat\l)}(\gamma^a\l)_\beta(\hat\l\gamma_a)^\a\ ,
\eea
with the following properties
\bea\label{kappapro}
(1-K)\gamma^a\lambda=0 \ ,\qquad K\gamma^a\gamma^b\l=0 \ ,\qquad K\nabla\l=0 \ ,\eea
and similar for $\hat\l$. Its traces over the spinor indices are $\tr K=5$ and $\tr(1-K)=11$. By means of the projector $K_\a^\b$ we can introduce the new gauge invariant quantity
\bea\label{newgauge}
w_\a(1-K)^\a_\b \ ,
\eea
The expression for the $AdS_5\times S^5$
antighost can be written in terms of the
$(1-K)^\a_\b$ projector as 
\bea\label{bghost}
b=&{(\hat\lambda\gamma_a J_3)J_2^a\over 2(\lambda\hat\lambda)}-w_\a(1-K)^\a_\b J_1^\b\ .
\eea
Using the BRST transformations in (\ref{brstrans}), it is easy to show 
that $\{Q,b\}=T$. We will be ignoring possible
normal-ordering corrections to the $b$ antighost and will only be considering the terms in $b$ which are lowest order in $\a'$.

The other crucial property of the $b$ antighost is
$$
\{\hat Q, b\}= 0 .
$$
Let us prove it. The variation of the first term in (\ref{bghost}) is
$$
{1\over 2(\l\hat\l)}\left((\hat\l\gamma_a \bar\nabla\hat\l)J_2^a-(\hat\l\gamma_a J_3)(\hat\l\gamma^a J_3)\right) \ ,
$$
which vanishes because of the pure spinor constraint and the properties of ten dimensional gamma matrices. The variation of the second term in (\ref{bghost}) is
$$
w_\a(1-K)^\a_\b(\gamma_a\hat\l)^\b J_2^a \ ,
$$
which vanishes due to the properties of the projector (\ref{kappapro}).

An analogous construction carries over to the right-moving sector.
The right-moving stress tensor and antighost are
\bea\label{tleft}
\hat T
=&-\half\bar J_2^a\bar J_2^b\eta_{ab}+\bar J_1^\a\bar J_{3\a}-\wh\bar\nabla\lh \ ,
\eea
\bea\label{bleft}
\hat b
=&-{1\over2\l\lh}(\l\gamma_a\bar J_1)\bar J_2^a-\wh^\a(1-K)_\a^\b (\bar J_3)_\b \ .
\eea
One can check that $\{\hat Q,\hat b\}=\hat T$ and
$\{Q,\hat b\}=0$.

We can apply the argument given in (\ref{nothol}) to show that the $b$ antighost is conserved up to BRST exact terms. The details are collected in \ref{appendix:bghost}.

{\bf Guide to the literature}

The AdS/CFT correspondence was introduced in the three seminal papers \cite{Maldacena:1997re,Gubser:1998bc,Witten:1998qj}. The amount of literature on this topic is huge and we will only mention three reviews. A comprehensive review of the early days of AdS/CFT correspondence is \cite{Aharony:1999ti}. A useful beginner review with some basic computations carried out in details is \cite{Petersen:1999zh}. An advanced review that focuses on the worldsheet aspects of the correspondence and exploits its integrability (discussed later in our Section \ref{section:integrability}) is \cite{Beisert:2010jr}.

The massless vertex operators of the pure spinor sigma model in AdS were analyzed in \cite{Berkovits:2000yr}. The radius changing operator is discussed in \cite{Berkovits:2007rj,Berkovits:2008qc,Berkovits:2008ga}, where a derivation of the pure spinor formalism in AdS from the gauge fixing of a classical $G/G$ coset model is also presented. The vertex operator for the beta-deformation was derived in \cite{Mikhailov:2009rx} and used to deform the $AdS_5\times S^5$ action in \cite{Bedoya:2010qz}. 

The massive string vertex operator dual to the Konishi multiplet has been studied in \cite{Vallilo:2011fj}, using the techniques developed in \cite{Callan:2003xr,Callan:2004uv} and \cite{Metsaev:2002re}. The same results have been obtained by semi-classical methods in \cite{Gromov:2011de} and \cite{Roiban:2011fe} and reproduce the earlier conjecure for the anomalous dimension of the Konishi multiplet proposed in \cite{Gromov:2009zb}, using integrability and the Y-system. The first two terms in (\ref{konishi}) were derived much earlier in \cite{Brower:2006ea}, in the context of the AdS/CFT approach to the Pomeron exchange in QCD.

There are a number of papers that address the computation of the string theory spectrum in $AdS_3$ backgrounds, which provide a somewhat different perspective from the one we took in this chapter. The construction of vertex operators in $AdS_3$ with NSNS flux was considered in \cite{Kutasov:1999xu}. More recently, in the series of papers \cite{Ashok:2009xx,Ashok:2009jw,Benichou:2010rk,Troost:2010zz,Troost:2011fd}, the $AdS_3$ background with RR flux has been considered from the point of view of the worldsheet current algebra and its massless sector and other interesting observables have been computed.

We have not mentioned the string spectrum in the BMN sector, namely small fluctuations around a point-like classical string moving fast around the equator of the five-sphere. The interested reader can find the GS sigma model on such background in \cite{Metsaev:2002re} and the pure spinor sigma model in \cite{Berkovits:2002zv,Berkovits:2008ga}. An alternative covariant  formulation with $\cN=2$ worldsheet supersymmetry can be found in \cite{Berkovits:2002vn}. The AdS/CFT analysis of the BMN sector is carried out in \cite{Berenstein:2002jq}.

The computation of string theory correlators in $AdS_5\times S^5$ is a relatively new topic. The three point functions of two semi-classical string states and a supergravity mode has been computed in the GS formalism in \cite{Janik:2010gc,Zarembo:2010rr,Costa:2010rz,Roiban:2010fe}. In the pure spinor formalism, the prescription for multiloop scattering amplitudes in AdS has been proposed in \cite{Berkovits:2008qc,Berkovits:2008ga} and the $b$ antighost has been constructed in \cite{Berkovits:2010zz}.

\clearpage


\section{Integrability}
\label{section:integrability}

The pure spinor superstring on $AdS_5\times S^5$ background is an integrable two-dimensional conformal field theory. It inherits its classical integrability from the GS sigma model. 

The pure spinor sigma model is an integrable model of a new kind. It displays the usual Yangian symmetry, which is encoded in the Lax representation of the string equations of motion. On top of this, there is a new structure, which is not present in other integrable sigma models: BRST symmetry. The interplay of Yangian and BRST symmetry leads to the quantum integrability of the sigma model, which has been proven both explicitly at one-loop and also algebraically at all orders in the sigma model perturbation theory. In this chapter we will explain all of these features.

\subsection{Classical integrability}

We will first derive the Lax equation. Consider a sigma model on a supercoset $G/H$, that has a $\ZZ_4$ automorphism, where the gauge symmetry acts on the right $g\to gh$. We can build two kind of currents: the {\it moving frame} currents $J=g^{-1}dg$ are invariant under the global symmetry acting on the left, but they are covariant with respect to the gauge symmetry acting on the right, namely $J_i\to h^{-1} J_ih$, for $i=1,2,3$ and $J_0\to h^{-1}J_0h +h^{-1}dh$; the {\it fixed frame} currents $j=dg g^{-1}$, which are invariant under the gauge symmetry, but transform in the adjoint representation of the global symmetry group $G$.

Suppose we can find a gauge invariant current $a(\mu)$, depending on a complex parameter $\mu$ that we will call the {\it spectral parameter}, such that it obeys the flatness condition
\bea\label{flata}
da+a\wedge a=0 \ .
\eea
Given a flat connection, the equation $(d+a)U=0$
is integrable and it is solved by the Wilson line of the flat connection around the string worldsheet
\bea\label{openu}
U=P\exp\left(-\int_C a\right)\ ,
\eea
where $C$ is contour on the worldsheet and $P$ denotes the path ordering of the Lie algebra. Because the connection is flat, the Wilson line is invariant under the continuous deformations of $C$ with fixed endpoints. Using (\ref{openu}) one can construct an infinite set of non-local conserved charges, which implies that the theory is classically integrable.

It is more convenient to work with upper case currents, so let us see how the conserved charges arise in the moving frame language.
Let us define a moving frame flat connection $A(\mu)$ as
$$
d+A(\mu)\equiv g^{-1}(d+a(\mu))g \ ,
$$
that yields
\bea\label{movingA}
A(\mu)=J+g^{-1} a(\mu) g \ ,
\eea
where $J=g^{-1}dg=J_0+J_1+J_2+J_3$ is our usual left-invariant current. Because of the flatness condition on $a(\mu)$ and the Maurer-Cartan equation on $J$, the left-invariant current $A(\mu)$, which is called a {\it Lax connection}, is flat as well\footnote{One can parameterize the left invariant currents as $\tilde A=g^{-1}ag$ instead, and derive from (\ref{flata}) the  equation $d\tilde A+J\wedge \tilde A+\tilde A\wedge J+\tilde A\wedge \tilde A=0$. This is how the pure spinor Lax connection was first derived, however note that the parametrization (\ref{movingA}) is more convenient, since we can construct a monodromy matrix directly in the moving frame.}
\bea\label{flatA}
dA+A\wedge A = 0\ .
\eea
If we can find an $A(\mu)$ that satisfies (\ref{flatA}), we can construct a {\it monodromy matrix}, that is the Wilson loop of the Lax connection around the closed string
\bea\label{monodromymatrix}
\Omega(\mu;Q)=P\exp\left(-\oint_C A(\mu)\right) \ ,
\eea
where $C$ is a non-contractible loop around the string worldsheet, bound to the point $Q$. Because of the flatness condition (\ref{flatA}), two monodromy matrices based at two different points $Q$ and $R$ only differ by a similarity transformation
$$
\Omega(\mu;Q)=u \Omega(\mu;R) u^{-1} \ ,
$$
where $u=P\exp\left(-\oint_\gamma A(\mu)\right)$ and $\gamma$ is a contour between $Q$ and $R$. It follows that the eigenvalues of the monodromy matrix do not depend on the reference point $Q$ and therefore are conserved quantities of the string sigma model.

Note that under any symmetry that transforms the Lax connection by conjugation
\bea\label{laxconj}
\delta A(\mu)=d\Lambda+\Lambda\wedge A(\mu) \ ,
\eea
the monodromy matrix (\ref{monodromymatrix}) transforms by similarity and therefore the eigenvalues are invariant. In particular, this holds for the gauge symmetry as well as for the BRST symmetry of the pure spinor sigma model, as we will see below.

{\it Pure Spinor Monodromy Matrix}

We are looking for a left invariant current $A(\mu)$, depending on the spectral parameter $\mu$, that satisfies the flatness equation (\ref{flatA}) and is built out of the currents of the pure spinor sigma model. Let us take a generic combination
$$
A=c_0 J_0+c_1J_1+c_2J_2+c_3J_3+c_NN \ ,
$$
$$
\bar A=\bar c_0 \bar J_0+\bar c_1\bar J_1+\bar c_2\bar J_2+\bar c_3\bar J_3+\bar c_N\hat N \ ,
$$
and impose (\ref{flatA}). By using the equations of motion and the Maurer-Cartan equations, we find conditions on the coefficients $c_i,\bar c_i$, that can be solved in terms of one parameter $\mu$ and yield the following Lax connection
\bea\label{laxps}
A=&J_0-N+{\mu}^{-1}J_3+\mu^{-2} J_2+ \mu^{-3} J_1+\mu^{-4}N \ ,\\
\bar A=&\bar J_0-\hat N+\mu\bar J_1+\mu^2\bar J_2+\mu^3\bar J_3+\mu^4 \hat N \ .\nonumber
\eea
Note that worldsheet
PT transformation on the Lax connection, namely $z\leftrightarrow\bar z$ and exchanging grading $1\leftrightarrow 3$, is equivalent to the inversion of the spectral parameter $\mu\leftrightarrow1/\mu$. Hence, the Lax connection is invariant under their combined action
\bea\label{ptsymmetry}
z\leftrightarrow \zb\ ,\qquad {\rm grad}\, 1\leftrightarrow {\rm grad}\,3 \ ,\qquad \mu\leftrightarrow 1/\mu \ .
\eea

Among the possible variation of the Lax connection (and hence of the monodromy matrix), the good ones are the ones that do not change the analytic structure in the spectral parameter $\mu$. In particular, given a function of the spectral parameter $\Phi(\mu)$, the dressing transformations are defined as
$\delta A(\mu)={\cal D}(\mu)\Phi(\mu)$, such that  ${\cal D}(\mu)\Phi(\mu)$ has the same analytic structure in $\mu$ as the Lax connection, namely a Laurent expansion in $\mu^0$ to $\mu^{-4}$ for $A(\mu)$ and from $\mu^0$ to $\mu^4$ for $\bar A(\mu)$. Here we introduced the covariant derivative of the Lax connection ${\cal D}(\mu)\Phi=d\Phi+A(\mu)\wedge\Phi$. The BRST variation of the Lax connection is an infinitesimal dressing transformation
\bea\label{dressingbrst}
[\epsilon Q, A(\mu)]={\cal D}(\mu)\epsilon\lambda(\mu) \ ,\qquad
[\epsilon Q,\bar A(\mu)]=\bar {\cal D}(\mu) \epsilon\lambda(\mu) \ ,
\eea
where we introduced the dressed pure spinors $\lambda(\mu)=\mu\lambda+{1\over \mu}\hat \lambda$. Note also that the BRST acts on the monodromy matrix by a similarity transformation
\bea\label{brstmonodromy}
[\epsilon Q,\Omega(\mu;Q)]=[\epsilon\lambda(\mu)|_Q,\Omega(\mu;Q)]\ .
\eea
hence the eigenvalues of the monodromy matrix are BRST invariant quantities.

The gauge invariant Lax connection is obtained by inverting the map (\ref{movingA}). We have that $a(\mu)=g(A(\mu)-J)g^{-1}$ explicitly
\bea\label{gaugeinva}a=&g[\mu^{-1}J_3+\mu^{-2} J_2+ \mu^{-3} J_1+(\mu^{-4}-1)N]g^{-1}\ ,\\
\bar a=&g[\mu\bar J_1+\mu^2\bar J_2+\mu^3\bar J_3+(\mu^4-1) \hat N]g^{-1} \ .\nonumber
\eea

\subsection{Integrability and BRST cohomology}

The existence of an infinite set of
BRST-invariant nonlocal charges
can be deduced from the absence of certain states
in the BRST cohomology at ghost-number two.
These ghost-number two states are $f^C_{AB} h^A h^B$ where
$h^A$ are the BRST-invariant ghost-number one states associated with
the global isometries and $f^C_{AB}$ are the structure constants.
Whenever $f^C_{AB} h^A h^B$ can be written as $Q \Omega^C$ for
some $\Omega^C$ (i.e. whenever $f^C_{AB} h^A h^B$ is not in the BRST
cohomology), one can construct an infinite set of BRST-invariant
nonlocal charges.

Suppose one has a BRST-invariant string theory with global physical
symmetries described by the charges $q^A = \int d\s ~j^A(\s)$.
Since these symmetries take physical states to physical states,
$q^A=\int d\s~j^A(\s)$ must satisfy $Q(q^A)=0$ where $Q$
is the BRST charge. Note that
$\{Q, b_0\}=H$ where $H$ is the Hamiltonian, so BRST invariance
implies charge conservation if $q^A$ commutes with the $b_0$
ghost, i.e. if $q^A$ can be chosen in Siegel gauge.
With the exception of the zero-momentum dilaton, it is expected that
all ghost-number zero states in the BRST cohomology
can be chosen in Siegel gauge.\footnote{In the pure spinor formalism for
the superstring, there
is no natural $b$ ghost. Nevertheless, it is
expected that for any ghost-number zero state in the pure-spinor BRST
cohomology, there exists a gauge in which the state is annihilated by $H$.}

Since $Q(\int d\s j^A(\s)) =0$,
$Q(j^A) = \p_\s h^A$
for some $h^A$ of ghost-number one. And $Q^2=0$ implies that
$Q( \p_\s h^A)=0$, which implies that $Q(h^A)=0$ since
there are no $\s$-independent worldsheet fields.

Consider the BRST-invariant ghost-number two states $f^C_{AB}:h^A h^B:$
where $f^C_{AB}$ are the
structure constants
and normal-ordering is defined in any BRST-invariant manner,
e.g. $:h^A(z) h^B(z): \equiv {1\over{2\pi i}}\oint dy (y-z)^{-1}
h^A(y) h^B(z)$ where the contour of $y$ goes around the point $z$.
It will now be shown that whenever $f^C_{AB} :h^A h^B :$ is not in the BRST
cohomology\footnote{This
BRST cohomology is defined in the ``extended'' Hilbert space which
includes the zero mode of the $x^m$ variables. As explained in \cite{Astashkevich:1995cv}
the inclusion of the $x^m$ zero mode in the Hilbert space
allows global isometries
to be described by ghost-number one states in the cohomology.}, i.e.
whenever there exists an operator
$\Omega^{C}$ satisfying $Q(\Omega^{C}) = f^C_{AB}:h^A h^B:$, one can
construct an infinite set of nonlocal BRST-invariant charges.

To prove this claim, consider the nonlocal charge
$$k^{C} = f^C_{AB} :\int_{-\infty}^\infty d\s ~j^A(\s)~
\int_{-\infty}^\s d\s' ~j^B(\s'): .$$
Using $Q( j^A) = \p_\s h^A$, one finds that
$Q( k^{C}) = \int d\s l^{C}(\s)$ where
$$l^{C} = -2 f^C_{AB} :h^A(\s)
j^B(\s):.$$
One can check that $Q(l^{C}) = f^C_{AB}\p_\s (:h^A h^B:)$, so
$Q(l^{C}-\p_\s\Omega^{C}) = 0$ where
$\Omega^{C}$ is the operator which is assumed to satisfy
$Q(\Omega^{C}) = f^C_{AB} : h^A h^B:$.

Since
$(l^{C}-\p_\s\Omega^{C})$ has $+1$ conformal weight and since
BRST cohomology is only nontrivial at zero conformal weight,
$l^{C}-\p_\s\Omega^{C} = Q(\Sigma^{C})$ for some $\Sigma^{C}$.
Using $\Sigma^{C}$, one can therefore construct the nonlocal
BRST-invariant charge
$$\widetilde q^{C} =f^C_{AB} :\int_{-\infty}^\infty d\s ~j^A(\s)~
\int_{-\infty}^\s d\s' ~j^B(\s'): -
\int_{-\infty}^{\infty} d\s \Sigma^{C}(\s) .$$

By repeatedly commuting $\widetilde q^{C}$ with
$\widetilde q^{D}$, one generates
an infinite set of nonlocal BRST-invariant charges. So as claimed,
$f^C_{AB}:h^A h^B:=
Q(\Omega^{C})$ implies the existence of
an infinite set of nonlocal BRST-invariant charges.

Let us know specialize to the pure spinor sigma model on $AdS_5\times S^5$. To prove the existence of an infinite set of BRST-invariant charges,
one needs to find an $\Omega =\Omega^{C} T_C$ satisfying
$Q\Omega = : \{ h, h\}:$ where $h= h^A T_A$, $Q (j) = \p_\s h$, and
$q^A= \int d\s j^A$ are the charges associated with the global
$PSU(2,2|4)$ isometries (\ref{noetherg}). It will be shown in (\ref{hexp}) that
$Q (j) =  \p_\s (g (\l -\lh) g^{-1}),$ so
\bea\label{later}h =  g (\l -\lh) g^{-1}.\eea

Note that $h$ is BRST-invariant since
\bea Q (h)= g \{(\l+\lh),(\l-\lh)\}g^{-1} =
g ((\l^\a\g^m_{\a\b}\l^\b) T_m -
(\lh^\ah\g^m_{\ah\bh}\lh^\bh) T_m) g^{-1} =0\nonumber\\\label{brstinv}\eea
because of the pure spinor constraint.
Consider the ghost-number two state
\bea\label{considerh}\{h,h\} = 2
g (\l-\lh)(\l-\lh) g^{-1} = -2 g \{\l,\lh\} g^{-1}.\eea
Since $\{\l+\lh, \l+\lh\} = 2\{\l,\lh\}$, one can write this state as
$Q\Omega$ where
\bea\label{omegae}\Omega = -  g(\l+\lh) g^{-1}.\eea
So $Q\Omega = \{h,h\}$, which implies
the existence of an infinite set of BRST-invariant charges.

To explicitly construct these BRST-invariant charges, suppose one has
a gauge invariant (i.e. lower case) current whose $\tau$-component $a$ satisfies
\bea\label{asat}Q a = \p_\s \Lambda + [a,\Lambda]\eea
for
some $\Lambda$. Then, as we discussed above, the charge
(\ref{openu})
transforms under BRST by conjugation, hence its eigenvalues are BRST invariant. It is easy to check that the gauge invariant Lax connection (\ref{gaugeinva}) transforms as in (\ref{asat}) with $\Lambda=g(\mu\l+\hat\l/\mu)g^{-1}$.

Note that the global charge $q=\int d\s j(\s)$ can be
obtained by expanding $a(\mu)$ near $\mu=1$. If $\mu=1+\epsilon$,
one finds that $a(\mu)= \epsilon j +
{\cal O}(\epsilon^2)$ and $\Lambda (\mu) = \epsilon h +
{\cal O}(\epsilon^2)$ where $Q (j) = h$.
So
one learns that
\bea\label{hexp}h = \lim_{\epsilon \to 0}
\epsilon^{-1}\Lambda (\mu) = g (\l-\lh) g^{-1},\eea
as was claimed in (\ref{later}).

\subsection{Quantum integrability}

In the previous section we established the classical integrability of the sigma model on $AdS_5\times S^5$. In this section we will prove that integrability survives at the quantum level, by explicit one-loop computation first and then by applying the all-loop argument based on BRST cohomology.

After proving quantum integrability, the next step is to study the algebra of the transfer matrices, that are the open Wilson lines of the Lax connection. At the leading order in the expansion around flat space, the transfer matrices satisfy a generalized version of the Yang-Baxter equation. This is the starting point to solve the spectrum of the sigma model by using integrability, which is still an open problem.

\subsection{Finiteness of the monodromy matrix}

In this section we will consider the one-loop finiteness of the monodromy matrix (\ref{monodromymatrix}) by direct evaluation of its UV logarithmic divergence first and then by proving that it is invariant under infinitesimal deformations of the contour.

{\it Log divergences}

Consider the open Wilson line of the Lax connection (\ref{laxps})
\bea\label{openwilson}
\Omega(C)=P\exp\left(-\int_C A(\mu)\right)
 \
 \eea
where $C$ is an open contour on the worldsheet. The flatness of the Lax connection implies that classically it does not depend on the choice of the contour. If we expand the path ordered exponential, we get terms of the kind
\bea\label{contour}
\int_{s_1<\ldots<s_n} A(s_1)\ldots A(s_n) \ .
\eea
Expectation values of such operators typically give linear divergences (that depend on the scheme) and logarithmic divergences, which are independent of the regularization scheme.

Let us show that if logarithmic divergences are present, then the Wilson line cannot be conformally invariant, which implies that the Wilson line would depend on the choice of the contour. Take two contours $C$ and $C'$, related by a dilatation: $C^\prime=\lambda C$, for $\lambda$ real. Both $\Omega(C)$ and $\Omega(C')$ contain divergences and need to be regularized. The renormalized Wilson line $\Omega^{ren}$ is equal to the regularized $\Omega_\epsilon$ plus counterterms $C_\epsilon$
$$
\Omega^{ren}(C)=\lim_{\epsilon\to0}\left(\Omega_\epsilon(C)+\Gamma_\epsilon (C)\right)\ .
$$
If the Wilson line does not depend on the choice of the contour, then we would have $\Omega^{ren}(C')=\Omega^{ren}(C)$. Conformal invariance implies that $\Omega_\epsilon(C)=\Omega_{\lambda\epsilon}(C')$, but if there are log divergences, then it is not true that $\Gamma_{\lambda\epsilon}(C')=\Gamma_\epsilon(C)$. Such divergences are of the form
$$
\int ds\,\partial_s X f(X) \ln\epsilon \ ,
$$
where $X$ are the bosonic fluctuations, and since $\epsilon$ has weight one they are not conformally invariant. Note that linear divergences $\int ds f(X) /\epsilon$ on the contrary are conformally invariant. We conclude that a necessary condition for the independence of the Wilson line on the contour is the absence of log divergences.

There are several possible sources of log divergences in the sigma model. First, the currents are composite objects and get internal log divergences: even in the theory of a free boson these give rise to the $k^2$ anomalous dimension for the vertex operator $e^{ikX}$ and they are certainly present in our interacting sigma model. In our case they come from the normal ordering quartic vertices and from the fish diagrams cubic vertices, but the they cancel in such a way that the currents $J_i$ are finite. A second source of log divergences arises when two or more points on the integration contour (\ref{contour}) coincide and the currents develop short distance singularities in their OPE.

If we choose the particular gauge $g=\exp{(X_1+X_3)\over R}\exp{X_2\over R}$ for the coset element, the effect of the multiple collisions of the currents generates non-vanishing log divergences, but it turns out that the full log divergence appears in the form of a total derivative. Hence, its contribution to the monodromy matrix, which is integrated on a closed contour, takes the form
\bea\label{logdiv}
\Omega_\epsilon(\mu,C)=f(\epsilon,\mu)\Omega^{finite}(\mu,C)f(\epsilon,\mu)^{-1} \ ,
\eea
where $f(\epsilon,\mu)$ is a gauge transformation that depends on the spectral parameter, whose one-loop expression is
\bea\label{fdiv}
f(\epsilon,\mu)=\exp\left(-{\ln \epsilon\over R^2}(\mu^2+\mu^{-2})\right) C_2.X\ .
\eea
As a result, the log divergences can be removed by a $\mu$-dependent gauge transformation. This ensures that there exists a regularization prescription for which the log divergences vanish. Note, however, that the log divergences do contribute to the path ordered exponential of an open Wilson line, giving divergent terms at the endpoints, which must be properly taken into account when studying the transfer matrix or the Wilson loop with operator insertions.

{\it Contour Deformations}

A second way to prove that the monodromy matrix does not depend on the contour is to look at its variation upon infinitesimal deformations of the contour
\bea\label{omegavariation}
{\delta\Omega(\mu,C)\over \delta \sigma^i(s)}=P\, F_{ij}{d\sigma^j(s)\over ds}\exp\left( -\oint_C A(\mu)
\right) \ ,\eea
where $i$ is the worldsheet vector index and $F_{ij}(\mu)=[\partial_i+A_i(\mu),\partial_j+A_j(\mu)]$ is the curvature of the Lax connection. Classically, $F_{ij}=0$, but there might be anomalies at the quantum level.
The most general form of such an anomaly is
\bea\label{laxanomaly}
\delta \Omega=P{\cal O}_{z\zb}\exp\left(-\oint_C A(\mu)\right) \ ,
\eea
for a generic insertion of a ghost number zero and weight $(1,1)$ local operator ${\cal O}_{z\zb}$. Requiring consistency with the BRST transformation of the monodromy matrix (\ref{brstmonodromy}), the BRST variation of such an operator must be
\bea\label{brsto}
[\epsilon Q,{\cal O}_{z\zb}]=[{\cal O}_{z\zb},\lambda(\mu)] \ ,
\eea
where the dressed pure spinor $\lambda(\mu)=\mu\lambda+{1\over \mu}\hat \lambda$. If we write down the most general expression for the operator ${\cal O}_{z\zb}$ and impose that it satisfies (\ref{brsto}) together with PT invariance (\ref{ptsymmetry}), we find that such operator must vanish at the tree level. This proves that the monodromy matrix is independent of the contour at leading order. By BRST cohomology arguments this procedure can be easily extended to all loops, as we will discuss in the next section from a different point of view.

\subsection{All-loop proof}

One can use arguments similar to those of section \ref{sectionquantum} to prove
that this construction is valid at the quantum level to all orders
in perturbation theory. For example, suppose that a non-local
BRST-invariant charge $q^A$ has been constructed to order $h^{n-1}$,
i.e. $\widetilde Q(q^C) = h^n \Omega^C +{\cal O}(h^{n+1})$
where $\Omega^C$ is some
integrated operator of ghost-number $+1$, $\widetilde Q = Q + Q_q$,
and $Q_q$ generates quantum corrections to the classical BRST
transformations of (\ref{brstrans}) generated by $Q$. Like
other types of quantum anomalies,
$\Omega^C$ must be a local integrated
operator since it comes from a short-distance
regulator in the operator product expansion $j^A(\s) j^B(\s')$.
So trivial cohomology implies that there exists a local
operator $\Sigma^C(\s)$ such that $\Omega^C= Q(\int_{-\infty}^\infty
d\s\Sigma^C(\s))$. Therefore, $q^C - h^n \int_{-\infty}^\infty
d\s\Sigma^C(\s)$ is BRST-invariant to order $h^n$.

To verify that the relevant cohomology class is trivial for the superstring
in an $AdS_5\times S^5$ background, it will be useful to recall that
for every integrated operator of ghost-number $+1$ in the
BRST cohomology, there exists a corresponding unintegrated operator
of ghost-number $+2$ and zero conformal weight in the BRST cohomology.
This is easy to prove since $Q(\int d\s W(\s))=0$ implies that
$Q(W(\s)) = \p_\s V(\s)$ where $V(\s)$ is a BRST-invariant operator
of zero conformal weight. And if $V$ is BRST-trivial, i.e. if
$V=Q\Lambda$ for some $\Lambda$, then $Q(W-\p_\s\Lambda)=0.$
Since the BRST cohomology is trivial for unintegrated operators of
nonzero conformal weight, $W-\p_\s\Lambda= Q\Sigma$ for some $\Sigma$.
So
\bea\label{fineq}\int d\s W(\s) = \int d\s (Q\Sigma(\s) + \p_\s\Lambda(\s)) =
Q(\int d\s \Sigma(\s)),\eea
which implies that $\int d\s W(\s)$ is BRST-trivial \cite{Berkovits:1991gj}.

At ghost-number $+2$, the only unintegrated operators of zero conformal
weight which transform in the adjoint representation of the global
$PSU(2,2|4)$ algebra are
\bea\label{twoun}V_1 = g~(\l^\a T_\a)(\lh^\bh T_\bh)~ g^{-1} \quad
{\rm and} \quad
V_2 = g~(\lh^\ah {\bf T}_\ah)(\l^\b {\bf T}_\b)~ g^{-1} ,\eea
where $g(x,\t,\th)$ transforms by left multiplication as
$\d g(x,\t,\th)= \Sigma g(x,\t,\th)$ under the global
$PSU(2,2|4)$ transformation parameterized by $\Sigma=\Sigma^A {\bf T}_A$.
One can easily verify that $Q(V_1-V_2)\neq 0$ and that
$V_1 + V_2 = Q\Omega$ where
\bea\label{veriom}
\Omega = \half g~ (\l^\a {\bf T}_\a + \lh^\ah {\bf T}_\ah)~ g^{-1}.\eea
So the cohomology is trivial, which implies the existence of an infinite
set of non-local BRST-invariant charges at the quantum level.

{\bf Guide to the literature}

Integrability of the GS action in AdS, discovered in  \cite{Bena:2003wd,Mandal:2002fs}, is reviewed in great details in \cite{Arutyunov:2009ga}. The recent review \cite{Beisert:2010jr} covers extensively the AdS/CFT application of integrability from both the gauge theory and the string theory side.

The Lax representation of the equations of motion of the pure spinor action in $AdS$ was derived in \cite{Vallilo:2003nx}. The relation between non-local conserved charges and BRST cohomology is discussed in \cite{Berkovits:2004jw}.

A classic discussion of quantum anomalies in the higher non-local conserved charges is \cite{Luscher:1977uq,Abdalla:1982yd}. Some quantum properties of the pure spinor monodromy matrix are studied \cite{Mikhailov:2007mr,Mikhailov:2007eg,Puletti:2008ym}. The algebraic proof of the BRST invariance of the non-local conserved charges at all loops appears in \cite{Berkovits:2004xu}.

More advanced topics, that we did not touch, are: The fusion rules and the classical Yang-Baxter equation satisfied by the open Wilson lines of the Lax connection \cite{Mikhailov:2007eg}; The Hirota equation (Y-system) satisfied by such Wilson lines in $PSL(n|n)$ sigma models \cite{Benichou:2010ts}.

\clearpage

{\bf Acknowledgements}

I would like to thank first and foremost my collaborators on pure spinor projects during the last ten years: I.~Adam, N.~Berkovits, A.~Dekel, P.A.~Grassi, M.~Matone, I.~Oda, Y.~Oz, P.~Pasti, D.~Sorokin, M.~Tonin, B.~Carlini Vallilo, and S.~Yankielowicz.
I am particularly indebted to Brenno Carlini Vallilo for his patient explanations of many aspects of the material contained in Section~\ref{section:effectiveaction} and \ref{section:spectrum}, and to Nathan Berkovits for his ever-present guidance. I would like to thank B.~Van Rees for his comments on the manuscript. Finally, it is a pleasure to thank Nathan Berkovits, Nikita Nekrasov, and Yaron Oz for pushing me to write this review, as part of a larger project that hopefully will materialize in the future. 


\clearpage


\appendix

\section{Notations}
\label{appendix:notations}

In this Appendix we collect the conventions and notations used in ten dimensional flat space, in a generic curved background and in an $AdS_5\times S^5$ background.

\subsection{Flat background}
\label{appendix:flat}

In the flat ten-dimensional background, we use lower case roman letters from the middle of the alphabet $m,n,\ldots=0,\ldots,9$ to denote vector indices and Greek letters $\alpha,\beta,\ldots=1,\ldots,16$ to denote sixteen-component Weyl spinor indices. 

The Dirac matrices in ten spacetime dimensions are thirty-two by thirty-two matrices and admit a real representation $\Gamma^m=\{\Id\otimes(i\tau_2),\sigma^\mu\otimes\tau_1,\chi\otimes\tau_1\}$, where $\sigma^\mu$ for $\mu=1,\ldots,8$ are the $SO(8)$ Dirac gamma matrices, which are symmetric sixteen by sixteen matrices; $\chi$ is the real diagonal $SO(8)$ chirality matrix; $\tau_i$ are the usual two-dimensional Pauli matrices. The ten-dimensional chirality matrix is $\Id\otimes \tau_3$ and the charge conjugation matrix $C$ is such that $C\Gamma^m=-(\Gamma^m)^TC$ and is equal to $\Gamma^0$. We can split a thirty-two component Dirac spinor into left and right moving sixteen component Weyl spinors $\Psi={\psi_L^\alpha\choose\psi_{R\bd}}$ and introduce the off-diagonal Pauli matrices 
\bea
\Gamma^m=\left(\ba{cc} 0 & (\sigma^m)^{\a\bd}\\
(\tilde \sigma^m)_{\bd\gamma} & 0 \ea\right) \ ,\qquad C=
\left(\ba{cc} 0 & c_\a{}^\bd\\
c^\bd{}_\g & 0 \ea\right) \ ,
\eea
where $\sigma^m=\{\Id,\sigma^\mu,\chi\}$, $\tilde \sigma^m=\{-\Id,\sigma^\mu,\chi\}$ and $c_\a{}^\bd=-c^\bd{}_\g=\Id$. The Dirac gamma matrices multiplied by the charge conjugation matrix can be split into off-diagonal Pauli matrices
\bea
C\Gamma^m=\left(\ba{cc} 0 & (\gamma^m)^{\ad\bd}\\
(\gamma^m)_{\b\g} & 0 \ea\right) \ ,
\eea
which are sixteen by sixteen symmetric matrices. Since Weyl spinors of different chirality are inequivalent $SO(1,9)$ irreps, we cannot raise and lower indices and therefore we will omit the dots on the upper indices and just use undotted indices $(\gamma^m)^{\a\b}, (\gamma^m)_{\a\b}$. The ten-dimensional Pauli matrices satisfy $\{\gamma^m,\gamma^n\}=2\eta^{mn}$.

We define the weighted antisymmetric product of gamma matrices as
\bea
\gamma^{m_1\ldots m_k}={1\over k!}\gamma^{[m_1}\ldots\gamma^{m_k]} \ ,
\eea
and $\gamma^{m},\gamma^{m_1\ldots m_5}$ are symmetric in their spinor indices, while $\gamma^{m_1m_2m_3}$ is antisymmetric.
Any bispinor can be decomposed along such basis as
\bea
f_\a g^\b=&{1\over16}\left(\delta_\a^\b(fg)+{1\over2!}(\gamma_{ab})_\a{}^\b(f\gamma^{ab}g)+{1\over4!}(\gamma^{abcd})_\a
{}^\b(f\gamma_{abcd}g\right)\ ,\\
f_\a g_\b=&{1\over16}\left(\gamma^a_{\a\b}(f\gamma_ag)+{1\over 3!}\gamma^{abc}_{\a\b}(f\gamma_{abc}g)+{1\over5!}\gamma^{abcde}_{\a\b}(f\gamma_{abcde}g)\right) \ .
\eea

{\it $SO(1,9)$ identities}

Some useful ten-dimensional gamma matrix identities are
\bea
\gamma^m_{(\a\b}(\gamma_{m})_{\g)\d}=&0 \ ,\nonumber\\
(\gamma_{ab})_\alpha{}^\beta(\gamma^{ab})_\gamma{}^\delta=&4(\gamma_a)^{\beta\delta}(\gamma_a)_{\alpha\gamma}-2 \delta_{\alpha}^\beta\delta_\gamma^\delta-8\delta_\alpha^\delta\delta_\gamma^\beta \ ,\nonumber\\
\gamma^{abc}_{\a\b}\gamma_{abc}^{\g\d}=&48 \delta^\g_{[\a}\d^\d_{\b]}\ ,\nonumber\\
\gamma^{a_1\ldots a_5}_{\a\b}\gamma_{a_1\ldots a_5}^{\g\d}=&5!\left(16\delta^\g_{(\a}\d^\d_{\b)}-2\gamma^a_{\a\b}\gamma_a^{\g\d}\right) \ ,\nonumber\\
\gamma^b\gamma_{a_1\ldots a_k}\gamma_b=&(-)^k(D-2k)\gamma_{a_1\ldots a_k} \ ,\nonumber\\
\gamma^{a_1\ldots a_k}\gamma_{a_1\ldots a_k}=&(-)^{k(k-1)/2}{D!\over (D-k)!} \ ,\label{gammaidentity}\\
\gamma^a\gamma^{b_1\ldots b_k}=&(-)^k\gamma^{b_1\ldots b_k}\gamma^a+k\eta^{a[b_1}\gamma^{b_2\ldots b_k]}\ ,\nonumber\\
\gamma^a\gamma^{b_1\ldots b_k}=&\gamma^{ab_1\ldots b_k}+\half k\eta^{a[b_1}\gamma^{b_2\ldots b_k]}\ ,\nonumber\\
\gamma_a\gamma^{ab_1\ldots b_k}=&(D-k+1)\gamma^{b_1\ldots b_k} \ ,\nonumber\\
\gamma^p\gamma^{mn}=&\gamma^{mn}\gamma^p+2\eta^{p[m}\gamma^{n]} \ ,\nonumber
\eea
where $D=10$. 

\subsection{Curved background}
\label{appendix:curved}

The curved ten-dimensional type II superspace coordinates  are described by indices from the middle of the alphabet $Z^M=(X^m,\theta^\mu,\hat\theta^{\hat\mu})$, where $\theta$ and $\hat\theta$ are sixteen component real Majorana-Weyl spinors of the same chirality $(\theta^\mu,\hat\theta^{\hat\mu})$ in the type IIB case and of opposite chirality 
$(\theta^\mu,\hat\theta_{\hat\mu})$ in the type IIA case. We denote the flat tangent superspace with indices from the beginning of the alphabet $A=(a,\alpha,\hat\alpha)$. The zehnbein $E^A_M$ convert curved into flat indices and satisfy
$$
G_{AB}=E^M_AG_{MN}E^N_B \ ,
$$
where the only non-zero component of $G_{AB}$ is the flat Minkowski metric $\eta_{ab}$. The pullback on the worldsheet of the target space super-zehnbein are denoted by
\bea
\Pi^A=E^A_MdZ^M \ .
\eea


\section{Super Yang-Mills in superspace}
\label{syminsuperspace}

We would like to show how pure spinors arise naturally in ten dimensions. We will discuss this from a hystorical point of view, which is the most pedagogical way to arrange the presentation. 

We will start by considering super Yang-Mills theory in ten dimensions \cite{Siegel:1978yi,Witten:1985nt}. To describe this gauge theory in a manifestly supersymmetric way we will introduce a ten-dimensional superspace with sixteen supersymmetries. One needs to impose a set of gauge invariant constraints on the superspace curvature to get rid of the unwanted degrees of freedom that get in once we go to superspace. We will show that in ten dimensions the relevant superspace constraint implies the equations of motion for the component fields. We will prove this by explicitly solving the Bianchi identity subject to the constraint.

In the next step, we will rewrite the constraint in a convenient form and then we will recast the super Yang-Mills equations of motion and gauge invariances in superspace as a BRST cohomology problem. At this point, we will introduce a new spacetime coordinate given by a bosonic Weyl spinor, which will play the role of the ghost in the BRST formalism and show that consistency requires that this bosonic spinor satisfy the pure spinor constraint. In this sense, the pure spinor constraint arises as an integrability condition on the cohomology \cite{Howe:1991mf,Howe:1991bx}.

\subsection{Ten-dimensional superspace}

Ten-dimensional super Yang-Mills theory consists of a gluon $a_m$ and a gluino $\chi^\a$, which is a Majorana-Weyl spinor in ten dimensions. We can form a covariant derivative $\nabla_m=\partial_m+[a_m,\cdot]$ and construct a  field strength $F_{mn}=[\nabla_m,\nabla_n]$, in terms of which the classical equations of motion are
\bea\label{symeoms}
\nabla_m F^{mn}+\half\gamma^n_{\a\b}\{\chi^\a,\chi^\b\}= 0 \ ,\qquad \gamma^m_{\a\b}\nabla_m \chi^\a = 0 \ .
\eea

We would like to describe this theory in a manifest supersymmetric way by using a ten dimensional superspace formalism. Ten-dimensional ${\cal N}=1$ superspace has sixteen supercharges and it is described in terms of the coordinates 
$(x^m,\t^\a)$, where $\t^\a$ are Grassmann-odd Majorana-Weyl spinors. 
Let us introduce the supersymmetric derivative 
$
D_\a={\partial\over\partial \theta^\a}-\half(\gamma^m\t)_\a\partial_m  $ that satisfies the supersymmetry algebra
\bea\label{susyalg}
\{D_\a,D_\b\} = -\gamma^m_{\a\b}\partial_m \ .
\eea

Let us introduce a super-connection one--form $A=E^B A_B$, where $E^B$ are the super--zehnbein and $A_B=(A_m,A_\alpha)$ is the super connection, while the super covariant derivatives are $\nabla_\a=D_\a+[A_\a,\cdot\}$ and $\nabla_a=\partial_m+[A_m,\cdot]$. The indices from the beginning of the alphabet are tangent space indices, $m=0,\ldots,9$ are bosonic directions and $\alpha=1,\ldots,16$ are fermionic directions. The superspace curvature is the two--form superfield 
$$
F=[\nabla,\nabla]=E^AE^B F_{BA} \ ,
$$
whose components are
\bea\label{curvature}
F_{\a\b}=&\{\nabla_\a,\nabla_\b\}+\gamma^m_{\a\b}\nabla_m \ ,\\ 
=&D_{(\a}A_{\b)}+\{A_\a,A_\b\}+\gamma^m_{\a\b}A_m \ ,\nonumber\\ 
F_{\a m}=&[\nabla_\a,\nabla_m]=D_\a A_m-\partial_m A_\a+[A_\a,A_m] \ ,\nonumber\\ 
F_{mn}=&[\nabla_m,\nabla_n]=\partial_{[m}A_{n]}+[A_m,A_n] \ .\nonumber
\eea
The curvatures are invariant under the gauge transformations
\bea\label{gauge}\left\{\begin{array}{c}\delta A_\a=\nabla_\a\Omega \ \\ 
\delta A_m=\nabla_m\Omega \end{array}\right.
\eea
The superconnections $(A_\a,A_m)$ contain too many degrees of freedom. We want to impose constraint on the gauge invariant curvatures (\ref{curvature}) such that we get rid of the unwanted degrees of freedom and at the same time we impose the equations of motion (\ref{symeoms}). We claim that both goals can be achieved by imposing the single contraint
\bea\label{constraint}
F_{\a\b}=0 \ .
\eea

Let us show that by plugging (\ref{constraint}) into the Bianchi identity we find the equations of motion (\ref{symeoms}). 
The first Bianchi identity reads
\bea\label{bianchione}
[\{\nabla_\a,\nabla_\b\},\nabla_\g]+
[\{\nabla_\b,\nabla_\g\},\nabla_\a]+[\{\nabla_\g,\nabla_\a\},\nabla_\b]= 0 \ ,
\eea
and by using the definitions of the curvature (\ref{curvature}) we can rewrite it as
\bea\label{bianchionea}
[F_{(\a\b},\nabla_{\g)}]-\gamma^m_{(\a\b}F_{\g) m} = 0 \ .
\eea
If we substitute the constraint (\ref{constraint}) into the Bianchi identity (\ref{bianchionea}) we end up with the equation
\bea\label{bianchioneb}
\g^m_{(\a\b} F_{\g)m} = 0 \ .
\eea
Because of the Fierz identities (\ref{gammaidentity}), the unique solution to (\ref{bianchioneb}) is in terms of a new superfield $W$
\bea\label{solbianchione}
F_{\g m}=(\gamma_m)_{\g\b} W^\b \ .
\eea
Note that the lowest component of $W$ is the gluino 
$$
W^\a|_{\t=0}=\chi^\a \ .
$$

The second Bianchi identity
\bea\label{bianchitwo}
[\{\nabla_\a,\nabla_\b\},\nabla_m]+\{[\nabla_m,\nabla_\a],\nabla_\b\}-\{[\nabla_\b,\nabla_m],\nabla_\a\} = 0 \ ,
\eea
can be rewritten, using the expressions for the curvature (\ref{curvature}), in the following form
\bea\label{bianchitwoa}
[F_{\a\b}-\gamma^p_{\a\b}\nabla_p,\nabla_m]+\{F_{m\a},\nabla_\b\}-\{F_{\b m},\nabla_\a\} = 0 \ .
\eea
Imposing the constraint (\ref{constraint}) and the solution of the first Bianchi identity (\ref{solbianchione}), the Bianchi identity (\ref{bianchitwoa}) becomes
\bea\label{bianchitwob}
(\gamma_m)_{\a\rho} \nabla_\b W^\rho+(\gamma_m)_{\b\rho}\nabla_\a W^\rho=\gamma^n_{\a\b}F_{nm} \ .
\eea
After multiplying this expression by $(\gamma^m)^{\b\g}(\g_h)_{\g\tau}(\g^h)^{\a\lambda}$ and making repeated use of ten--dimensional gamma matrix identities we eventually find
\bea\label{bianchitwoc}
\nabla_\a W^\b=-{1\over 4}(\gamma^{mn})_\a{}^\b F_{mn} \ ,
\eea
and by further taking the trace obtain
\bea\label{tracetwo}
\nabla_\a W^\a= 0 .
\eea
This last equation will be useful when comparing the ten--dimensional onshell superspace with the four--dimensional offshell superspace.

Consider now the third Bianchi identity
\bea\label{bianchithree}
[[\nabla_m,\nabla_n],\nabla_\a]+
[[\nabla_\a,\nabla_m],\nabla_n]+
[[\nabla_n,\nabla_\a],\nabla_m]=0 \ .
\eea
By plugging the curvature (\ref{curvature}) and the previous solution (\ref{solbianchione}) we can recast (\ref{bianchithree}) in the following form
\bea\label{bianchithreea}
\nabla_\a F_{mn}=\nabla_{[m}(\g_{n]} W)_\a \ .
\eea

We will now see that (\ref{bianchitwoc}) and (\ref{bianchithreea}) imply the gluino equation of motion. Let us act on (\ref{bianchitwoc}) with $\nabla_\g$ and symmetrize on the $(\a\g)$ indices. By the curvature definition (\ref{curvature}) and the contraint (\ref{constraint}) we find
\bea\label{intermediate}
-\gamma^m_{\a\g}\nabla_m W^\b = -{1\over 4}(\g^{mn})_{(\a}{}^\b\nabla_{\g)}F_{mn} \ .
\eea
Now let us plug (\ref{bianchithreea}) into (\ref{intermediate}) 
\bea\label{intertwo}
\g^m_{\a\g} \nabla_m W^\b=-\half(\g^{mn})_{(\g}{}^\b(\g_m\nabla_n W)_{\a)} \ ,
\eea
and finally take the $\delta_\b^\a$ trace and find the equations of motion for the gluino in (\ref{symeoms})
\bea\label{gluinoeom}
\gamma^m_{\a\g}\nabla_m W^\g= 0 \ .
\eea
By further applying $(\gamma_n\nabla)^\a$ to (\ref{gluinoeom}) and using (\ref{solbianchione}) we arrive at
\bea\label{gluoneom}
\nabla^m F_{mn}=(\gamma_n)_{\a\b} W^\b W^\a \ ,
\eea
which is nothing but the gluon equation of motion (\ref{symeoms}).

In summary, by using the Bianchi identities (\ref{bianchione}), (\ref{bianchitwo}) and (\ref{bianchithree}), we have shown that the superspace constraint 
$$
F_{\a\b}=0\ ,
$$
is equivalent to the ten-dimensional super Yang-Mills equations of motion (\ref{symeoms}).
The expansion in components of the fields encountered in this analysis is given by
\bea\label{components}
W^\a=\chi^\a-{1\over 4}(\gamma^{mn}\t)^\a F_{mn}+\ldots 
\eea

\subsubsection{Comparison to four dimension}

In four dimensional ${\cal N}=1$ superspace, spinors have dotted and undotted indices. One can impose the following set of constraints on the superspace curvatures
\bea\label{fourdcon}
F_{\a\b}=F_{\a\dot\b}=F_{\dot\a\b}=0 \ ,
\eea
and plug them into the Bianchi identity as in the previous section. The Bianchi identities can then be solved in terms of the two complex conjugate superfields $W^\a,\bar W^\ad$ satisfying the constraints
\bea\label{fourdW}
\left\{ 
\begin{array}{c}\bar D_{\ad}W^\a=0=D_\a \bar W^\ad \ ,\\ 
\bar D_\ad\bar W^\ad- D_\a W^\a=0 \ ,\end{array}\right.
\eea
whose component expansion is
\bea\label{fourdcomp}
W^\a=\chi^\a+\t^\a D+(\sigma^{mn}\t)^\a F_{mn}+\ldots 
\eea
where the dots are terms proportional to the equations of motion. The constraints (\ref{fourdW}) imply that the $D$ auxiliary field sitting in $W$ is real. The constraints (\ref{fourdcon}) allow for a description of the offshell four--dimensional super Yang-Mills multiplet, that is including auxiliary fields. We can then write down an action whose equations of motion are
\bea\label{eomsfourd}
D_\a W^\a+\bar D_\ad \bar W^\ad= 0 
\ ,
\eea
which, together with the constraints (\ref{fourdW}), imply the equation $D=0$ that puts the theory onshell. 

In ten dimensions, there is no dotted spinor representation, hence the curvature constraints immediately set the theory onshell.

\subsection{Pure spinors}

The constraint (\ref{constraint}) can be equivalently written as
a condition on the spinorial connection
\bea\label{fivecon}
\gamma_{mnpqr}^{\a\b}\left(D_\a A_\b+\{A_\a,A_\b\}\right) = 0 \ .
\eea
This can be seen by projecting the curvature $F_{\a\b}$ in (\ref{curvature}) on the five--form product of gamma matrices and using the fact that $\tr \gamma_{mnpqr}\gamma_s=0$. Conversely, the solution to the equation (\ref{fivecon}) defines the connection $A_m$ in terms of the spinorial connection as
$$
A_m={\g_m^{\a\b}\over 10}\left(D_\a A_\b+\{A_\a,A_\b\}\right) \ .
$$
Hence, the contraints (\ref{constraint}) and (\ref{fivecon}) are completely equivalent. As a result, we can consider (\ref{fivecon}) as an alternative way to write the ten--dimensional super Yang Mills equations of motion (\ref{symeoms}).

We would like now to recover the equations of motion (\ref{fivecon}) and gauge invariances (\ref{gauge}) as a solution to an auxiliary cohomology problem. Let us consider the linearized theory only, namely the super Maxwell theory, in which case we have
\bea\label{supermax}
\g_{mnpqr}^{\a\b}D_\a A_\b = 0 \ ,\\ 
\delta A_\a = D_\a \Omega \ .
\eea
Let us augment the target space coordinates $(X,\t)$ by introducing a bosonic spinorial direction $\l^\a$, given by a complex Weyl spinor with sixteen components. Such objects are usually referred to as twistors. This new variable will carry an abelian charge that we call ``ghost number." Let us introduce two more operators. The first is a BRST charge
\bea\label{qbrst}
Q=\l^\a D_\a \ .
\eea
The second is a ``vertex operator," namely a ghost number one operator
\bea\label{vertexop}
{\cal U}^{(1)}=\l^\a A_{\a}(X,\t) \ .
\eea
We would like to solve the cohomology problem
\bea\label{cohomology}
\left\{\begin{array}{c} Q{\cal U}^{(1)}=0\ , \\ 
\delta {\cal U}^{(1)}=Q\Omega \ ,
\end{array}\right.
\eea
where the gauge superfield $\Omega(X,\t)$ has ghost number zero. 

For the cohomology problem to be well defined, the BRST charge must be nilpotent. The supersymmetry algebra (\ref{susyalg}) implies that
\bea\label{qsquare}
\{ Q,Q\}=-\l^\a\l^\b \gamma^m_{\a\b}\partial_m \ ,
\eea
hence we find that the BRST charge is nilpotent if and only if the bosonic spinors $\l^\a$ satisfy the constraint
\bea\label{purecon}
\l^\a\gamma^m_{\a\b}\l^\b=0 \ .
\eea
Such spinors are called "pure spinors."

Let us go back to the cohomology problem (\ref{cohomology}). The first condition implies
$$
Q{\cal U}^{(1)}=0 \quad\Rightarrow\quad \l^\a\l^\b D_\a A_\b(X,\t)=0 \ .
$$
The symmetric bispinor $\l^\a\l^\b$ can be decomposed along the basis of odd--form gamma matrices
$$
16\l^\a\l^\b=\gamma_m^{\a\b}(\l\gamma^m\l)+{1\over 3!}\g_{mnp}^{\a\b}(\l\g^{mnp}\l)+{1\over 5!}\g_{mnpqr}(\l\g^{mnpqr}\l) \ ,
$$
but $\g_{mnp}$ is antisymmetric in the spinor indices and the $\l$'s satisfy (\ref{purecon}), so the product of two pure spinors is proportional to the product of five gamma matrices
$$
\l^\a\l^\b={1\over16 \cdot 5!}\g_{mnpqr}(\l\g^{mnpqr}\l) \ .
$$
Hence the cohomology equation (\ref{cohomology}) gives
\bea\label{cohotwo}
\left\{\ba{c}\g_{mnpqr}^{\a\b}D_\a A_\b=0 \ ,\\ 
\delta A_\a=D_\a\Omega \ ,
\ea\right.
\eea
which are precisely the equations of motion and gauge invariance of ten--dimensional super Maxwell theory.

The pure spinors are nothing but the way to rewrite the ten--dimensional onshell super Yang--Mills theory as a cohomology problem. This is why they provide a natural way to describe superstring theory in superspace.

\section{Supergroups}
\label{appendix:supergroups}

In this Appendix we list the details of the superalgebras we need
to realize the various backgrounds in the text. We constructed our
superalgebras according to .

\subsection{Notations}

The superalgebra
satisfies the following commutation relations:
 $$
 [{\bf T}_{m},{\bf T}_{n}]=f^{p}_{mn}{\bf T}_{p}
 $$
 $$
 [{\bf T}_{m},{\bf Q}_{\a}]=F^{\b}_{m\a}{\bf Q}_{\b}
 $$
 $$
 \{{\bf Q}_{\a},{\bf Q}_{\b}\}=A^{m}_{\a\b}{\bf T}_{m}
 $$
where the ${\bf T}$'s are the bosonic (Grassman even) generators of a
Lie algebra and the ${\bf Q}$'s are the fermionic (Grassman odd)
elements. The indices are $m=1,...,d$ and $\a=1,...,D$. The
generators satisfy the following super-Jacobi identities:
 $$
f^{p}_{nr}f^{q}_{mp}+f^{p}_{rm}f^{q}_{np}+f^{p}_{mn}f^{q}_{rp}=0
 $$
 $$
F^{\g}_{n\a}F^{\d}_{m\g}-F^{\g}_{m\a}F^{\d}_{n\g}-f^{p}_{mn}F^{\d}_{p\a}=0
 $$
 $$
F^{\d}_{m\g}A^{n}_{\b\d}+F^{\d}_{m\b}A^{n}_{\g\d}-f^{n}_{mp}A^{p}_{\b\g}=0
 $$
 $$
A^{p}_{\b\g}F^{\d}_{p\a}+A^{p}_{\g\a}F^{\d}_{p\b}+A^{p}_{\a\b}F^{\d}_{p\g}=0
 $$

Generally we can define a bilinear form
 $$
 <{\bf X}_{ M },{\bf X}_{ N }>={\bf X}_{ M }{\bf X}_{ N }-(-1)^{g({\bf X}_{ M })g({\bf X}_{ N })}{\bf X}_{ N }{\bf X}_{ M }=C^{P}_{NM}{\bf X}_{P}
 $$
where ${\bf X}$ can be either ${\bf T}$ or ${\bf Q}$ and $P=1,...,d+D$ (say the
first $d$ are ${\bf T}$'s and the rest $D$ are ${\bf Q}$'s). $g({\bf X}_{M})$ is the
Grassmann grading, $g({\bf T})=0$ and $g({\bf Q})=1$ and $C^{P}_{NM}$ are the
structure constants. The latter satisfy the graded antisymmetry
property
 $$
 C^{P}_{NM}=-(-1)^{g({\bf X}_{ M }) g({\bf X}_{ N })}C^{P}_{MN}
 $$

We define the super-metric on the super-algebra as the supertrace
of the generators in the fundamental representation
 \bea\label{supermetric}
 g_{MN}=\Str {\bf X}_M {\bf X}_N,
 \eea
We can further define raising and lowering rules when the metric
acts on the structure constants
 $$
C_{ M  N P}\equiv g_{ M  S }C^{ S }_{ N P} \ .
 $$
 
For a semi-simple super Lie algebra ($|g_{ M  N }|\neq0$ and
the bosonic part $|h_{mn}|\neq0$) we can define a contravariant metric tensor
through the relation
 $$
g_{ M P}g^{PN }=\d^{ N }_{ M } \ .
 $$
The totally graded antisymmetric structure constants are defined as
$$
f_{ABC}=g_{AD}f^D_{BC} \ ,\qquad f^{AB}_C=f^B_{CD}g^{DA} \ .
$$

The Killing form is defined as the supertrace of the generators in
the adjoint representation
 $$
K_{ M  N }\equiv (-1)^{g({\bf X}_{P})}C^{ S }_{P M }C^{P}_{ S  N
}=(-1)^{g({\bf X}_{ M })g({\bf X}_{ N })}K_{ N  M }
 $$
(while on the (sub)Lie-algebra we define the metric
$K_{mn}=f^{p}_{mq}f^{q}_{np}$). Explicitly we have
 $$
K_{mn}=h_{mn}-F^{\b}_{m\a}F^{\a}_{n\b}=K_{nm}
 $$
 $$
K_{\a\b}=F^{\g}_{m\a}A^{m}_{\b\g}-F^{\g}_{m\b}A^{m}_{\a\g}=-K_{\b\a}
 $$
 $$
K_{m\a}=K_{\a m}=0
 $$
The Killing form is proportional to the supermetric up to the
second Casimir $C_2(G)$ of the supergroup, which is also called
the dual Coxeter number
 $$
 K_{MN}=-\,C_2(G)\,g_{MN}.
 $$

In the main text, we have computed the one-loop beta-functions in
the background field method. It turns out that the sums of one-loop
diagrams with fixed external lines are proportional to the Ricci
tensor $R_{MN}$ of the supergroup. The super Ricci tensor of a
supergroup is defined as
 \bea\label{superi}
 R_{MN}(G)=-{1\over 4} f^P_{MQ}f^{Q}_{NP}(-)^{g(X_Q)},
 \eea
and we immediately see that $R_{MN}=-K_{MN}$, in particular, we
can write it as
 $$
 R_{MN}(G)={C_2(G)\over4}g_{MN},
 $$

\subsection{Gamma matrices in AdS}
\label{appendix:gammahat}

In ten dimensions, we have $32\times32$ Dirac gamma matrices $\Gamma^m$. We use an off-diagonal representation which is well adapted to the $SO(1,4)\times SO(5)$ subgroup of the local Lorentz group of the $AdS_5\times S^5$ background
 \bea\label{gammama}
 \Gamma^a=
\sigma^a\otimes \Id_4\otimes\tau^1\ ,\qquad
 \Gamma^{a'}= \Id_4 \otimes \sigma^{a'}\otimes\tau^2 \ ,
\eea
where $\sigma^a$, for $a=0,\ldots,4$, are $4\times4$ dimensional $SO(1,4)$ gamma matrices and $\sigma^{a'}$, for $a'=5,\ldots,9$, are $4\times4$ dimensional $SO(5)$ gamma matrices and $\tau^i$ are the usual two dimensional Pauli matrices. In the text we used the $SO(1,9)$ Pauli matrices $(\gamma^m)^{\alpha\beta}$ and $(\gamma^m)_{\alpha\beta}$, that are $16\times16$ off-diagonal blocks of the Dirac matrices in (\ref{gammama}). Note that, because of the $\tau^2$ in (\ref{gammama}), we have that numerically
 \bea\label{gammaads}
 (\gamma^a)^{\alpha\beta}=\gamma^a_{\alpha\beta},\qquad (\gamma^{a'})^{\alpha\beta}=-\gamma^{a'}_{\alpha\beta} \ .
 \eea

The five-form flux in bispinor notation
$\gamma^{01234}_{\alpha\bh}\equiv \eta_{\alpha\bh}$ is numerically equal to the identity matrix and in our conventions it is antisymmetric in the spinor indices $\eta_{\alpha\ah}=-\eta_{\ah\alpha}$. Since, in the pure spinor formalism, left and right moving spinors transform with independent Lorentz parameters, we introduce two sets of gamma matrices, $(\gamma^m_{\alpha\beta},(\gamma^m)^{\alpha\beta})$ acting on the left moving spinors, and $(\gamma^m_{\ah\bh},(\gamma^m)^{\ah\bh})$ acting on the right moving spinors. In the $AdS_5\times S^5$ backgrounds, the two sets are related by the RR bispinor $\eta_{\alpha\ah}$ in the following way
 \bea\label{gammahat}
 \gamma^m_{\ah\bh}=\eta_{\alpha\ah}\eta_{\beta\bh}(\gamma^m)^{\alpha\beta} \ , \qquad (\gamma^m)^{\ah\bh}=\eta^{\alpha\ah}\eta^{\beta\bh}\gamma^m_{\alpha\beta} \ .
 \eea
Because of the decomposition (\ref{gammama}), we have that numerically $\gamma^a_{\ah\bh}=\gamma^a_{\alpha\beta}$ along the $AdS_5$ directions, however $\gamma^{a'}_{\ah\bh}=-\gamma^{a'}_{\alpha\beta}$ along the $S_5$ directions, and the same holds for the gamma matrices with upper spinor indices. We have also
 $$
 (\gamma_{ab})_\ah{}^\bh
= \eta_{\ah\delta} (\gamma_{ab})^\delta{}_\kappa
\eta^{\kappa\bh}
$$
and numerically we find $ (\gamma_{ab})_\ah{}^\bh=-(\gamma_{ab})_{\alpha\beta}$, with the same sign for all ten directions.

\subsection{$\mathfrak{psu}(2,2|4)$}
\label{appendix:psu}

We list the structure constants of the $\mathfrak{psu}(2,2|4)$ superalgebra in their ten-dimensional form. The notation is adapted to the $\ZZ_4$ automorphism of the superalgebra. The underlined index $\underline{a}=1,\ldots,10$ is a ten dimensional vector, while $a,a'=1,\ldots,5$ denote $AdS_5$ and $S^5$ vectors respectively. The spinor indices $\a,\ah=1,\ldots,16$ denote ten dimensional Majorana-Weyl spinors of the same chirality, for the type IIB superstring.
\bea\label{structconstpsu}
f_{\a\b}^{\underline{c}} =\g^{\underline{c}}_{\a\b},\quad
f_{\ah\bh}^{\underline{c}} =\g^{\underline{c}}_{\ah\bh},
\eea
$$
f_{\a \bh}^{[ef]}=
\half(\g^{ef})_\a{}^\g \eta_{\g\bh},\quad
f_{\a \bh}^{[e'f']}=-\half
(\g^{e'f'})_\a{}^\g \eta_{\g\bh},$$
$$f_{\a \underline{c}}^\bh
=-(\g_{\underline c})_{\a\b}
\eta^{\b\bh},\quad
f_{\ah \underline{c}}^\b =
(\g_{\underline c})_{\ah\bh} \eta^{\b\bh}=-\gamma_c^{\a\b}\eta_{\a\ah},$$
$$f_{c d}^{[ef]}= \d_c^{[e} \d_d^{f]},
\quad f_{c' d'}^{[e'f']}= -\d_{c'}^{[e'} \d_{d'}^{f']},$$
$$f_{[\underline{cd}][\underline{ef}]}^{[\underline{gh}]}=\half (
\eta_{\underline{ce}}\d_{\underline{d}}^{[\underline{g}}
\d_{\underline{f}}^{\underline{h}]}
-\eta_{\underline{cf}}\d_{\underline{d}}^{[\underline{g}}
\d_{\underline{e}}^{\underline{h}]}
+\eta_{\underline{df}}\d_{\underline{c}}^{[\underline{g}}
\d_{\underline{e}}^{\underline{h}]}
-\eta_{\underline{de}}\d_{\underline{c}}^{[\underline{g}}
\d_{\underline{f}}^{\underline{h}]})$$
$$f_{[\underline{cd}] \underline{e}}^{\underline{f}} = \eta_{\underline{e}
\underline{[c}} \d_{\underline d]}^{\underline{f}},\quad
f_{[\underline{cd}] \a}^{\b} = \half(\g_{\underline{cd}})_\a{}^\b,\quad
f_{[\underline{cd}] \ah}^{\bh} = \half(\g_{\underline{cd}})_\ah{}^\bh.$$
The components of the supermetric $\eta_{AB}$ are
$$
\str {\bf T}_{\underline {a}} {\bf T}_{\underline{b}}=\eta_{\underline{ab}} ,\qquad \str {\bf T}_{\underline{ab}}{\bf T}_{\underline{cd}}=\eta_{[\underline{ab}][\underline{cd}]} \ ,
$$
$$
\str {\bf T}_\alpha {\bf T}_\ah=-\str {\bf T}_\ah {\bf T}_\alpha= \eta_{\alpha\ah} \ ,
$$
where we defined $\eta_{\underline{a}\underline{b}}=(\eta_{ab},\delta_{a'b'})$ and $\eta_{[\underline{ab}][\underline{cd}]}=(\half\eta_{a[c}\eta_{d]b},-\half\delta_{a'[c'}\delta_{d']b'})$.

{\it Gamma matrix identities}

From the Jacobi identities for $\mathfrak{psu}(2,2|4)$ we derive some useful ten-dimensional gamma matrix identities
\bea\label{jacob}
(\gamma^a)_{\b\g}(\gamma^a)^{\d\rho}+{1\over4}(\gamma^{ef})_{(\g}{}^\rho(\gamma_{ef})_{\b)}^\d
-{1\over4}(\gamma^{e'f'})_{(\g}{}^\rho(\gamma_{e'f'})_{\b)}^\d=0 \ , \\
\gamma^{ef}\gamma_{\underline{mnpqr}}\gamma_{ef}-\gamma^{e'f'}\gamma_{\underline{mnpqr}}\gamma_{e'f'}=0 \ ,\\
\gamma^{\underline{a}}={1\over80}\left(\gamma^{ef}\gamma^{\underline{a}}\gamma_{ef}-
\gamma^{e'f'}\gamma^{\underline{a}}\gamma_{e'f'}\right) \ .
\eea

\subsection{Alternative form of the algebra}
\label{appendix:alternative}

For the computation of the massive string spectrum we used an alternative form of the $\mathfrak{psu}(2,2|4)$ superalgebra, adapted to the classical string configuration sitting at the center of AdS space. Let us first rewrite the ten-dimensional Pauli matrices $(\gamma^m)^{\a\b}$ and $(\gamma^m)_{\a\b}$ as
$$
(\gamma^0)^{\a\b}=-(\gamma^0)_{\a\b}=\left(\begin{array}{cc}1&0\\0&1\end{array}\right)\ , \qquad (\gamma^5)^{\a\b}=(\gamma^5)_{\a\b}=\left(\begin{array}{cc}1&0\\0&-1\end{array}\right) \ ,
$$
$$
(\gamma^i)^{\a\b}=(\gamma^i)_{\a\b}=\left(\begin{array}{cc}0&\sigma_{a\dot b}^i\\ \sigma_{\dot a b}^i&0\end{array}\right) \ ,
$$
where $\sigma^i$'s are the $SO(8)$ Pauli matrices, the vector index $i=1,2,3,4,6,7,8,9$ and the spinor indices $a,\dot a=1,\ldots,8$. The five-form flux reads
$$
\eta_{\a\ah}=(\gamma^{01234})_{\a\ah}=\left(\begin{array}{cc}\Pi_{a b}&0\\ 0&\Pi_{\ad\dot b}\end{array}\right) \ ,
$$
$$
\eta^{\a\ah}=-(\gamma^{01234})^{\a\ah}=\left(\begin{array}{cc}\Pi_{a b}&0\\ 0&\Pi_{\ad\dot b}\end{array}\right) \ ,
$$
where $\Pi$ is symmetric and $\Pi^2=1$ and
$$
\Pi_{a b}=-(\sigma^1\s^2\s^3\s^4)_{a b} \ ,\qquad 
\Pi_{\ad \bd}=-(\sigma^1\s^2\s^3\s^4)_{\ad \bd} \ .
$$
Note that we dropped the hat from the gamma matrices to avoid having nasty superscripts, but we will keep a hat in the grading-three generators appearing in the commutators below.

Let us give now the $\mathfrak{psu}(2,2|4)$ commutation relations, instead of writing the structure functions explicitly. The eight-dimensional index $i$ above gets split into $i=(A,J)$ where now $A,J=1,\ldots,4$ are four directions along AdS and along the five-sphere respectively.  We group the generators according to their gradings as
\bea
\cH_0=\{{\bf M}_{AB},{\bf M}_A,{\bf M}_{IJ},{\bf M}_I\} \ ,&\qquad& \cH_2=\{{\bf T},{\bf P}_A,{\bf J},{\bf P}_J\} \nonumber\\
\cH_1=\{{\bf Q}_a, {\bf Q}_{\aad}\} \ ,&\qquad& \cH_3=\{{\bf \widehat Q}_\aad,\widehat {\bf Q}_a\} \ ,
\eea
where the supertraces of the generators are
\bea
\str {\bf T}{\bf T}=-\str{\bf J}{\bf J}=-1 \ ,\nonumber\\ \str {\bf P}_A{\bf P}_B=\delta_{AB} \ ,\quad \str {\bf P}_I{\bf P}_J=\d_{IJ}\nonumber\\
\str {\bf Q_a}{\bf \widehat Q_b}=\Pi_{ab} \ ,\quad
\str {\bf Q_\aad}{\bf \widehat Q_\bbd}=\Pi_{\aad\bbd} \ ,\nonumber\\
\eea
and the commutation relations involving spinor indices are \bea
\{{\bf Q}_a,{\bf Q}_b\}=({\bf J}-{\bf T})\delta_{ab}=\{\widehat{\bf Q}_a,\widehat{\bf Q}_b\} \ ,\\
\{{\bf Q}_\ad,{\bf Q}_\bd\}=({\bf J}-{\bf T})\delta_{\ad\bd}=\{\widehat{\bf Q}_\ad,\widehat{\bf Q}_\bd\} \ ,\nonumber
\eea 
$$
\{{\bf Q}_a,{\bf Q}_\bd\}=\sigma^i_{a\bd}{\bf P}_i=\{\widehat{\bf Q}_a,\widehat{\bf Q}_\bd\} \ ,
$$
$$
\{{\bf Q}_a,\widehat{\bf Q}_b\}=-\half \sigma^{AB}_{ac}\Pi_{cb}{\bf M}_{AB}+\half \sigma^{IJ}_{ac}\Pi_{cb}{\bf M}_{IJ} \ ,
$$
$$
\{{\bf Q}_\ad,\widehat{\bf Q}_\bd\}=-\half \sigma^{AB}_{\ad\dot c}\Pi_{\dot c \bd}{\bf M}_{AB}+\half \sigma^{IJ}_{\ad\dot c}\Pi_{\dot c\bd}{\bf M}_{IJ} \ ,
$$
$$
\{{\bf Q}_a,\widehat{\bf Q}_\bd\}=\half \sigma^{A}_{a\ad}\Pi_{\ad \bd}{\bf M}_{A}-\half \sigma^{I}_{a\ad}\Pi_{\ad\bd}{\bf M}_{I} \ ,
$$
$$
[{\bf Q}_a,{\bf T}]=[{\bf Q}_a,{\bf J}]=-\half \Pi_{ab}\widehat{\bf Q}_b \ ,\qquad [\widehat{\bf Q}_a,{\bf T}]=[\widehat{\bf Q}_a,{\bf J}]=\half \Pi_{ab}{\bf Q}_b \ ,
$$
$$
[{\bf Q}_\ad,{\bf T}]=-[{\bf Q}_\ad,{\bf J}]=-\half \Pi_{\ad\bd}\widehat{\bf Q}_\bd \ ,\qquad [\widehat{\bf Q}_\ad,{\bf T}]=-[\widehat{\bf Q}_\ad,{\bf J}]=\half \Pi_{\ad\bd}{\bf Q}_\bd \ ,
$$
$$
[{\bf Q}_a,{\bf P}^i]=-\half\sigma^i_{a\ad}\Pi_{\ad\bd}\widehat{\bf Q}_\bd \ ,\qquad [\widehat{\bf Q}_a,{\bd P}^i]=\half\sigma^i_{a\ad}\Pi_{\ad\bd}{\bf Q}_\bd \ ,
$$
$$
[{\bf Q}_\ad,{\bf P}^i]=-\half\sigma^i_{\ad a}\Pi_{ab}\widehat{\bf Q}_b \ ,\qquad [\widehat{\bf Q}_\ad,{\bf P}^i]=\half\sigma^i_{\ad a}\Pi_{ab}{\bf Q}_b \ ,
$$
$$
[{\bf M}^A,{\bf Q}_a]=-\half \sigma^A_{a\bd} {\bf Q}_\bd\ ,\qquad [{\bf M}^A,{\bf Q}_\ad]=-\half \sigma^A_{\ad b} {\bf Q}_b \ ,
$$
$$
[{\bf M}^I,{\bf Q}_a]=\half \sigma^I_{a\bd}{\bf Q}_\bd \ ,\qquad
[{\bf M}^I,{\bf Q}_\ad]=\half \sigma^I_{\ad b}{\bf Q}_b \ ,
$$
$$
[{\bf M}^A,\widehat{\bf Q}_a]=-\half \sigma^A_{a\bd} \widehat{\bf Q}_\bd\ ,\qquad [{\bf M}^A,\widehat{\bf Q}_\ad]=-\half \sigma^A_{\ad b} \widehat{\bf Q}_b \ ,
$$
$$
[{\bf M}^I,\widehat{\bf Q}_a]=\half \sigma^I_{a\bd}\widehat {\bf Q}_\bd \ ,\qquad
[{\bf M}^I,\widehat {\bf Q}_\ad]=\half \sigma^I_{\ad b}\widehat {\bf Q}_b \ ,
$$
$$
[{\bf M}^{ij},{\bf Q}_a]=-\half \sigma^{ij}_{ab} {\bf Q}_b\ ,\qquad [{\bf M}^{ij},{\bf Q}_\ad]=-\half \sigma^{ij}_{\ad\bd} {\bf Q}_\bd \ ,
$$
$$
[{\bf M}^{ij},\widehat{\bf Q}_a]=-\half \sigma^{ij}_{ab} \widehat{\bf Q}_b\ ,\qquad [{\bf M}^{ij},\widehat{\bf Q}_\ad]=-\half \sigma^{ij}_{\ad\bd} \widehat{\bf Q}_\bd \ ,
$$

while the purely bosonic ones are the same as in (\ref{structconstpsu}).

\section{Some worldsheet results}
\label{appendix:worldsheet}

\subsection{Integrals}

The following useful integrals, are extensively used in Section~\ref{section:effectiveaction}
\bea\label{integrals}
\int {d^2\sigma \over (\bar \sigma-\bar w)(\sigma -z)}=&-2\pi\ln|z-w|^2 \ ,\\
\int {d^2\sigma \over (\bar \sigma-\bar w)^2(\sigma -z)}=&{2\pi \over\bar z -\bar w}\ ,\\
\int {d^2\sigma \over (\bar \sigma-\bar w)(\sigma -z)^2}=&-{2\pi \over z - w}\ ,\\
\int {d^2\sigma \over (\sigma-w)^2(\sigma -z)}=&-2\pi{\bar z-\bar w \over(z -w)^2}\ ,
\eea

\subsection{Action from BRST invariance}
\label{secpurespinorsigmamodelfromBRST}

Using the fact that $\langle A B \rangle \neq 0$ only for $A \in
\cH_r$ and $B \in \cH_{4 - r}$, $r = 0, \dots, 3$
\cite{Berkovits:1999zq}, the most general matter part which has a
global symmetry under left multiplication by elements of $G$ and
is invariant under the gauge symmetry $g \simeq g h$, where $h \in
H$, is
$$
  \int d^2 z \langle \alpha J_2 \bar J_2 + \beta J_1 \bar J_3 + \gamma
  J_3 \bar J_1 + \delta J_3 \bar d + \epsilon \bar J_1 d - f d \bar d
  \rangle \ ,
$$
where we used the Lie-algebra valued field $d$, $\bar d$ defined
by $d = d_\alpha \eta^{\alpha \hat \alpha} {\bf T}_{\hat \alpha}$, $\bar
d = \bar d_{\hat \alpha} \eta^{\alpha \hat \alpha} {\bf T}_\alpha$ and
$f$ is the RR-flux. While in flat background the $d$'s are
composite fields, in curved backgrounds they can be treated as
independent fields.

The pure spinor part includes the kinetic terms $\langle w \bar
\partial \lambda \rangle$ and $\langle \hat w \partial \hat \lambda
\rangle$ for the pure spinor $\beta \gamma$-systems. Since these
terms are not gauge invariant, they must be accompanied by terms
coupling the pure spinor gauge generators with the matter gauge
currents $\langle N \bar J_0 + \hat N J_0 \rangle$ in order to
compensate. The backgrounds we are considering also require
additional terms which must be gauge invariant under the pure
spinor gauge transformation of $w$ and $\hat w$ and hence must be
expressed in terms of the Lorentz currents and the ghost currents
$J_{gh} = \langle w \lambda \rangle$ and $\bar
J_{gh} = \langle \hat w \hat \lambda \rangle$. The additional term
required is $\langle N \hat N \rangle$.

Therefore the sigma-model is of the form
\bea\label{sigmaform}S  = & \int d^2 z \langle \alpha J_2 \bar J_2 + \beta J_1 \bar J_3 +
  \gamma J_3 \bar J_1 + \delta J_3 \bar d + \epsilon \bar J_1 d - f d
  \bar d + \\ \nonumber
  & {} + w \bar \partial \lambda + \hat w \partial \hat \lambda + N
  \bar J_0 + \hat N J_0 + a N \hat N \rangle
\eea
and the accompanying BRST-like operator is
\bea\label{equnintegratedBRST}
  Q_B = \oint \langle dz \lambda d - d\bar z \hat  \lambda \bar d
  \rangle \ .
\eea
By integrating out $d$ and $\bar d$ and redefining $\gamma \to
\gamma + {\epsilon \delta\over f}$ one gets
\bea S = \int d^2 z \langle \alpha J_2 \bar J_2 + \beta J_1 \bar J_3 +
  \gamma J_3 \bar J_1 + w \bar \partial \lambda + \hat w \partial \hat
  \lambda + N \bar J_0 + \hat N J_0 + a N \hat N \rangle \nonumber \\
 \label{actionone}  \eea
After rescaling $\lambda \to {\delta\over f} \lambda$, $w \to
{f\over \delta} w$, $\hat \lambda \to {\epsilon\over f} \hat
\lambda$, $\hat w \to { \epsilon\over f} \hat w$ the BRST currents
are $j_B = \langle \lambda d \rangle = \langle \lambda J_3
\rangle$ and $\bar j_B = \langle \hat \lambda \bar d \rangle =
\langle \hat \lambda \bar J_1 \rangle$. The BRST charge
(\ref{equnintegratedBRST}) now reads
$$
  Q_B = \oint \langle dz \lambda J_3 + d\bar z \hat \lambda \bar J_1
  \rangle \ .
$$

The coefficients of the various terms will be determined by
requiring the action to be BRST invariant, i.e. the BRST
currents are holomorphic and the corresponding charge is
nilpotent.

From the action (\ref{actionone}) we derive the following equations
of motion
 \bea  (\beta + \gamma) \bar \nabla J_3  = & (2 \beta - \alpha) [J_1,
  \bar J_2] + (\alpha + \beta - \gamma) [J_2, \bar J_1] + [N, \bar
  J_3] + [\hat N, J_3] \ , \nonumber \\\
  (\beta + \gamma) \nabla \bar J_1  = & (\alpha - 2 \beta) [J_2, \bar
  J_3] + (\gamma - \alpha - \beta) [J_3, \bar J_2] + [N, \bar J_1] +
  [\hat N, J_1] \ , \nonumber \\
  \bar \nabla \lambda  = & - a [\hat N, \lambda] \ , \quad
  \nabla \hat \lambda  = - a [N, \hat  \lambda] \ . \nonumber
\eea
After one takes into account that $[N, \lambda] = 0$ because of
the pure spinor condition $\{ \lambda, \lambda \} = 0$, requiring $\bar \partial j_B = 0$  leads
to the equations
$$
  \beta + \gamma = 1 \ , \quad \alpha = 2 \beta \ , \quad \alpha +
  \beta = \gamma \ , \quad a = -1 \ ,
$$
whose solution is
$$
  \alpha = \half\ , \quad \beta = {1\over4}\ , \quad
  \gamma  = {3\over4}\ , a = -1 \ .
$$
With this solution it is easy to check that
$$
  \partial \bar j_B = \langle [\hat \lambda, \hat N] J_1 \rangle \ ,
$$
which again vanishes because of the constraint $\{ \hat \lambda,
\hat \lambda \} = 0$. The proof of the nilpotence of the BRST
charge then follows.

Hence the pure spinor sigma-model is
$$
  S = \int d^2 z \left\langle \half J_2 \bar J_2 + {1\over4} J_1
  \bar J_3 + {3\over4}J_3 \bar J_1 + w \bar\partial  \lambda + \hat w
  \partial \hat \lambda + N \bar J_0 + \hat N J_0 -N \hat N
  \right\rangle
$$
for all dimensions and this of course matches the critical case as
well.

\subsection{Ghost number one cohomology}
\label{empti}

In this appendix we will study the BRST cohomology at ghost number one and prove two facts.  First we will prove the claim made in Section~\ref{quantumbrst} that the classical BRST cohomology of integrated
vertex operators $\int d^2z \langle {\cal O}^{(1)}_{z\bar
z}\rangle$ at ghost number one is empty. Seconly we will prove the claim in Section \ref{allloopconf} that the BRST cohomology for conserved charges at ghost number one is empty.

The most general ghost number one gauge-invariant integrated vertex operator
is
 \bea\label{ghostone}
 \langle {\cal O}_{z\bar z}^{(1)} \rangle=&
 \langle a_1\bar J_2[J_3,\e\bar\l]+\bar a_1 J_2[\bar
 J_1,\e\l]+a_2\bar J_2[J_1,\e\l]+\bar a_2 J_2[\bar J_3,\bar
 \l]\\
 &+a_3J_3[\bar N,\e\l]+\bar a_3\bar
 J_1[N,\e\bar\l]+a_4J_3\bar\nabla(\e\l)+\bar a_4\bar
 J_1\nabla(\e\bar\l)\rangle, \nonumber
 \eea
where we have written all the independent terms up to integrating
by parts on the Maurer-Cartan equations. The BRST variation of the  operator (\ref{ghostone})
consists of three different kind of terms
 $$
 \e'Q\langle {\cal O}_{z\bar z}^{(1)}
 \rangle=\Omega_1+\Omega_2+\Omega_3+{\rm e.o.m.'s}+{\rm pure gauge},
 $$
where we have omitted terms proportional to the ghost equations of
motion and to the gauge transformations
parameterized by $\{\l,\bar\l\}$. We have to impose that the three
terms $\Omega_i$ vanish separately. The first term is
 $$
 \Omega_1=(a_3+a_4-\bar a_3-\bar a_4)\langle \bar
 \nabla(\e\l)\nabla(\e'\bar\l)\rangle,
 $$
so we demand
 \bea\label{coedue}
 a_3+a_4=\bar a_3+\bar a_4.
 \eea
Imposing the vanishing of the second term
 \bea\label{vanise}
 \Omega_2=&\langle(a_1-\bar a_1+a_3+a_4-\bar a_3-\bar
 a_4)[J_3,\e\bar\l]+(a_2-\bar a_2)[\bar
 J_3,\e'\bar\l][J_1,\e\l]\\
 &+(a_1-\bar a_1+a_2-\bar
 a_2)[J_2,\e'\l][\bar J_2,\e\bar\l]\rangle,\nonumber
\eea
we find the additional conditions
\bea\label{coeone}
 a_1=\bar a_1,\quad a_2=\bar a_2.
 \eea
Finally, the third term reads
 \bea\label{termini}
 \Omega_3=&\langle(a_2+\bar a_1)[\bar J_1,\e'\l][J_1,\e\l]+ (a_1+\bar a_2)[J_3,\e\bar\l][\bar
 J_3,\e'\bar\l]\\
 &-a_4[ J_3,\e\l][\bar J_3,\e'\l]-\bar a_4[\bar
 J_1,\e\bar\l][J_1,\e'\bar\l]\rangle.\nonumber
\eea
If we expand on the supergroup generators, the first term on the
right hand side is proportional to $ \l^\a\l^\b
\langle[{\bf T}_\delta,{\bf T}_\a][{\bf T}_\rho,{\bf T}_\b]\rangle$, where we summarized
with a greek letter the various spinor properties of the
supercharges and the pure spinors in the various dimensions. Due to the supersymmetry algebra, the term inside
the supertrace is proportional to
$(\gamma^a)_{\delta\a}(\gamma_a)_{\b\rho}$. Since the product of two pure spinors is
proportional to the middle dimensional form
$\gamma^{a_1\ldots a_n}_{\a\b}$, therefore the terms in
(\ref{termini}) are all proportional to $ \gamma^a\gamma^{a_1\ldots a_n}\gamma_a$,
but this expression vanishes due to the
properties of the gamma matrix algebra. We
find that $\Omega_3=0$ identically. As a result, imposing that
$\int d^2z\langle{\cal O}_{z\bar z}^{(1)}\rangle$ is BRST closed
requires that the coefficients $a_i,\bar a_i$ satisfy (\ref{coedue})
and (\ref{coeone}).

On the other hand, the following operator
 \bea\label{folloop}
 \Sigma^{(0)}_{z\bar z}=&-a_2\bar J_2J_2+(a_1-a_2)\bar J_1 J_3+(a_3-\bar
 a_4+a_2-a_1)N\bar N\\&+(a_4+a_1-a_2)w\bar\nabla\l+(\bar
 a_4+a_1-a_2)\bar w\nabla\bar\l,\nonumber
 \eea
is such that
 $$
 Q\int d^2z\langle\Sigma^{(0)}_{z\bar z}\rangle=\int
 d^2z\langle{\cal O}^{(1)}_{z\bar z}\rangle,
 $$
so the cohomology for integrated vertex operators at ghost number
one is empty.

\subsection{Background field expansion}
\label{extraterms}

In this Appendix we collect the extra terms in the background field expansion of the action, that depend on two background currents but do not contribute to the beta function. It contains the terms in the action for the matter fields where the grading of the two background currents does not sum up to zero
\bea\label{secondnonzero}
 S_{II}'=&\int d^2z\langle \half [J_3,X_1][X_1,\bar J_3]+\half [J_1,X_3,X_3\bar J_1]\nonumber\\
&+{3\over8}[J_2,X_2][X_1,\bar J_3]-{3\over8}[J_2,X_1][X_2,\bar J_3]\nonumber\\&+{3\over8}[J_3,X_2][X_1,\bar J_2]+{5\over8}[J_3,X_1][X_2,\bar J_2]\nonumber\\
&+{3\over8}[J_2,X_3][X_2,\bar J_1]+{5\over8}[J_2,X_2][X_3,\bar J_1]\nonumber\\&+{3\over8}[J_1,X_3][X_2,\bar J_2]-{3\over8}[J_1,X_2][X_3,\bar J_2]\rangle \ .\nonumber
\eea
There are contribution from the action for ghosts as well
\bea\label{nonzerogh}
S_{gh}'=&\int \,\str\Bigl\{\half N_{(0)}\bigl(\bigl[[\bar J_1,X_2]+[\bar J_2,X_1],X_1\bigr]+\bigl[[\bar J_1,X_1]+[\bar J_3,X_3],X_2\bigr]\nonumber\\ 
&+\bigl[[\bar J_2,X_3]+[\bar J_3,X_2],X_3\bigr]\bigr)
+\half \hat N_{(0)}\bigl(\left[[J_1,X_2]+[ J_2,X_1],X_1\right]\nonumber\\&+\left[[J_1,X_1]+[ J_3,X_3],X_2\right]+\left[[J_2,X_3]+[ J_3,X_2],X_3\bigr]\right)\nonumber\\ &
+N_{(1)}\left([\bar J_1,X_3]+[\bar J_2,X_2]+[\bar J_3,X_1]\right)+\hat N_{(1)}\left([J_1,X_3]+[ J_2,X_2]+[ J_3,X_1]\right)\Bigr\}\nonumber
\eea
These term would give rise to counterterms in the effective action that would not be gauge invariant and one can show that such terms are all zero at one-loop.

\subsection{OPE}
\label{appendix:OPE}

Consider the OPE of $J_2^a$ with itself. We expand it in background currents plus quantum fluctuations as in (\ref{gradingex}) and evaluate the OPE (\ref{opescheme}). The first term on the r.h.s. is (\ref{examples}), which in this case reads
\bea\label{firstope}
\langle \partial X^{\underline{a}}(z)\partial X^{\underline{b}}(0)\rangle=\partial X^{\underline{a}}(z)\partial X^{\underline{b}}(0)-
\eea
$$
-{1\over 2\pi} \partial X^{\underline{a}}(z)\partial X^{\underline{b}}(0)\int d^2\sigma\str \half (N_{(0)} [\bar\partial X_2, X_2]+\hat N_{(0)} [\partial X_2,X_2] )
$$
$$
=-{\eta^{\underline{a}\underline{b}}\over z}+\half\int d^2\sigma\Bigl[N_{(0)}^{[ef]}(\sigma)\left(\eta^{ac}\eta^{bd}{\delta^{(2)}(z-\sigma)\over w-\sigma}+\eta^{ad}\eta^{bc}{\delta^{(2)}(w-\sigma)\over z-\sigma}\right)
$$
$$
+\hat N^{[ef]}_{(0)}(\sigma)\left(\eta^{ac}\eta^{bd}{1\over (z-\sigma)^2(w-\sigma)}+\eta^{ad}\eta^{bc}{1\over (w-\sigma)^2(z-\sigma)}\right)\Bigr] \ .
$$
Now we can expand $N(\sigma)$ and $\hat N(\sigma)$ around $\sigma=z$ or $\sigma=w$ and perform the integrals (\ref{integrals}), arriving at the final result (\ref{opes}).

\subsection{Conservation of $b$}
\label{appendix:bghost}

Let us rewrite (\ref{bghost}) in the convenient form
$$
b={\hat\lambda_\a\over (\lambda\hat\lambda)}G^\alpha
$$
$$
G^{\alpha}=-\half(\gamma_aJ_3)^\a J_2^a-\lambda^\a (wJ_1)+\half (\gamma^a w)^\a(\lambda\gamma_a J_1) \ .
$$
Since the $b$ antighost is a Lorentz scalar, we have that $\bar\partial b=\bar\nabla b$ and 
\bea\label{debi}
\bar\nabla b=\bar \nabla \left({\hat\lambda_\a\over (\lambda\hat\l)}\right) \left(-\half(\gamma_aJ_3)^\a J_2^a+\half (\gamma^a w)^\a(\lambda\gamma_a J_1)\right)+{\hat\lambda_\a\over (\lambda\hat\l)}\bar \nabla G^\a \ .
\eea
Let us look at the second term in (\ref{debi}). By using the equations of motion (\ref{purespinoreoms}) and the Maurer-Cartan equations (\ref{eqMaurerCartan}) we find
\bea
{\hat\lambda_\a\over (\lambda\hat\l)}\bar \nabla G^\a=&b_0+b_w+b_{w\hat w}+b_{ww} \ ,\nonumber\\
b_0=&{1\over2(\l\hat\l)}(\hat\l\gamma_a J_3)(J_3\gamma^a\bar J_3)\ ,\nonumber\\
b_w=&[w(1-K)\gamma_a]^\a\left( (J_3)_\a\bar J_2^a-(\bar J_3)_\a J_2^a\right) \nonumber\\
&+{1\over 2(\l\hat\l)}\left(\half (\bar J_3\gamma_{ab}\gamma_c\hat\l)N^{ab}J_2^c+2(\hat\l\gamma_a J_3) (\bar J_2)_bN^{ab}\right) \ ,\nonumber\\
b_{ww}=& -\half[w(1-K)\gamma_{ab}\bar J_1]]N^{ab} \ ,\nonumber
\eea
where the subscript indicates the number of $w$'s and $\wh$'s present in each term. The term $b_{w\hat w}$ is proportional to $\eta_{[ab][cd]}\hat N^{ab} (\hat\l\gamma^{cd})_\a$ which vanishes on the pure spinor constraint.

Let us show that $\bar \nabla b$ is BRST exact. Consider the operator ${\cal O}$ of weight $(2,1)$, defined as the coefficient of the single pole in the OPE of the hatted and unhatted antighosts  
\bea\label{method}
\hat b(z,\bar z) b(0)=\ldots+{{\cal O}_{zz\bar z }(0)\over \bar z}+\ldots \ .
\eea
Since $\{\hat Q,\hat b\}=\hat T$ and $\{\hat Q , b\}= 0$,
by applying $\hat Q$ to (\ref{method}) we conclude that
\bea\label{hatqb}
\{\hat Q,{\cal O}\}=\bar\nabla b \ .
\eea

Since the pure spinor superstring in $AdS_5\times S^5$ is an interacting two--dimensional conformal field theory, the OPE (\ref{method}) has to be computed in the worldsheet perturbation theory. We are only interested in the leading order result that we obtain using the tree level algebra of OPE's between the left invariant currents (\ref{opes}). One finds
\bea\label{omeg}
{\cal O}=A_0+A_w+A_{ww} \ ,
\eea
where
\bea\label{omegaa}
A_0=&{1\over2\l\lh}\left(\bar J_2^a(J_3\gamma_a KJ_3)-J_2^a (\bar J_3(1-K)\gamma_a J_3)\right)\cr&+{2\over(2\l\lh)^2 }\left(-\bar J_2^a(\l\bar J_3)(\lh\gamma_aJ_3)+J_2^a(\l\bar J_3)(\lh\gamma_a J_3)\right) \ ,
\eea
\bea\label{omegaw}
A_w=&{1\over (2\l\lh)^2}\Bigl(\half(\l\gamma_a\gamma_{ef}\gamma_b\lh)\bar J_2^aJ_2^b N^{ef}-2(\l\gamma_a\bar J_1)(\lh\gamma_b J_3)N^{ab}\Bigr)\cr&+{1\over 2\l\lh}\Bigl(-(w\gamma_a\gamma_b\l)J_2^a\bar J_2^b-(\l\gamma_a\bar J_1)[w(1-K)\gamma^a J_3]\cr&+[w\gamma^a(1-K)\bar J_3](\l\gamma_aJ_1)-2(\l\bar J_3)(wKJ_1)\Bigr) \ ,
\eea
\bea\label{omegaww}
A_{ww}=&\half[w(1-K)\gamma_{ef}(1-K)\wh]N^{ef} \ .
\eea

\clearpage


\bibliographystyle{elsarticle-num}
\bibliography{purespinor}

\end{document}